\documentclass{article}

\usepackage[english]{babel}

\usepackage[letterpaper,top=2cm,bottom=2cm,left=3cm,right=3cm,marginparwidth=1.75cm]{geometry}

\usepackage{amsthm}
\usepackage{amsmath,amssymb}
\usepackage{graphicx}
\usepackage[colorlinks=true, allcolors=blue]{hyperref}
\theoremstyle{definition}
\newtheorem{definition}{Definition}
\usepackage{braket}
\usepackage{algorithm}
\usepackage{algpseudocode}
\usepackage{booktabs}
\usepackage{multirow}
\usepackage{float}
\usepackage{tikz}
\usepackage[below]{placeins}
\usetikzlibrary{arrows.meta,positioning,fit}
\usepackage[most]{tcolorbox}

\newtcolorbox{highlightbox}{
    colback=gray!8,
    colframe=black!50,
    boxrule=0.6pt,
    arc=2pt,
    left=6pt,
    right=6pt,
    top=5pt,
    bottom=5pt
}

\theoremstyle{plain}
\newtheorem{theorem}{Theorem}

\theoremstyle{remark}

\title{Beyond Error Mitigation: Using Boosting for Noisy Quantum Machine Learning}
\author{Anastasiia Lukianenko, Yash J. Patel, Nurbolat Kenbayev and Vedran Dunjko}
\begin{document}
\maketitle

\begin{abstract}
Quantum error mitigation (QEM) can recover information about ideal quantum computations from noisy circuit executions, but this may require a large sampling overhead. Fundamental lower bounds show that recovering noiseless observables without using problem-specific information can require a superpolynomial number of samples, with exponential sampling overhead arising for broad classes of noisy circuits~\cite{QuekEtAl2024,Takagi2022}. However, additional information about a problem can reduce the number of samples needed to obtain the required answer. We investigate this possibility in supervised quantum machine learning, where labeled examples provide information about the function to be learned. This motivates asking whether the ideal reconstruction is necessary when the learning task depends only on a specific property of the circuit output. Binary classification provides such a setting, since correct prediction depends on the sign and margin of the classifier output rather than on accurate reconstruction of every noiseless expectation value.

We develop a theoretical and empirical framework for comparing boosting of noisy parameterized quantum circuits with probabilistic error cancellation (PEC), requiring both methods to reach the same learning target. For the specified circuit architectures under Pauli noise, we establish a common hypothesis-class complexity bound: matching empirical margin loss gives the same upper bound on expected classification risk. We derive resource criteria for the two approaches and evaluate so-called AdaBoost$\mathbb{R}$ and Real AdaBoost in finite-shot teacher--student simulations across circuit depth and physical noise strength. Under the same classification-accuracy target, boosting requires fewer circuit executions as depth and noise increase. Boosting also achieves higher test accuracy than the corresponding reference. For the supervised-learning problems studied, these results demonstrate a task-dependent resource advantage: exploiting information retained in noisy outputs can avoid the cost of ideal-output reconstruction without sacrificing observed classification performance.
\end{abstract}

\section{Introduction}

Parameterized quantum circuits (PQCs) are widely studied as trainable quantum models for supervised learning, e.g. binary classification~\cite{Preskill2018,BenedettiEtAl2019}. Classical inputs and trainable parameters enter a sequence of quantum gates, possibly through repeated data encoding, and measurements of the resulting state provide real-valued prediction scores. The functions represented by such circuits depend on the circuit architecture, depth, and encoding strategy~\cite{SchuldEtAl2021}.

On near-term hardware, the noisy circuit executions are used during training to estimate prediction scores, losses, gradients, and other quantities required by the optimizer. Noise can therefore alter both the final circuit evaluations and the optimization trajectory that produces the classifier, so the procedure intended to reproduce the corresponding ideal learning process must account for it.

These training quantities are typically computed from circuit expectation values. Quantum error mitigation (QEM) offers a way to estimate the corresponding noiseless expectation values from noisy circuit executions~\cite{CaiEtAl2023}. Methods include zero-noise extrapolation~\cite{Temme2017,LiBenjamin2017}, virtual distillation~\cite{HugginsEtAl2021}, and probabilistic error cancellation (PEC)~\cite{Temme2017,Endo2018}. PEC is particularly convenient for a resource comparison because it produces an unbiased estimator of an ideal expectation value through a quasiprobability decomposition of the inverse noise channel ~\cite{Temme2017,Endo2018}. The resulting variance amplification leads to a sampling overhead proportional to the squared quasiprobability norm, which can grow exponentially with the number of noisy circuit locations ~\cite{Temme2017,Endo2018,Takagi2022}. More generally, lower bounds show that recovery of noiseless observables can require superpolynomial sampling resources in broad noisy-circuit settings ~\cite{Takagi2022,QuekEtAl2024}.

These results motivate a different question. They concern recovery of the ideal circuit output, but a learning task need not require that entire quantity. In binary classification, for example, the relevant objective is to assign the correct label with sufficient margin.  Accurate reconstruction of every ideal expectation value is therefore a stronger requirement than correct classification. This task structure suggests that it may be possible to use information that remains available in noisy circuit outputs without first reconstructing the corresponding ideal outputs. Boosting provides a well-established method for obtaining accurate predictions by combining less accurate classifiers~\cite{Schapire1990,FreundSchapire1997}.

More specifically, we investigate this connection by treating noisy PQCs as so-called weak learners (as called in computational learning theory, and defined later) and combining them using classical boosting. We compare this strategy with PEC for the same binary-classification problem and ask how many quantum-circuit executions each method requires to attain a common learning criterion. The PEC strategy applies quasiprobability sampling to the expectation-value evaluations of a target-reaching noiseless reference procedure, whereas the boosting strategy trains and combines noisy PQCs directly under the physical noise model.

A resource comparison of this type is meaningful only when the ideal circuit output is not already efficiently available by classical computation. For efficiently classically simulable circuit families, such as Clifford circuits, the ideal quantity can instead be computed directly~\cite{AaronsonGottesman2004}.

To compare PEC and boosting on common statistical terms, we use a margin-based generalization bound.  Both strategies are required to attain the same empirical margin-loss target on the same training sample and are using the same reference hypothesis class. Consequently, attaining the common empirical target gives the same high-probability upper bound on expected classification risk, allowing the circuit resources required to reach that target to be compared directly.

The two methods incur qualitatively different resource costs. For PEC, the sampling cost is governed by the so-called squared quasiprobability overhead $\gamma^2$, which is determined by the circuit and noise model. For boosting, the required resource depends on the sequence of weak learners obtained under the adaptively reweighted training distributions. For Real AdaBoost~\cite{SchapireSinger1999}, we derive an upper bound on the number of boosting rounds needed to reach the common learning target, provided that every weak learner meets a performance requirement determined by the chosen margin. For AdaBoost$\mathbb{R}$~\cite{NockNielsen2007}, the corresponding condition also depends on the weights assigned to the weak learners during training. These requirements are stated precisely in Section~\ref{sec:boosting_resources}. The performance of the weak learners cannot, however, be determined from the physical noise strength alone.

Consequently, we could not determine the exact PEC--boosting resource comparison. We therefore perform finite-shot numerical experiments to measure the boosting resource directly and determine how the ordering of the two methods changes with circuit depth and physical noise strength. In these experiments, we compare AdaBoost$\mathbb{R}$ and Real AdaBoost with \emph{Projected PEC}, obtained by applying the quasiprobability sampling overhead to the target-reaching reference trajectory. We additionally simulate explicit stochastic \emph{Sampled PEC} in selected settings to test the projected resource model. Boosting requires fewer circuit executions than Projected PEC in $39$ of the $56$ reported comparisons. For the classification target, its advantage generally grows with physical noise strength and circuit depth, although the depth dependence is not monotonic. In all $36$ comparisons the median difference in test accuracy also favours boosting over the corresponding reference.

\paragraph{Related work.} Ensembles of noisy variational quantum classifiers have already been studied. Li \emph{et al.} combine shallow classifiers using bagging and adaptive boosting and compare their predictive performance with error-mitigated primitive classifiers~\cite{LiEtAl2024Ensemble}. Wang \emph{et al.} derive a supervised-learning guarantee for quantum AdaBoost and report settings in which a noisy boosted classifier outperforms a noiseless unboosted classifier \cite{WangEtAl2024QuantumAdaBoost}. Chen \emph{et al.} introduce AdaBoost.Q and experimentally demonstrate improved quantum-classifier performance on a programmable superconducting processor~\cite{ChenEtAl2025QuantumEnsemble}. In contrast, the focus here is the circuit-execution cost of reaching matched learning targets.

Our main contributions are:
\begin{enumerate}
    \item
    \textbf{A common margin-based framework for comparing PEC and boosting.}
    We place both strategies under the same margin-based generalization criterion, so reaching the same empirical margin-loss target implies the same high-probability upper bound on expected classification risk.

    \item
    \textbf{Resource criteria for reaching the common learning target.}
    We express the PEC cost through its quasiprobability sampling overhead and derive the corresponding boosting conditions in terms of the weak-learner edges and, for AdaBoost$\mathbb{R}$, the realized coefficient trajectory.

    \item
    \textbf{A finite-shot numerical comparison across circuit depth and
    physical noise.}
    We compare AdaBoost$\mathbb{R}$ and Real AdaBoost with Projected PEC under both the theoretical positive-margin criterion and a matched classification criterion, and validate the Projected-PEC resource model using Sampled PEC in selected settings.
\end{enumerate}
Throughout this work, we restrict the analysis to binary classification. The paper is organized as follows:
\begin{itemize}
    \item \textbf{Background (Section~\ref{sec:background}).}
    We introduce PEC, the two boosting algorithms, and the margin-based generalization tools used in the comparison.

    \item \textbf{Learning setup (Section~\ref{learning_setup}).}
    We specify the binary-classification problem, the common learning target, and the noiseless-reference resource budget. This fixes what the methods must achieve and which resources are compared.

    \item \textbf{PEC resources (Section~\ref{sec:pec_resources}).}
    We estimate the cost of training with PEC by accounting for the additional measurements needed to recover noiseless circuit outputs. We explain how this cost changes with circuit depth and noise strength.

    \item \textbf{Boosting resource bounds
    (Section~\ref{sec:boosting_resources}).}
    We derive sufficient conditions for reaching the empirical margin loss target for Real AdaBoost and AdaBoost$\mathbb R$.

    \item \textbf{Common theoretical comparison
    (Section~\ref{sec:comparative_analysis}).}
    We bring the preceding results together in our central theoretical contribution: a theorem comparing the resources needed by PEC and boosting to meet the same learning target. Under the stated assumptions, it gives conditions under which combining noisy predictors can require fewer circuit executions than the PEC budget derived earlier.

    \item \textbf{Numerical comparison (Section~\ref{sec:numerics}).}
    Finite-shot simulations compare resource costs to reach learning targets. We identify depth--noise regimes favouring each method, examine test performance and the tightness of the theoretical bounds.

    \item \textbf{Discussion and conclusions
    (Sections~\ref{sec:discussion} and~\ref{sec:conclusion}).}
    We summarize the observed resource advantages and identify open questions.
\end{itemize}

\section{Background}
\label{sec:background}
\subsection{PAC Learning Framework}
\label{pac_learning}

We briefly recall the Probably Approximately Correct (PAC) learning framework in the realizable setting \cite{Valiant1984,KearnsVazirani1994}, which formalizes the notions of weak and strong learnability underlying boosting theory.

Let $n$ denote a problem size (e.g. number of qubits, input dimension, circuit size), $\mathcal{X} \subseteq \mathbb{R}^{N}$ define the input space and $\mathcal{C}\subseteq \{c : \mathcal{X} \to \{-1,+1\}\}$ be a concept class consisting of binary labeling functions, where $N(n)$ is polynomially bounded in $n$. Training examples are generated by drawing inputs $x \sim \mathcal{D}$ independently from an unknown but fixed distribution $\mathcal{D}$ over $\mathcal{X}$ and assigning labels deterministically according to $y = c(x)$.

Given the above and a labeled sample $S = \{(x_i, c(x_i))\}_{i=1}^{m(n)}$, the goal of a learning algorithm is to output a hypothesis $h : \mathcal{X} \to \{-1,+1\}$ that generalizes well to new samples drawn from the same distribution $\mathcal{D}$.

\begin{definition}\textbf{(Efficient Weak Learner)}\label{def:weak_learner}
Let $\mathcal{C}$ be a concept class over $\mathcal{X} \subseteq \mathbb{R}^{N}$. A (possibly randomized) learning algorithm $\mathcal{A}$ is called an efficient weak learner for $\mathcal{C}$ if there exist polynomials $p$ and $q$ such that for every distribution $\mathcal{D}$ over $\mathcal{X}$, every target concept $c \in \mathcal{C}$, and every confidence parameter $\delta \in (0,1)$, the following holds:

Given $m = p\!\left(n, \frac{1}{\delta}\right)$ i.i.d.\ labeled samples \(S = \{(x_i, c(x_i))\}_{i=1}^m\) with $x_i \sim \mathcal{D}$ on input, the algorithm $\mathcal{A}$ runs in time polynomial in $n$ and $1/\delta$ and outputs, with probability at least $1-\delta$, a hypothesis $h$ such that
\begin{equation} \label{eq:weak_hypothesis}
\Pr_{x \sim \mathcal{D}}\!\left[ h(x) = c(x) \right]
\;\ge\;
\frac{1}{2} + \eta_{\mathrm{PAC}},
\qquad
\eta_{\mathrm{PAC}} = \frac{1}{q(n)}.
\end{equation}
The quantity $\eta_{\mathrm{PAC}}$ is the advantage in classification accuracy over random guessing.
\end{definition}

\begin{definition}\textbf{(Efficient Strong Learner)}\label{def:strong_learner}
Let $\mathcal{C}$ be a concept class over $\mathcal{X} \subseteq \mathbb{R}^N$. A (possibly randomized) learning algorithm $\mathcal{A}$ is called an efficient strong learner for $\mathcal{C}$ if there exist polynomial $p$ such that for every accuracy parameter $\epsilon \in (0,1)$, every confidence parameter $\delta \in (0,1)$, every distribution $\mathcal{D}$ over $\mathcal{X}$, and every target concept $c \in \mathcal{C}$, the following holds:

Given $m = p\!\left(n, \frac{1}{\epsilon}, \frac{1}{\delta}\right)$ i.i.d.\ labeled samples \(S = \{(x_i, c(x_i))\}_{i=1}^m\) with $x_i \sim \mathcal{D}$ on input, the algorithm $\mathcal{A}$ runs in time polynomial in $n$, $1/\epsilon$, and $1/\delta$ and outputs, with probability at least $1-\delta$, a hypothesis $h$ such that
\begin{equation}\label{eq:strong_hypothesis}
\Pr_{x \sim \mathcal{D}} \big[ h(x) = c(x) \big] \ge 1-\epsilon
\end{equation}
\end{definition}

The relevant distinction between definitions~\ref{def:weak_learner} and \ref{def:strong_learner} is that a strong learner can reach an arbitrarily small classification error $\epsilon$ (expected risk), whereas a weak learner is guaranteed only an inverse-polynomial advantage over random guessing. A foundational result showed that any concept class that is efficiently weakly learnable in the PAC model is also efficiently strongly learnable~\cite{Schapire1990}. Boosting algorithms give constructive procedures for combining repeated outputs of a weak learner into a strong classifier~\cite{FreundSchapire1997}.

These PAC definitions provide the weak-learning terminology used in our boosting analysis. In this work, however, the resource comparison is carried out for fixed classification tasks rather than by proving distribution-free efficient learnability of the PQC hypothesis class. In the theoretical analysis, we assume that the noisy PQC weak learners attain a positive edge under the successive reweighted training distributions. In the numerical study, we then train these weak learners explicitly and measure the circuit resources required to reach the same task-specific learning target as the reference procedure. PEC needs additional circuit executions to approximate the expectation values produced by the noiseless reference procedure. Boosting, on the other hand, uses circuit executions to train and combine noisy classifiers that satisfy the weak-learning condition. This distinction forms the basis of the resource comparison developed later in Section~\ref{sec:comparative_analysis}.

\subsection{Probabilistic Error Cancellation}
\label{subsec:pec}
Quantum error mitigation (QEM) techniques aim to reduce the effect of noise in near-term quantum devices without implementing full quantum error correction. Several QEM methods aim to estimate quantities associated with an ideal computation from noisy circuit executions. Here we focus on probabilistic error cancellation (PEC), which constructs unbiased estimators of ideal expectation values through quasiprobability sampling~\cite{Temme2017,Endo2018}.

Let $h_{\theta}^{\mathrm{noiseless}}(x)=\operatorname{tr}\!\left(O\,\rho_{\theta}^{\mathrm{noiseless}}(x)\right)$ denote the noiseless PQC expectation for an observable satisfying $\lVert O\rVert_{\infty}\leq1$. PEC represents the noiseless channel as a linear combination of implementable noisy channels $\{\mathcal O_\nu\}$:
\begin{equation}
\mathcal U(\cdot)
=
\sum_{\nu} \zeta_\nu\, \mathcal O_\nu(\cdot),
\qquad
\gamma := \sum_{\nu} |\zeta_\nu| \ge 1, \qquad \zeta_\nu \in \mathbb{R}
\label{eq:pec_decomposition}
\end{equation}
where $\gamma$ is the $\ell_1$ norm of the quasiprobability coefficients in the circuit-level decomposition. In a gatewise implementation of PEC, each noiseless gate channel $\mathcal{U}_g$ at location $g\in[G]$ is decomposed as
\[
\mathcal U_g
=
\sum_{\nu_g}\zeta_{g,\nu_g}\mathcal O_{g,\nu_g},
\qquad
\gamma_g=\sum_{\nu_g}|\zeta_{g,\nu_g}|.
\]
The complete sampled circuit is indexed by $\boldsymbol\nu=(\nu_1,\ldots,\nu_G)$, where $G$ is the number of noisy gate locations. Its quasiprobability norm is
\begin{equation}
\gamma=\prod_{g=1}^{G}\gamma_g.
\label{eq:gamma_product}
\end{equation}
A single PEC sample is constructed by drawing a complete circuit index $\boldsymbol{\nu}$ with probability
\[
p_{\boldsymbol{\nu}}
=
\frac{
\left|\zeta_{\boldsymbol{\nu}}\right|
}{\gamma},
\qquad
\zeta_{\boldsymbol{\nu}}
=
\prod_{g=1}^{G}\zeta_{g,\nu_g},
\]
executing the corresponding noisy circuit $\mathcal{O}_{\boldsymbol{\nu}}$, and measuring $O$ to obtain an outcome $Z_{\boldsymbol{\nu}}\in[-1,1]$. The reweighted outcome is
\begin{equation}
\widehat Z_{\mathrm{PEC}}
=
\gamma\,
\operatorname{sign}
\!\left(\zeta_{\boldsymbol{\nu}}\right)
Z_{\boldsymbol{\nu}}.
\label{eq:pec_single_shot}
\end{equation}
The empirical PEC estimator based on $N_{\mathrm{sh}}$ shots is
\begin{equation}
\widehat h_{\mathrm{PEC}}(x)
=
\frac{1}{N_{\mathrm{sh}}}
\sum_{k=1}^{N_{\mathrm{sh}}} \widehat Z_{\mathrm{PEC}}^{(k)}
\label{eq:pec_estimator}
\end{equation}
Provided that the quasiprobability decomposition is constructed from an accurate description of the physical noisy operations, the estimator is unbiased:
\begin{equation}
\mathbb E\!\left[\widehat h_{\mathrm{PEC}}(x)\right]
=
h_{\theta}^{\mathrm{noiseless}}(x)
\label{eq:pec_unbiased}
\end{equation}
Since $Z \in [-1,1]$, the reweighted PEC outcomes satisfy $\widehat Z_{\mathrm{PEC}} \in [-\gamma,\gamma]$. Thus, PEC enlarges the range of a single measurement outcome by a factor $\gamma$ and the variance scale by $\gamma^2$:
\begin{equation}
\mathbb{E}\!\left[
\left(\widehat Z_{\mathrm{PEC}}\right)^2
\right]
\leq \gamma^2,
\qquad
\operatorname{Var}
\!\left(
\widehat Z_{\mathrm{PEC}}
\right)
\leq \gamma^2.
\label{eq:pec_variance_bound}
\end{equation}

Because the PEC samples are independent and bounded, Hoeffding's inequality applies. For any $\epsilon>0$ and confidence parameter $\delta_{sh}\in(0,1)$,
\begin{equation}
\Pr\!\left(
\bigl|\widehat h_{\mathrm{PEC}}(x) - h_{\theta}^{\mathrm{noiseless}}(x)\bigr|
\ge \epsilon
\right)
\le
2\exp\!\left(
-\frac{2N_{\mathrm{sh}}\epsilon^2}{(2\gamma)^2}
\right)
\label{eq:hoeffding_pec}
\end{equation}
It therefore suffices to choose
\begin{equation}
N_{\mathrm{sh}}^{\mathrm{PEC}}
\ge
\frac{2\gamma^2}{\epsilon^2}
\log\frac{2}{\delta_{sh}}.
\label{eq:shots_pec}
\end{equation}
shots to estimate each expectation value within accuracy $\epsilon$ with confidence $1-\delta_{sh}$.

The number of circuit executions required to achieve fixed additive precision and confidence therefore scales as $\gamma^2$. Under the gate-wise construction,
\begin{equation}
\gamma^2=\prod_{g=1}^{G}\gamma_g^2.
\label{eq:gamma_squared_product}
\end{equation}
Consequently, if the gate-level norms are bounded below by a constant greater than one, the PEC sampling cost grows exponentially with the number of noisy gates and hence, for fixed-width layered circuits, with circuit depth ~\cite{Temme2017,Endo2018,Takagi2022}.

Importantly, the $\gamma^2$ factor is the sampling overhead incurred when the ideal expectation value is estimated using PEC. In some special circuit families, the same ideal expectation value can instead be computed efficiently by a different method. For example, if the circuit is efficiently classically simulable, one may compute the ideal expectation value directly and bypass PEC altogether. This does not make PEC itself require fewer samples; it removes the need to use PEC for that problem.

In the context of supervised learning with PQCs, PEC can therefore be interpreted as using noisy circuit executions to estimate the expectation values of the ideal PQC hypothesis class, with a multiplicative shot overhead determined by $\gamma^2$. This perspective will serve as the reference point for our subsequent comparison with boosting-based strategies.

\subsection{Boosting Framework}
\label{sec:boosting_framework}

Let $\mathcal S=\{(x_i,y_i)\}_{i=1}^m$ be a fixed training dataset drawn i.i.d.\ from a distribution $\mathcal D$ over $\mathcal X\subseteq\mathbb R^N$. Boosting constructs an ensemble predictor by repeatedly invoking a weak learner on reweighted versions of the fixed training sample. According to Section~\ref{pac_learning}, we use \emph{weak learner} for the learning procedure and \emph{weak hypothesis} for the function returned by that procedure. We first introduce the notation and ingredients common to both boosting methods considered in this work; their algorithm-specific coefficients and reweighting rules are defined separately below.

The procedure operates in rounds $t=1,2,\ldots,T$. At each round, it maintains a distribution $D^t$ over the indices of the fixed training sample $\mathcal S$. Initially,
\begin{equation}
D^1(i)
=
\frac{1}{m},
\qquad
i=1,\ldots,m.
\label{eq:D1}
\end{equation}

As boosting proceeds, the distribution is updated so that training examples that are handled poorly by the current ensemble receive greater importance in subsequent rounds. The distribution $D^t$ is therefore a discrete reweighting of the fixed sample $\mathcal S$ and should not be confused with the underlying data-generating distribution $\mathcal D$, which remains fixed throughout boosting.

At round $t$, the weak learner is trained according to the distribution $D^t$ and returns a weak hypothesis
\[
h_t:\mathcal X\to[-1,1].
\]
In our setting, $h_t$ is realized by a noisy parameterized quantum circuit (PQC) of fixed ansatz depth trained for a bounded number of optimization epochs. For fixed trained parameters, the circuit defines the mean prediction
\begin{equation}
h_t(x)
:=
\mathbb{E}_{\mathrm{shots}}\!\left[
\widehat h_t(x)
\right]
\in[-1,1],
\label{eq:noisy_pqc_mean_hypothesis}
\end{equation}
where $\widehat h_t(x)$ is the corresponding finite-shot estimate. The bound $h_t(x)\in[-1,1]$ follows from measuring an observable $O$ satisfying $\lVert O\rVert_\infty\leq1$.

The boosting analysis below is formulated in terms of the deterministic mean hypotheses $h_t$, whereas $\widehat h_t$ is random because it is obtained from a finite number of measurement shots. The finite-shot accuracy required to estimate the quantities entering the boosting procedure is treated separately in the resource analysis.

Following the confidence-rated prediction formalism~\cite{SchapireSinger1999}, a real-valued weak hypothesis $h_t:\mathcal X\to[-1,1]$ retains both the predicted class and the magnitude of its prediction: $\operatorname{sign}(h_t(x))$ determines the predicted label, while $|h_t(x)|$ is the confidence rating used by the boosting rule. Thus, the real-valued hypothesis is combined directly rather than being thresholded to $\{-1,+1\}$ before constructing the ensemble.

We define the empirical edge of $h_t$ under the current distribution $D^t$ as
\begin{equation}
r_t
:=
\sum_{i=1}^{m}
D^t(i)\,y_i h_t(x_i).
\label{eq:edge_condition}
\end{equation}
For each training example, $y_i h_t(x_i)$ is positive when the predicted sign agrees with the label and negative when it does not. Hence $r_t$ is the $D^t$-weighted average of these signed prediction values.

To connect this definition with the PAC weak-learning edge introduced in Section~\ref{pac_learning}, consider the special case in which the weak hypothesis is binary-valued on the training sample, $h_t(x_i)\in\{-1,+1\}$. Define its weighted classification error by
\begin{equation}
\varepsilon_t
:=
\sum_{i=1}^{m}
D^t(i)\,
\mathbb{I}\!\left\{
h_t(x_i)\neq y_i
\right\},
\end{equation}
where $\mathbb{I}\{A\}=1$ when $A$ is true and $0$ otherwise. In this binary case, $y_i h_t(x_i)=+1$ on correctly classified examples and $y_i h_t(x_i)=-1$ on incorrectly classified examples. Therefore,
\begin{equation}
r_t
=
(1-\varepsilon_t)-\varepsilon_t
=
1-2\varepsilon_t.
\end{equation}
Consequently,
\[
\frac{r_t}{2}=\frac{1}{2}-\varepsilon_t,
\]
which is the fixed-sample, round-dependent analogue of the PAC accuracy advantage $\eta_{\mathrm{PAC}}$. In particular, $r_t>0$ is equivalent to weighted classification accuracy greater than $1/2$ in this binary special case.

For the real-valued hypotheses used in this work, $h_t(x_i)\in[-1,1]$, this equivalence no longer holds because $r_t$ depends also on the magnitudes of the predictions. We therefore work directly with the real-valued edge in Eq.~\eqref{eq:edge_condition}.

Both boosting methods below use the same sign convention. If a trained weak hypothesis has $r_t<0$, we replace $h_t$ by $-h_t$, changing its edge from $r_t$ to $-r_t>0$. For the noisy-PQC hypotheses considered here, this operation is purely classical post-processing of already evaluated predictions and requires neither retraining nor additional circuit executions. The convergence analyses below concern the positive-edge regime $0<r_t<1$.

After obtaining the weak hypothesis, the boosting algorithm assigns it a coefficient and updates the distribution over the training examples. These steps are algorithm dependent: Real AdaBoost uses the confidence-rated hypothesis directly, whereas AdaBoost$\mathbb{R}$ first normalizes the weak hypothesis on the training sample. Their respective coefficients and reweighting rules are defined in the following subsections.

After $T$ rounds, the ensemble score and associated classifier are
\begin{equation}
F_T(x)
=
\sum_{t=1}^{T}
\alpha_t h_t(x),
\qquad
H_T(x)
=
\operatorname{sign}\!\left(F_T(x)\right).
\label{eq:ensemble_classifier}
\end{equation}

We now specialize this common framework to two boosting algorithms for real-valued weak hypotheses: Real AdaBoost~\cite{SchapireSinger1999} and AdaBoost$\mathbb{R}$~\cite{NockNielsen2007}. Both are compatible with bounded PQC prediction functions $h_t:\mathcal X\to[-1,1]$, which arise naturally as expectation values of bounded observables.

\paragraph{Real AdaBoost with Confidence-Rated Predictions.}
\label{subsec:real_adaboost_schapire_singer}

We first consider Real AdaBoost with confidence-rated predictions in the formulation~\cite{SchapireSinger1999}. In the general boosting framework above, the coefficient $\alpha_t$ assigned to the new weak hypothesis and the update of the sample weights were left unspecified. Real AdaBoost chooses these quantities so that adding the new hypothesis reduces the exponential loss of the ensemble on the training sample.

Let
\begin{equation}
F_{t-1}(x)
=
\sum_{s=1}^{t-1}\alpha_s h_s(x)
\end{equation}
denote the ensemble score before round $t$, with $F_0(x)=0$. Real AdaBoost measures its training performance through the exponential potential
\begin{equation}
\Phi_{t-1}
:=
\sum_{i=1}^{m}
\exp\!\left(
-y_iF_{t-1}(x_i)
\right).
\label{eq:real_ab_exponential_potential}
\end{equation}
An example with a negative or small ensemble margin $y_iF_{t-1}(x_i)$ contributes more strongly to $\Phi_{t-1}$ than an example that is classified correctly with high confidence.

The distribution used to train the next weak hypothesis is obtained by normalizing these contributions:
\begin{equation}
D^t(i)
=
\frac{
\exp\!\left(
-y_iF_{t-1}(x_i)
\right)
}{
\Phi_{t-1}
}.
\label{eq:real_ab_weights_from_potential}
\end{equation}
For $F_0=0$, one has $\Phi_0=m$ and therefore $D^1(i)=1/m$.

At round $t$, the weak learner returns a confidence-rated hypothesis $h_t:\mathcal{X}\to[-1,1]$. Let $r_t$ denote its empirical edge under $D^t$, as defined in Eq.~\eqref{eq:edge_condition}. Suppose that $h_t$ is added to the ensemble with a candidate coefficient $\alpha$. The resulting exponential potential is
\begin{equation}
\Phi_t(\alpha)
:=
\sum_{i=1}^{m}
\exp\!\left(
-y_i\bigl(F_{t-1}(x_i)+\alpha h_t(x_i)\bigr)
\right)
=
\Phi_{t-1}
\sum_{i=1}^{m}
D^t(i)
\exp\!\left(
-\alpha y_i h_t(x_i)
\right).
\label{eq:real_ab_potential_factorization}
\end{equation}
This motivates the definition
\begin{equation}
Z_t(\alpha)
:=
\sum_{i=1}^{m}
D^t(i)
\exp\!\left(
-\alpha y_i h_t(x_i)
\right).
\label{eq:Zt_def}
\end{equation}
Equation~\eqref{eq:real_ab_potential_factorization} gives
\begin{equation}
Z_t(\alpha)
=
\frac{\Phi_t(\alpha)}{\Phi_{t-1}}.
\label{eq:Zt_potential_ratio}
\end{equation}
Thus, $Z_t(\alpha)$ is the factor by which the exponential training potential changes after adding $\alpha h_t$ to the current ensemble. In particular, $Z_t(\alpha)<1$ means that the new hypothesis decreases the exponential training potential.

For the positive-edge regime $0<r_t<1$, used in the convergence analysis, \cite{SchapireSinger1999} show that
\begin{equation}
Z_t(\alpha)
\leq
\frac{1-r_t}{2}e^{\alpha}
+
\frac{1+r_t}{2}e^{-\alpha}.
\label{eq:Zt_upper}
\end{equation}
Although $r_t$ is computed from the current weights $D^t$, the labels, and the weak-hypothesis values $h_t(x_i)$, the bound above depends on these quantities only through $r_t$. The right-hand side is minimized by
\begin{equation}
\alpha_t
=
\frac{1}{2}
\log\!\left(
\frac{1+r_t}{1-r_t}
\right),
\qquad
0<r_t<1.
\label{eq:alpha_closed_form}
\end{equation}
Since $0<r_t<1$, the denominator $1-r_t$ is positive and
\[
\frac{1+r_t}{1-r_t}>1,
\]
so $\alpha_t>0$. Substituting \eqref{eq:alpha_closed_form} into \eqref{eq:Zt_upper} yields
\begin{equation}
Z_t(\alpha_t)
\leq
\sqrt{1-r_t^2}
<
1.
\label{eq:Zt_edge_bound}
\end{equation}
Hence, every accepted positive-edge hypothesis reduces the exponential training potential by a factor of at most $\sqrt{1-r_t^2}$.

For binary-valued hypotheses $h_t(x_i)\in\{-1,+1\}$, the inequality in \eqref{eq:Zt_upper} is attained exactly, and therefore
\begin{equation}
Z_t(\alpha_t)
=
\sqrt{1-r_t^2}.
\label{z_t_real}
\end{equation}

Using the coefficient \eqref{eq:alpha_closed_form}, the sample weights are updated according to
\begin{equation}
D^{t+1}(i)
=
\frac{
D^t(i)
\exp\!\left(
-\alpha_t y_i h_t(x_i)
\right)
}{
Z_t(\alpha_t)
}.
\label{eq:real_ab_update}
\end{equation}
To interpret the update, define the signed score $s_{t,i}:=y_i h_t(x_i)$. Equation~\eqref{eq:real_ab_update} gives
\begin{equation}
\frac{D^{t+1}(i)}{D^t(i)}
=
\frac{
\exp(-\alpha_t s_{t,i})
}{
Z_t(\alpha_t)
}.
\label{eq:real_ab_relative_weight_factor}
\end{equation}

Since $\alpha_t>0$, the factor $\exp(-\alpha_t s_{t,i})$ decreases as the signed score increases. Therefore, examples with smaller signed scores are emphasized relative to examples with larger signed scores.

More precisely, define
\begin{equation}
\vartheta_t
:=
-\frac{1}{\alpha_t}
\log Z_t(\alpha_t).
\label{eq:real_ab_weight_threshold}
\end{equation}
Because $Z_t(\alpha_t)<1$, one has $\vartheta_t>0$. It follows from \eqref{eq:real_ab_relative_weight_factor} that
\begin{equation}
\begin{cases}
D^{t+1}(i)>D^t(i),
&
s_{t,i}<\vartheta_t,
\\[1mm]
D^{t+1}(i)=D^t(i),
&
s_{t,i}=\vartheta_t,
\\[1mm]
D^{t+1}(i)<D^t(i),
&
s_{t,i}>\vartheta_t.
\end{cases}
\label{eq:real_ab_weight_change_cases}
\end{equation}
Thus, every wrong-sign prediction, for which $s_{t,i}<0$, receives greater weight. A correctly signed but low-confidence prediction may also receive greater weight when its signed score lies below $\vartheta_t$, whereas a sufficiently confident correct prediction receives less weight.

The complete procedure is summarized in Algorithm~\ref{alg:real_adaboost}. The update rule below is stated for the regime $0<r_t<1$ used in the theoretical analysis. Numerical handling of the endpoint $r_t=1$ is specified separately in the implementation protocol.
\begin{algorithm}[H]
\caption{Real AdaBoost with confidence-rated predictions}
\label{alg:real_adaboost}
\textbf{Input:}
weak learner $\mathcal{A}$,
sample $\mathcal{S}=\{(x_i,y_i)\}_{i=1}^{m}$,
number of rounds $T$
\\
\textbf{Initialize:}
$D^1(i)=1/m$ for every $i\in[m]$ and $F_0(x)=0$.
\begin{algorithmic}[1]
\For{$t=1$ to $T$}
    \State Invoke $\mathcal{A}$ with respect to $D^t$ to obtain
    $h_t:\mathcal{X}\to[-1,1]$.
    \State Compute the raw edge
    \[
    r_t^{\mathrm{raw}}
    =
    \sum_{i=1}^{m}
    D^t(i)y_i h_t(x_i).
    \]
    \If{$r_t^{\mathrm{raw}}<0$}
        \State Replace $h_t=-h_t$.
    \EndIf
    \State Set
    \[
    r_t
    =
    \left|r_t^{\mathrm{raw}}\right|.
    \]
    \State Set
    \[
    \alpha_t
    =
    \frac{1}{2}
    \log\!\left(
    \frac{1+r_t}{1-r_t}
    \right).
    \]
    \State Update, for every $i\in[m]$,
    \[
    D^{t+1}(i)
    =
    \frac{
    D^t(i)
    \exp\!\left(
    -\alpha_t y_i h_t(x_i)
    \right)
    }{
    Z_t(\alpha_t)
    }.
    \]
    \State Set
    \[
    F_t(x)
    =
    F_{t-1}(x)
    +
    \alpha_t h_t(x).
    \]
\EndFor
\State \textbf{Return}
\[
F_T(x)
=
\sum_{t=1}^{T}\alpha_t h_t(x),
\qquad
H_T(x)
=
\operatorname{sign}\!\left(F_T(x)\right).
\]
\end{algorithmic}
\end{algorithm}

\paragraph{AdaBoost$\mathbb{R}$ with Rational Reweighting.}
\label{subsec:adaboostr}

We next consider the rational-update boosting algorithm denoted by AdaBoost$\mathbb{R}$ throughout this work. AdaBoost$\mathbb{R}$ first normalizes the real-valued weak hypothesis on the training sample and then updates the weights using a rational function of its signed scores.

At round $t$, the weak learner returns a real-valued hypothesis $h_t:\mathcal{X}\to[-1,1]$. To separate the direction of its predictions from their overall scale on the training sample, define
\begin{equation}
h_t^\star
:=
\max_{i\in[m]}
\left|h_t(x_i)\right|.
\label{eq:adaboostr_scale}
\end{equation}
If $h_t^\star=0$, the hypothesis assigns zero score to every training example and is therefore rejected. For $h_t^\star>0$, define the normalized score
\begin{equation}
\overline h_t(x)
:=
\frac{h_t(x)}{h_t^\star}.
\label{eq:adaboostr_normalized_hypothesis}
\end{equation}
By construction,
\[
\left|\overline h_t(x_i)\right|
\leq 1
\qquad
\text{for every }i\in[m].
\]
The normalization does not change the predicted labels on the training sample, since $\operatorname{sign}\!\left(\overline h_t(x_i)\right)=\operatorname{sign}\!\left(h_t(x_i)\right)$, but it makes the normalized scores, the normalized edge, and the resulting ensemble increment invariant under a positive rescaling of $h_t$ on the training sample. The normalized empirical edge is
\begin{equation}
r_t
:=
\sum_{i=1}^{m}
D^t(i)\,y_i\overline h_t(x_i)
\in[-1,1].
\label{eq:adaboostr_edge_norm}
\end{equation}
As in the general framework, $r_t$ measures the weighted alignment between the normalized scores and the labels. For the positive-edge regime $0<r_t<1$ considered in the convergence analysis, AdaBoost$\mathbb{R}$ assigns the normalized hypothesis the coefficient
\begin{equation}
\beta_t
:=
\frac{1}{2}
\log\!\left(
\frac{1+r_t}{1-r_t}
\right),
\qquad
0<r_t<1.
\label{eq:adaboostr_beta}
\end{equation}
Since $r_t>0$, one has $\beta_t>0$. The coefficient multiplying the original, unnormalized hypothesis is therefore
\begin{equation}
\alpha_t
:=
\frac{\beta_t}{h_t^\star}
=
\frac{1}{2h_t^\star}
\log\!\left(
\frac{1+r_t}{1-r_t}
\right).
\label{eq:adaboostr_alpha_fixed}
\end{equation}
Consequently,
\begin{equation}
\alpha_t h_t(x)
=
\beta_t\overline h_t(x).
\label{eq:adaboostr_scaled_increment}
\end{equation}
Thus, the contribution of the new hypothesis to the ensemble is determined by its normalized score and its normalized edge, rather than by the arbitrary overall magnitude of $h_t$ on the training sample.

The distribution over the training examples is updated according to
\begin{equation}
D^{t+1}(i)
=
D^t(i)
\frac{
1-r_t y_i\overline h_t(x_i)
}{
1-r_t^2
},
\qquad
i\in[m].
\label{eq:adaboostr_weight_fixed}
\end{equation}
The denominator $1-r_t^2$ normalizes the updated weights. Indeed, using the definition of $r_t$,
\begin{equation}
\sum_{i=1}^{m}
D^t(i)
\left(
1-r_t y_i\overline h_t(x_i)
\right)
=
1-r_t
\sum_{i=1}^{m}
D^t(i)y_i\overline h_t(x_i)
=
1-r_t^2.
\label{eq:adaboostr_normalization}
\end{equation}
Moreover, since
$|y_i\overline h_t(x_i)|\leq1$ and $0<r_t<1$,
\[
1-r_t y_i\overline h_t(x_i)
\geq
1-r_t
>
0.
\]
Thus, the updated weights are nonnegative and sum to one.

To determine how the weight of an individual example changes, define its signed score by $s_{t,i}:=y_i\overline h_t(x_i)$. Then
\begin{equation}
\frac{D^{t+1}(i)}{D^t(i)}-1
=
\frac{
r_t\left(r_t-s_{t,i}\right)
}{
1-r_t^2
}.
\label{eq:adaboostr_relative_weight_change}
\end{equation}
Because $0<r_t<1$, the weight of example $i$ increases if $s_{t,i}<r_t$, remains unchanged if $s_{t,i}=r_t$, and decreases if $s_{t,i}>r_t$. In particular, every wrong-sign example satisfies $s_{t,i}<0<r_t$ and therefore receives greater weight. A correctly classified example may also receive greater weight when its positive signed score lies below the round edge, whereas examples with signed scores above the round edge receive less weight.

When the weak hypothesis is binary-valued, $h_t(x_i)\in\{-1,+1\}$, one has
\[
h_t^\star=1,
\qquad
\overline h_t=h_t,
\qquad
r_t=1-2\varepsilon_t,
\]
where $\varepsilon_t$ is the weighted classification error. The coefficient then becomes
\begin{equation}
\alpha_t
=
\frac{1}{2}
\log\!\left(
\frac{1-\varepsilon_t}{\varepsilon_t}
\right),
\end{equation}
and the rational update \eqref{eq:adaboostr_weight_fixed} reduces to the normalized weight update of discrete AdaBoost. Thus, AdaBoost$\mathbb{R}$ differs from discrete AdaBoost only when the weak hypotheses take intermediate confidence-rated values in $(-1,1)$~\cite{FreundSchapire1997,NockNielsen2007}. The complete procedure is summarized in Algorithm~\ref{alg:adaboostr}.
\begin{algorithm}[H]
\caption{AdaBoost$\mathbb{R}$ with rational reweighting}
\label{alg:adaboostr}
\textbf{Input:}
weak learner $\mathcal{A}$, sample $\mathcal{S}=\{(x_i,y_i)\}_{i=1}^{m}$, number of rounds $T$
\\
\textbf{Initialize:}
$D^1(i)=1/m$ for every $i\in[m]$ and $F_0(x)=0$.
\begin{algorithmic}[1]
\For{$t=1$ to $T$}
    \State Invoke $\mathcal{A}$ with respect to $D^t$ to obtain
    $h_t:\mathcal{X}\to[-1,1]$.
    \State Compute
    \[
    h_t^\star
    =
    \max_{i\in[m]}
    |h_t(x_i)|.
    \]
    \If{$h_t^\star=0$}
    \State Reject the hypothesis and continue to the next weak-learner fit.
    \EndIf
    \State Define
    \[
    \overline h_t(x)
    =
    \frac{h_t(x)}{h_t^\star}.
    \]
    \State Compute the raw normalized edge
    \[
    r_t^{\mathrm{raw}}
    =
    \sum_{i=1}^{m}
    D^t(i)y_i\overline h_t(x_i).
    \]
    \If{$r_t^{\mathrm{raw}}<0$}
        \State Replace
        $h_t=-h_t$
        and
        $\overline h_t=-\overline h_t$.
    \EndIf
    \State Set $r_t=\left|r_t^{\mathrm{raw}}\right|$.
    \State Set
    \[
    \beta_t
    =
    \frac{1}{2}
    \log\!\left(
    \frac{1+r_t}{1-r_t}
    \right),
    \qquad
    \alpha_t
    =
    \frac{\beta_t}{h_t^\star}.
    \]
    \State Update, for every $i\in[m]$,
    \[
    D^{t+1}(i)
    =
    D^t(i)
    \frac{
    1-r_t y_i\overline h_t(x_i)
    }{
    1-r_t^2
    }.
    \]
    \State Set
    \[
    F_t(x)
    =
    F_{t-1}(x)
    +
    \alpha_t h_t(x).
    \]
\EndFor
\State \textbf{Return}
\[
F_T(x)
=
\sum_{t=1}^{T}\alpha_t h_t(x),
\qquad
H_T(x)
=
\operatorname{sign}\!\left(F_T(x)\right).
\]
\end{algorithmic}
\end{algorithm}

\subsection{Generalization Bounds via Rademacher Complexity}
\label{subsec:generalization_bounds}

Fundamentally, learning selects a hypothesis from \(\mathcal H\) based on a finite training sample, but the main objective is good performance on new examples drawn from the same distribution $\mathcal D$. Generalization theory allows to relate empirical performance to expected classification risk.

In practice, generalization performance can be estimated using data that are not used to fit the model, for example through a separate test set or cross-validation. Here, instead, we seek an explicit theoretical upper bound on the expected risk in terms of quantities measured on the training sample and the complexity of the hypothesis class.

In what follows, we derive such a generalization bound using standard uniform complexity arguments. In particular, we employ Rademacher complexity as our primary measure of hypothesis class capacity~\cite{Mohri2018,BartlettMendelson2002}. Rademacher complexity provides uniform bounds over the entire hypothesis class and is well suited to bounded real-valued predictors in $[-1,1]$. Moreover, as we explain shortly, it is invariant under taking convex hulls of a hypothesis class, a property that will be relevant when analyzing ensemble constructions.

This framework allows us to decompose the expected risk into empirical, complexity, and finite-sample contributions, which will serve as the basis for our comparison of noiseless training, PEC-based mitigation, and boosting with noisy PQCs.

We use this standard bound because the compared methods will later be placed under the same reference hypothesis class. Consequently, for fixed sample size, margin threshold, and confidence level, the complexity and finite-sample terms in this bound are common to the compared methods. The resource comparison can therefore be based on the empirical margin-loss term.

Before defining risk formally, we emphasize that although the task is binary classification, all theoretical analysis is carried out on the hypotheses that are real-valued functions $h : \mathcal{X} \to [-1,1]$, since margins and generalization bounds depend on the magnitude of the score rather than only its sign. The associated classifier is then obtained by applying a $sign$ ($signum$) function:
\[
\hat y = \mathrm{sign}(h(x))
\]
In the realizable PAC setting introduced in Section~\ref{pac_learning}, the labels are determined by an unknown target concept $c:\mathcal X\to\{-1,+1\}$, so that $y_i=c(x_i)$.
\begin{definition}\textbf{(Empirical risk)} Given a finite training sample 
\(
S = \{(x_i, y_i)\}_{i=1}^m
\)
with $y_i \in \{-1,+1\}$, the empirical risk of a hypothesis $h$ is defined as
\begin{equation} \label{eq:empirical_risk}
    \widehat{R}_S(h) 
    =
    \frac{1}{m} 
    \sum_{i=1}^m 
    \mathbb{I}\{ y_i h(x_i) \le 0 \}.
\end{equation}
\end{definition}
\begin{definition}\textbf{(Expected risk)}
The \emph{expected risk} (or generalization error) of a classifier \(h\) is defined as the expected classification error with respect to the data distribution \(\mathcal D\):
\begin{equation}
    R(h) = \Pr_{x \sim \mathcal{D}} [yh(x) \le 0].
\end{equation}
\end{definition}
This quantity captures the true generalization performance of the classifier, but is not directly accessible in practice since \(\mathcal D\) is unknown. The empirical risk is observable on the training sample, whereas the expected risk is the quantity of interest for generalization. Statistical learning theory therefore seeks to control the difference between these quantities in terms of properties of the hypothesis class and the number of training examples. We use Rademacher complexity for this purpose.
\begin{definition}\textbf{(Empirical Rademacher complexity)}
Let $\mathcal X \subseteq \mathbb R^N$ denote the input space and 
$\mathcal{H} \subseteq \{h: \mathcal X \to [-1, +1]\}$ be a family of functions. Let $S = (x_1, \ldots, x_m)$ be a sample of size $m$ drawn i.i.d.\ from a distribution $\mathcal{D}$ over $\mathcal X$. 
The empirical Rademacher complexity of $\mathcal{H}$ with respect to $S$ is defined as
\begin{equation}\label{eq:empirical_rademacher_complexity}
    \widehat{\mathfrak{R}}_S(\mathcal{H}) 
    =
    \mathbb{E}_{\boldsymbol{\sigma} \sim \mathrm{Unif}(\{-1,+1\}^m)}
    \Bigg[
    \sup_{h \in {\mathcal{H}}}
    \frac{1}{m}\sum_{i=1}^m \sigma_i h(x_i)
    \Bigg],
\end{equation}
where $\mathrm{Unif}(\{-1, +1\}^m)$ is the uniform distribution on $\{-1, +1\}^m$. The i.i.d. random variables $\boldsymbol{\sigma} = (\sigma_i)_{i \in [m]}$ are often called Rademacher random variables, that is, $\Pr(\sigma_i=+1)
=
\Pr(\sigma_i=-1)
=
\frac{1}{2}$.
\end{definition}

The (distribution-dependent) Rademacher complexity of $\mathcal{H}$ is defined as
\begin{equation}\label{eq:rademacher_complexity}
    \mathfrak{R}_m(\mathcal{H})
    =
    \mathbb{E}_{S \sim \mathcal{D}^m}
    \big[
    \widehat{\mathfrak{R}}_S(\mathcal{H})
    \big]
\end{equation}

In many learning settings, one additionally seeks hypotheses that assign the correct label with a large confidence margin $y h(x)$: it is positive for a correctly signed prediction and its magnitude measures the confidence rating assigned by the predictor. In our analysis, margin considerations play an explicit role in the theoretical comparison, as boosting algorithms are naturally analyzed through margin-based guarantees.

\begin{definition}\textbf{(Empirical margin loss)}
The empirical margin loss of $h$ given $S$ and $\mu \in (0, 1]$ is defined as 
\begin{equation}\label{eq:empirical_margin_loss}
    \widehat{R}_{S, \mu}(h) = \frac{1}{m} \sum_{i=1}^m \Phi_{\mu}(y_i h(x_i)) \quad \text{where} \quad \Phi_{\mu}(x) = \min\Bigg\{1, \max\Big\{0, 1 - \frac{x}{\mu}\Big\} \Bigg\} = \begin{cases}
        1, \quad x \leq 0 \\
        1 - \frac{x}{\mu}, \quad 0 \leq x \leq \mu \\
        0, \quad \mu \leq x
    \end{cases}
\end{equation}
\end{definition}
It assigns full loss to wrong predictions, partial loss to correct but low-margin predictions, and zero loss to predictions with margin at least $\mu$.
\begin{definition}\textbf{(Empirical margin error)}
For a predictor $h:\mathcal X\to\mathbb R$ and margin parameter $\mu>0$, we define the empirical margin error
\begin{equation}\label{eq:empirical_margin_error}
\widehat R^{\mathrm{mar}}_{S,\mu}(h)
=
\frac{1}{m}
\sum_{i=1}^m
\mathbb{I}\{y_i h(x_i)\le \mu\}.
\end{equation}
This quantity counts the fraction of examples whose margin falls below the threshold $\mu$.
\end{definition}

Having introduced the empirical margin loss in Eq.~\eqref{eq:empirical_margin_loss}, we now use it to obtain a generalization bound that depends explicitly on the margin. The bound applies uniformly to a class of hypotheses and depends on the Rademacher complexity of that class.

To apply a Rademacher-complexity bound to the boosting ensemble, we need a hypothesis class that contains the ensemble while retaining a complexity term that can be related directly to that of the base hypotheses. We therefore show that a normalized boosting ensemble belongs to the convex hull of the signed base class. The key property used below is that taking a convex hull does not increase Rademacher complexity.

As specified in Section \ref{sec:boosting_framework}, if a trained weak hypothesis has a negative edge, its output is replaced by $-h_t$ through classical post-processing. The hypotheses used by the ensemble may therefore belong either to $\mathcal H$ or to $-\mathcal H$. We consequently define the signed closure:
\begin{equation}
\mathcal H_{\pm}
:=
\mathcal H\cup(-\mathcal H),
\qquad
-\mathcal H
:=
\{-h:h\in\mathcal H\}.
\label{eq:signed_base_class}
\end{equation}
If $\mathcal H$ is already closed under sign reversal, then $\mathcal H_{\pm}=\mathcal H$. The convex hull of $\mathcal H_{\pm}$ is
\begin{equation}
\operatorname{conv}(\mathcal H_{\pm})
:=
\left\{
\sum_{k=1}^{p}\lambda_k h_k
\;:\;
p\in\mathbb N,\;
h_k\in\mathcal H_{\pm},\;
\lambda_k\geq0,\;
\sum_{k=1}^{p}\lambda_k=1
\right\}.
\label{eq:convex_hull}
\end{equation}
Here, $p$ is the finite number of base hypotheses in the combination and $\lambda_1,\ldots,\lambda_p$ are their nonnegative weights. The condition $\sum_{k=1}^{p}\lambda_k=1$ makes the resulting predictor a convex combination.

Recall that the boosting score after $T$ rounds is
\begin{equation}
F_T(x)
=
\sum_{t=1}^{T}\alpha_t h_t(x),
\qquad
\alpha_t\geq0.
\end{equation}
The boosting coefficients $\alpha_t$ do not generally sum to one, so the raw score $F_T$ need not belong to $\operatorname{conv}(\mathcal H_{\pm})$. To place the ensemble in this class without changing its predicted labels, we normalize the coefficients by their sum. Define
\begin{equation}
A_T
:=
\sum_{t=1}^{T}\alpha_t,
\qquad
\widetilde F_T(x)
:=
\frac{F_T(x)}{A_T},
\label{eq:normalized_boosting_ensemble}
\end{equation}
assuming $A_T>0$. Then
\begin{equation}
\widetilde F_T(x)
=
\sum_{t=1}^{T}
\frac{\alpha_t}{A_T}h_t(x),
\qquad
\frac{\alpha_t}{A_T}\geq0,
\qquad
\sum_{t=1}^{T}\frac{\alpha_t}{A_T}=1,
\end{equation}
and hence
\begin{equation}
\widetilde F_T
\in
\operatorname{conv}(\mathcal H_{\pm}).
\label{eq:normalized_ensemble_in_convex_hull}
\end{equation}

Since $A_T>0$, normalization does not change the predicted label:
\begin{equation}
\operatorname{sign}\!\left(\widetilde F_T(x)\right)
=
\operatorname{sign}\!\left(F_T(x)\right).
\label{eq:normalization_preserves_classifier}
\end{equation}
It only fixes the scale on which the ensemble margins and the threshold $\mu$ are measured.

Taking the convex hull does not increase Rademacher complexity:
\begin{equation}
\mathfrak R_m
\!\left(
\operatorname{conv}(\mathcal H_{\pm})
\right)
=
\mathfrak R_m(\mathcal H_{\pm}).
\label{eq:rademacher_convex_hull_invariance}
\end{equation}
Indeed, for fixed Rademacher signs and any
$f=\sum_{k=1}^{p}\lambda_k h_k
\in\operatorname{conv}(\mathcal H_{\pm})$,
\begin{align}
\frac{1}{m}
\sum_{i=1}^{m}\sigma_i f(x_i)
&=
\sum_{k=1}^{p}
\lambda_k
\left(
\frac{1}{m}
\sum_{i=1}^{m}\sigma_i h_k(x_i)
\right)
\nonumber
\leq
\sup_{h\in\mathcal H_{\pm}}
\frac{1}{m}
\sum_{i=1}^{m}\sigma_i h(x_i).
\end{align}
Conversely,
$\mathcal H_{\pm}\subseteq
\operatorname{conv}(\mathcal H_{\pm})$,
so the reverse inequality between the corresponding suprema is immediate. This proves Eq.~\eqref{eq:rademacher_convex_hull_invariance}~\cite{BartlettMendelson2002,Mohri2018}. We can now state the margin-based generalization guarantee.

\begin{theorem}[\textbf{Ensemble Rademacher margin bound}~\cite{Mohri2018}]
\label{thm:ensemble_margin_bound}
Let $\mathcal H_{\pm}\subseteq\{h:\mathcal X\to[-1,1]\}$,
let $\mathcal D$ be a probability distribution over \(
\mathcal X \subseteq\ \mathbb{R}^N\), and let
$c:\mathcal X\to\{-1,+1\}$ be the target concept. Fix a margin threshold
$\mu\in(0,1]$ and a confidence parameter $\delta\in(0,1)$.

Then, with probability at least $1-\delta$ over a sample
\(
S
=
\{(x_i,c(x_i))\}_{i=1}^{m},
x_i\overset{\mathrm{i.i.d.}}{\sim}\mathcal D,
\)
the following holds simultaneously for every $f\in\operatorname{conv}(\mathcal H_{\pm})$:
\begin{equation}
R(f,c)
\leq
\widehat R_{S,\mu}(f)
+
\frac{2}{\mu}
\mathfrak R_m(\mathcal H_{\pm})
+
\sqrt{
\frac{\log(1/\delta)}{2m}
}.
\label{eq:rademacher_margin_bound}
\end{equation}
\end{theorem}
Applying Theorem~\ref{thm:ensemble_margin_bound} to the normalized boosting score $f=\widetilde F_T$ gives
\begin{equation}
R(\widetilde F_T,c)
\leq
\widehat R_{S,\mu}(\widetilde F_T)
+
\frac{2}{\mu}
\mathfrak R_m(\mathcal H_{\pm})
+
\sqrt{
\frac{\log(1/\delta)}{2m}
}.
\label{eq:boosting_ensemble_margin_bound}
\end{equation}
Because $F_T$ and $\widetilde F_T$ define the same classifier, $R(F_T,c)=R(\widetilde F_T,c)$.

The bound separates three contributions. The first term is the empirical margin loss $\widehat R_{S,\mu}$ of the normalized ensemble and penalizes both incorrect and low-margin predictions. The second term measures the capacity of the signed base hypothesis class, while the final term accounts for statistical fluctuations due to the finite training sample. Thus, a small empirical margin loss together with controlled Rademacher complexity yields a small upper bound on the expected classification risk.

The empirical risk (Eq.~\ref{eq:empirical_risk}), empirical margin loss (Eq.~\ref{eq:empirical_margin_loss}), and empirical margin error (Eq.~\ref{eq:empirical_margin_error}) satisfy
\begin{equation}
\widehat R_S(h)
\leq
\widehat R_{S,\mu}(h)
\leq
\widehat R^{\mathrm{mar}}_{S,\mu}(h).
\label{eq:classification_error_margin_relation}
\end{equation}
Indeed, the three quantities coincide on examples with nonpositive margin. For an example with margin $0<y_i h(x_i)<\mu$, the empirical risk is zero, the empirical margin loss lies strictly between zero and one, and the empirical margin error is one. Consequently, a bound $\widehat R_{S,\mu}(h)\leq\varepsilon'$ implies $\widehat R_S(h)\leq\varepsilon'$. The converse is not true. Hence small classification error alone does not imply small empirical margin loss. The empirical margin error $\widehat R^{\mathrm{mar}}_{S,\mu}(f)$ remains useful as an auxiliary quantity. Whenever an algorithm-specific analysis yields an upper bound on $\widehat R^{\mathrm{mar}}_{S,\mu}$, it also provides a sufficient upper bound on the empirical margin loss $\widehat R_{S,\mu}(f)$.

We emphasize that the quantity of interest in this work is the expected $0$--$1$ classification risk $R(h)$. Theorem~\ref{thm:ensemble_margin_bound} provides the condition required for a common resource comparison. Suppose two learning strategies produce predictors that can both be bounded using the same reference class $\mathcal H_\pm$, and suppose that the sample size $m$, margin threshold $\mu$, and confidence level $\delta$ are fixed across the comparison. Then the last two terms in Eq.~\eqref{eq:rademacher_margin_bound} are common to both strategies. Consequently, comparing the resources required to attain the same empirical margin-loss level
\begin{equation}
\widehat R_{S,\mu}(f)
\leq
\varepsilon_{\mathrm{mar}}
\label{eq:generic_common_margin_target}
\end{equation}
allows to compare the resources required to attain the same high-probability upper bound on expected classification risk.

This provides the generic comparison principle used for the analysis. The following sections instantiate it for noiseless PQCs, PEC, and boosting. In particular, we will identify an appropriate common reference class and then compare the circuit runs (shots) as resource cost required by PEC and boosting to attain the same empirical margin-loss target.

\section{Learning Setup}
\label{learning_setup}

\subsection{Binary Classification with PQCs}

We study a standard binary classification problem in the realizable PAC learning framework. Let $\mathcal{C}$ be a concept class consisting of labeling functions $c: \mathcal{X} \to \{-1,+1\}$, where $\mathcal{X} \subseteq \mathbb{R}^N$ denotes the input space. There exists an unknown target concept $c \in \mathcal{C}$ that determines the labels.

Data points $x \in \mathcal{X}$ are drawn i.i.d.\ from an unknown but fixed distribution $\mathcal{D}$ over $\mathcal{X}$, and each label is given by $y = c(x)$. Given a training dataset $\mathcal{S} = \{(x_i, c(x_i))\}_{i=1}^m$ of a fixed size $m$, the goal is to learn a classifier that generalizes well to unseen inputs drawn from the same distribution $\mathcal{D}$.

We consider real-valued hypotheses $h : \mathcal X \to [-1,1]$, where the predicted label is given by \(\hat y = \mathrm{sign}(h(x))\). In our setting, hypotheses are realized by parameterized quantum circuits (PQCs): for fixed circuit architecture and parameters, the output \(h(x)\) corresponds to the expectation value of a bounded observable (e.g., a Pauli operator or Pauli string) measured on the quantum state prepared by the circuit. The restriction \(h(x)\in[-1,1]\) reflects the bounded spectrum of such observables. We model a PQC hypothesis by specifying a family of input- and parameter-dependent quantum circuits, together with the physical noise model and the final measurement observable.

Given an input feature vector $x\in\mathcal X\subseteq\mathbb R^N$ and trainable parameters $\theta$, we consider a circuit family in which the operations may depend on both $x$ and $\theta$. This includes architectures with a single data-encoding stage as well as data-reuploading architectures in which the input enters at multiple layers. We use $d$ for circuit depth (the number of repeated layers) and $\ell\in[d]=\{1,\ldots,d\}$ for a layer index. Concretely, the ideal state prepared by the circuit can be written as
\[
\rho_\theta(x)
=
U_d(\theta_d, x)\cdots U_2(\theta_2, x) U_1(\theta_1, x)\,\rho_{\mathrm{in}}\,
U_1^\dagger(\theta_1, x)\cdots U_d^\dagger(\theta_d, x),
\]
where $\rho_{\mathrm{in}}$ is a fixed initial state and each $U_\ell(\theta_\ell,x)$ may depend both on trainable parameters $\theta_\ell$ and on the input $x$. This formulation includes standard single-encoding architectures as a special case, as well as data re-uploading schemes in which classical information is injected at multiple layers. The chosen circuit family and parameter space determine the corresponding PQC hypothesis class.

We model physical noise by a local Markovian completely positive trace-preserving map (CPTP) channel applied after each circuit layer. The channel may depend on the layer, so the maps $\mathcal N_\ell$ need not be identical, but in this model they are taken to be independent of the particular input $x$ and parameter values $\theta$:
\[
\rho^{\mathrm{noisy}}_{0,\theta}(x)
:=
\rho_{\mathrm{in}},
\]
and for $\ell\in[d]$,
\begin{equation}
\rho^{\mathrm{noisy}}_{\ell,\theta}(x)
:=
\mathcal N_\ell
\left(
U_\ell(\theta_\ell,x)
\rho^{\mathrm{noisy}}_{\ell-1,\theta}(x)
U_\ell^\dagger(\theta_\ell,x)
\right),
\label{eq:noisy_pqc_state_recursive}
\end{equation}
where each $\mathcal N_\ell$ is a CPTP map describing the noise associated with the $\ell$th circuit layer. The final noisy state is
\begin{equation}
\rho^{\mathrm{noisy}}_\theta(x)
:=
\rho^{\mathrm{noisy}}_{d,\theta}(x).
\label{eq:final_noisy_pqc_state}
\end{equation}
The same construction applies at gate level by associating a CPTP noise channel with each individual gate location rather than with each layer. The numerical experiments below use this gate-local form.

Let $O$ be a fixed Hermitian observable acting on the $n$-qubit Hilbert space and satisfying $\lVert O\rVert_\infty\leq1$. The ideal prediction score is $h_\theta(x):=\operatorname{tr}\left[O\rho_\theta(x)\right]$. The operator-norm condition ensures that $h_\theta(x)\in[-1,1]$. The physical circuit noise is incorporated into the final noisy state $\rho_\theta^{\mathrm{noisy}}(x)$. The corresponding mean noisy prediction score is
\begin{equation}
h_\theta^{\mathrm{noisy}}(x)
:=
\operatorname{tr}
\left[
O\rho_\theta^{\mathrm{noisy}}(x)
\right].
\label{eq:mean_noisy_pqc_score}
\end{equation}
The same observable $O$ is used in the ideal and noisy prediction scores. In the present model, noise is represented by channels acting on the quantum state before an ideal measurement of $O$. We do not consider measurement-assignment or readout errors.

For a fixed $d$-layer architecture with parameter space $\Theta_d$, the ideal and noisy hypothesis classes are
\begin{equation}
\mathcal H_d^{\mathrm{noiseless}}
:=
\left\{
h_\theta:\mathcal X\to[-1,1]
\;\middle|\;
h_\theta(x)
=
\operatorname{tr}
\left[
O\rho_\theta(x)
\right],
\;
\theta\in\Theta_d
\right\},
\label{eq:hypothesis_class_noiseless}
\end{equation}
and
\begin{equation}
\mathcal H_d^{\mathrm{noisy}}
:=
\left\{
h_\theta^{\mathrm{noisy}}:\mathcal X\to[-1,1]
\;\middle|\;
h_\theta^{\mathrm{noisy}}(x)
=
\operatorname{tr}
\left[
O\rho_\theta^{\mathrm{noisy}}(x)
\right],
\;
\theta\in\Theta_d
\right\}.
\label{eq:hypothesis_class_noisy}
\end{equation}

The functions $h_\theta$ and $h_\theta^{\mathrm{noisy}}$ denote exact mean expectation values. In the numerical experiments, they are estimated using a finite number of measurement shots. For the Pauli observables considered here, the finite-shot estimator satisfies
\begin{equation}
\mathbb E_{\mathrm{shots}}
\left[
\widehat h_\theta^{\mathrm{noisy}}(x)
\right]
=
h_\theta^{\mathrm{noisy}}(x).
\label{eq:finite_shot_unbiased_estimator}
\end{equation}
Thus, circuit noise changes the mean prediction $h_\theta^{\mathrm{noisy}}(x)$, whereas finite-shot sampling introduces statistical fluctuations around that mean, which is treated separately from the physical noise defining $\mathcal H^{\mathrm{noisy}}_d$.

\subsection{Common comparison target and assumptions}
\label{sec:assumptions}

For the resource comparison, we fix a realized training sample $S$, the base circuit architecture and depth $d$, the parameter space $\Theta_d$, the observable $O$, and the physical noise condition $\mathcal N$.

We also fix the parameters entering the generalization bound: a margin threshold $\mu\in(0,1]$, an empirical margin-loss target $\varepsilon_{\mathrm{mar}}\in(0,1)$, and a confidence parameter $\delta\in(0,1)$. Here, $\mu$ specifies the positive margin used in the empirical loss, while $\delta$ controls the probability, over the random draw of the training sample, with which the generalization bound holds.

Our goal is to compare the resources required by PEC and boosting under the same generalization criterion. As established in Section~\ref{subsec:generalization_bounds}, any predictor $f$ in the common reference class satisfies, with probability at least $1-\delta$ over the draw of $S$,
\begin{equation}
R(f,c)
\leq
\widehat R_{S,\mu}(f)
+
\Delta_{\mathrm{gen}},
\end{equation}
where $\Delta_{\mathrm{gen}}$ represents the complexity and finite-sample terms. For fixed $m$, $\mu$, $\delta$, and reference hypothesis class, $\Delta_{\mathrm{gen}}$ is common to the compared methods. We therefore require both methods to attain the same empirical margin-loss target,
\begin{equation}
\widehat R_{S,\mu}(f)
\leq
\varepsilon_{\mathrm{mar}}.
\label{eq:comparison_margin_target}
\end{equation}
Any method reaching this target then satisfies the same upper bound $R(f,c)\leq\varepsilon_{\mathrm{mar}}+\Delta_{\mathrm{gen}}$.

To make the resource comparison meaningful, we also fix training ingredients that are common by construction, including the optimizer, initialization distribution, batching rule, and per-evaluation shot policy. Method-specific operations remain different: PEC replaces ideal expectation-value evaluations by mitigated estimates, whereas boosting trains and combines noisy predictors and therefore introduces its own round-dependent updates and stopping conditions.

The resource analysis below additionally requires assumptions concerning the physical noise model, attainability of the common target, and the cost of reaching that target with the noiseless reference learner.

\paragraph{Assumption A1: local Markovian circuit noise.}
\label{ass:a1}
The maps in $\mathcal N$ are assumed to be local Markovian CPTP channels acting at prescribed circuit locations. Here, \emph{local} means that each noise map acts only on the qubits associated with that circuit location, while \emph{Markovian} means that the noise at a given location acts on the current state without dependence on earlier noise events. In the numerical experiments, this model is specialized to gate-local depolarizing noise with the channel strength specified separately for each experimental condition.

This assumption specifies the physical-noise model used in the resource analysis and numerical implementation. Additional structure on the noise model is introduced later when establishing the common reference hypothesis class.

\paragraph{Assumption A2: noiseless margin-target attainability.}
For the given training dataset $S$, the fixed margin threshold $\mu$, and the selected circuit architecture of depth $d$, there exists a parameter vector $\theta\in\Theta_d$ such that
\begin{equation}
\widehat R_{S,\mu}(h_\theta)
\leq
\varepsilon_{\mathrm{mar}}.
\label{eq:ideal_margin_attainability}
\end{equation}
Thus, the common empirical margin-loss target Eq.~\eqref{eq:comparison_margin_target} is attainable by the ideal PQC hypothesis class.
\paragraph{Assumption A3: finite-evaluation reference optimization reachability.}
We fix a reference training procedure for the noiseless PQC model. This is a specified learning procedure, not an optimal one: it fixes the optimizer, initialization rule, batching scheme, and other training choices used to define the noiseless baseline. The procedure may still be randomized; for example, parameter initialization, mini-batch selection, or finite-shot measurement outcomes. Let $\xi$ denote this internal randomness. Define the random variable $N_{\mathrm{hit}}^{\mathrm{ref}}(\mu,\varepsilon_{\mathrm{mar}};\xi)$ as the total number of expectation-value evaluations performed by the reference training procedure up to the first iterate whose deterministic mean predictor satisfies $\widehat R_{S,\mu}(h_\theta)\leq\varepsilon_{\mathrm{mar}}$.

We assume that there exists a finite reference evaluation budget $N_{\mathrm{noiseless}}$ such that
\begin{equation}
\Pr_\xi
\left[
N_{\mathrm{hit}}^{\mathrm{ref}}
(\mu,\varepsilon_{\mathrm{mar}};\xi)
\leq
N_{\mathrm{noiseless}}
\right]
\geq
1-\delta_{\mathrm{opt}},
\label{eq:ideal_evaluation_reachability}
\end{equation}
where $\delta_{\mathrm{opt}}\in[0,1/2)$ is the optimization failure probability.

Assumptions~A2 and A3 define the noiseless reference learning problem. Assumption~A2 states that the selected PQC family contains a hypothesis that attains the common empirical margin-loss target, whereas Assumption~A3 states that the fixed reference training procedure reaches such a hypothesis within the finite evaluation budget $N_{\mathrm{noiseless}}$ with probability at least $1-\delta_{\mathrm{opt}}$ over its internal randomness. These assumptions are conditional on the realized training sample $S$: the probability in Assumption~A3 is over the internal training randomness $\xi$. This is distinct from the confidence parameter $\delta$ in the generalization bound, whose probability statement is over the random draw of $S$. We do not assume a uniform optimization guarantee over all possible training samples.

\subsection{Resource Cost in the Noiseless Case}
\label{sec:resource_ideal}

We distinguish two levels of resource accounting. An \emph{expectation-value evaluation} is the estimation of one circuit expectation value for a fixed parameter--input pair $(\theta,x)$ using a prescribed shot allocation. The physical resource required for such an evaluation is the number of circuit executions (shots) used to construct that estimate. Different components of the training procedure, such as loss evaluation, gradient estimation, or stopping checks, may require multiple expectation-value evaluations. Under Assumption~A3, we use the finite reference budget $N_{\mathrm{noiseless}}$ as the evaluation-level cost of the noiseless reference procedure:
\begin{equation}
\mathrm{Cost}_{\mathrm{noiseless}}^{(\mathrm{eval})}
=
N_{\mathrm{noiseless}}.
\label{eq:cost_ideal_steps}
\end{equation}
This coarse-grained accounting is sufficient for the resource comparison because the details of the optimizer, batching scheme, and stopping procedure are already reflected in the total number of expectation-value evaluations.

We now convert this evaluation count into the physical circuit-execution cost. Even in the noiseless reference model, each expectation value must be estimated from finitely many measurement shots. Thus, ``noiseless'' refers to ideal circuit evolution and does not imply exact access to expectation values. If each expectation-value evaluation uses $N_{\mathrm{sh}}$ shots, then the corresponding physical circuit-execution budget is
\begin{equation} \label{eq:shots_ideal}
\mathrm{Shots}_{\mathrm{noiseless}}
=
N_{\mathrm{noiseless}} \times N_{\mathrm{sh}}.
\end{equation}
For a Pauli observable with outcomes in $\{-1,+1\}$, Hoeffding’s inequality implies that estimating an expectation value within additive error $\varepsilon_{\mathrm{eval}}$ with confidence $1-\delta_{sh}$ requires
\begin{equation} \label{n_shots}
N_{\mathrm{sh}} \ge \frac{2}{\varepsilon_{\mathrm{eval}}^2}
\log\frac{2}{\delta_{sh}}.
\end{equation}

In what follows, we express the total cost of PEC and boosting-based trainings relative to reference baseline. Specifically, additional multiplicative factors will account for: (i) quasiprobability variance amplification in PEC, and (ii) repeated training of weak learners in boosting. This allows a direct comparison of strategies in terms of the overhead required to reach the same target empirical margin loss $\varepsilon_{\mathrm{mar}}$.

\section{PEC as a Cost Multiplier of the Ideal Case}
\label{sec:pec_resources}
\subsection{Effect of PEC on Optimization and Risk}
\label{sec:pec_effect_risk}

In the PEC-based strategy (Section~\ref{subsec:pec}), each ideal expectation-value evaluation required by the reference training procedure is replaced by a quasiprobability estimator constructed from noisy circuit executions. This construction requires classical knowledge of the elementary noisy operations. To relate this finite-sampling procedure to the common learning target and to the reference resource budget, we now state the additional assumptions specific to PEC.
\paragraph{Assumption P1: known elementary noisy operations.}
PEC is assumed to have classical descriptions of the elementary noisy operations used by the device, and these descriptions are treated as exact within the theoretical model. Channel-characterization and calibration errors, as well as the resources required to obtain the channel descriptions, are not included in the PEC training cost.

Under this assumption, the quasiprobability representation is constructed for the actual noisy operations, and the resulting PEC estimator is unbiased for the corresponding ideal expectation value. This channel-knowledge requirement is specific to PEC. The boosting strategy does not require an explicit classical description of the physical noise channels, although it relies on its own weak-learning assumptions.

Now we establish the deterministic mean function represented by PEC, then relate finite-shot PEC estimates to the common empirical margin-loss target introduced in Section~\ref{sec:assumptions}, and finally quantify the additional sampling cost required to attain a fixed expectation-value precision.

For every fixed parameter vector $\theta\in\Theta_d$ and input $x\in\mathcal X$, let $\widehat h_\theta^{\mathrm{PEC}}(x)$ denote the finite-sample PEC estimator of the ideal prediction score. By the unbiasedness guarantee of PEC,
\begin{equation}
\mathbb E_{\mathrm{PEC}}
\left[
\widehat h_\theta^{\mathrm{PEC}}(x)
\right]
=
h_\theta(x),
\label{eq:pec_score_unbiasedness}
\end{equation}
where the expectation is taken over quasiprobability sampling and finite-shot measurement randomness.

PEC therefore preserves the ideal hypothesis class at the level of mean prediction functions. More precisely, define
\begin{equation}
\mathcal H_d^{\mathrm{PEC,mean}}
:=
\left\{
x\mapsto
\mathbb E_{\mathrm{PEC}}
\left[
\widehat h_\theta^{\mathrm{PEC}}(x)
\right]
\;\middle|\;
\theta\in\Theta_d
\right\}.
\label{eq:pec_mean_hypothesis_class_definition}
\end{equation}
Equation~\eqref{eq:pec_score_unbiasedness} implies
\begin{equation}
\mathcal H_d^{\mathrm{PEC,mean}}
=
\mathcal H_d^{\mathrm{noiseless}}.
\label{eq:pec_mean_hypothesis_class}
\end{equation}
A finite-shot PEC estimate is nevertheless random and is not itself a deterministic element of $\mathcal H_d^{\mathrm{noiseless}}$. PEC does not enlarge the deterministic mean-function class; instead, finite sampling produces a random estimator whose expectation is the ideal prediction.

Since the two deterministic mean-function classes coincide, their Rademacher complexities are equal:
\begin{equation}
\mathfrak R_m
\left(
\mathcal H_d^{\mathrm{PEC,mean}}
\right)
=
\mathfrak R_m
\left(
\mathcal H_d^{\mathrm{noiseless}}
\right).
\label{eq:pec_mean_rademacher_equality}
\end{equation}

\paragraph{Assumption P2: PEC finite-shot margin condition.}
The generalization analysis applies to the deterministic mean predictor $h_\theta$, whereas in practice its prediction scores are accessed through finite-shot PEC estimates. We therefore relate the empirical margin loss computed from the finite-shot PEC estimates to the empirical margin loss of the underlying deterministic mean predictor. For a fixed parameter vector $\theta$, define the event
\begin{equation}
\mathcal E_{\varepsilon_{\mathrm{eval}}}^{\mathrm{PEC}}(\theta)
:=
\bigcap_{i=1}^{m}
\left\{
\left|
\widehat h_\theta^{\mathrm{PEC}}(x_i)
-
h_\theta(x_i)
\right|
\leq
\varepsilon_{\mathrm{eval}}
\right\}.
\label{eq:pec_margin_certification_event}
\end{equation}
We assume that the PEC shot allocation used for the margin evaluation is chosen such that
\begin{equation}
\Pr
\left(
\mathcal E_{\varepsilon_{\mathrm{eval}}}^{\mathrm{PEC}}(\theta)
\right)
\geq
1-\delta_{\mathrm{est}}^{\mathrm{PEC}},
\qquad
\delta_{\mathrm{est}}^{\mathrm{PEC}}\in(0,1).
\label{eq:pec_good_event_probability}
\end{equation}

On the event $\mathcal E_{\varepsilon_{\mathrm{eval}}}^{\mathrm{PEC}}(\theta)$, multiplication by $y_i\in\{-1,+1\}$ preserves the deviation as $|y_i|=1$, so
\begin{equation}
\left|
y_i\widehat h_\theta^{\mathrm{PEC}}(x_i)
-
y_i h_\theta(x_i)
\right|
\leq
\varepsilon_{\mathrm{eval}}.
\label{eq:pec_signed_margin_deviation}
\end{equation}

The ramp function $\Phi_\mu$ defined in Eq.~\eqref{eq:empirical_margin_loss} is $1/\mu$-Lipschitz. Therefore,
\begin{equation}
\left|
\Phi_\mu
\left(
y_i\widehat h_\theta^{\mathrm{PEC}}(x_i)
\right)
-
\Phi_\mu
\left(
y_i h_\theta(x_i)
\right)
\right|
\leq
\frac{\varepsilon_{\mathrm{eval}}}{\mu}.
\label{eq:pec_pointwise_ramp_deviation}
\end{equation}
Averaging over the training sample gives
\begin{equation}
\left|
\widehat R_{S,\mu}
\left(
\widehat h_\theta^{\mathrm{PEC}}
\right)
-
\widehat R_{S,\mu}
\left(
h_\theta
\right)
\right|
\leq
\frac{\varepsilon_{\mathrm{eval}}}{\mu}.
\label{eq:pec_margin_loss_transfer}
\end{equation}

Consequently,
\begin{equation}
\widehat R_{S,\mu}
\left(
h_\theta
\right)
\leq
\widehat R_{S,\mu}
\left(
\widehat h_\theta^{\mathrm{PEC}}
\right)
+
\frac{\varepsilon_{\mathrm{eval}}}{\mu}.
\label{eq:pec_mean_margin_certification}
\end{equation}
Hence, provided $\varepsilon_{\mathrm{mar}}>\varepsilon_{\mathrm{eval}}/\mu$, the finite-shot stopping condition
\begin{equation}
\widehat R_{S,\mu}
\left(
\widehat h_\theta^{\mathrm{PEC}}
\right)
\leq
\varepsilon_{\mathrm{mar}}
-
\frac{\varepsilon_{\mathrm{eval}}}{\mu}
\label{eq:pec_buffered_margin_target}
\end{equation}
is sufficient, on $\mathcal E_{\varepsilon_{\mathrm{eval}}}^{\mathrm{PEC}}(\theta)$, to meet the deterministic mean target $\widehat R_{S,\mu}(h_\theta)\leq\varepsilon_{\mathrm{mar}}$. Thus, an additive prediction error of at most $\varepsilon_{\mathrm{eval}}$ changes the empirical margin loss by at most $\varepsilon_{\mathrm{eval}}/\mu$. The PEC sampling cost required to achieve this prediction accuracy is part of the resource analysis and numerical protocol.

\paragraph{Assumption P3: PEC evaluation stability.}
Although PEC expectation-value estimators are unbiased, their finite-shot fluctuations propagate into loss and gradient estimates and may therefore change the optimization trajectory. Unbiasedness alone does not imply that PEC training reaches the target using the same number of expectation-value evaluations as the ideal reference procedure. Let $\xi^\star$ collect all internal randomness of PEC training, including parameter initialization, mini-batch selection, quasiprobability sampling, and finite-shot measurement outcomes. Define $N_{\mathrm{hit}}^{\mathrm{PEC}}(\mu,\varepsilon_{\mathrm{mar}};\xi^\star)$ to be the total number of expectation-value evaluations performed by PEC training up to the first training iterate whose deterministic mean predictor satisfies $\widehat R_{S,\mu}\left(h_{\theta_{\mathrm{PEC}}}\right)\leq\varepsilon_{\mathrm{mar}}$. We assume that the PEC shot allocation is sufficiently large that
\begin{equation}
\Pr_{\xi^\star}
\left[
N_{\mathrm{hit}}^{\mathrm{PEC}}
(\mu,\varepsilon_{\mathrm{mar}};\xi^\star)
\leq
N_{\mathrm{noiseless}}
\right]
\geq
1-\delta_{\mathrm{opt}}^{\mathrm{PEC}},
\label{eq:pec_evaluation_stability}
\end{equation}
where $N_{\mathrm{noiseless}}$ is the ideal reference evaluation budget defined in Section~\ref{sec:resource_ideal}.

Assumption~P3 therefore bounds the number of expectation-value evaluations required by PEC to reach the common deterministic margin-loss target by the reference evaluation budget. Consequently, the PEC optimization cost has the form
\begin{equation}
\mathrm{Cost}_{\mathrm{PEC}}^{(\mathrm{eval})}
\leq
N_{\mathrm{noiseless}}
=
\mathrm{Cost}_{\mathrm{noiseless}}^{(\mathrm{eval})}.
\label{eq:n_pec}
\end{equation}
Assumption~P3 concerns the number of expectation-value evaluations and does not yet account for the number of physical circuit executions required for each evaluation. That additional cost is determined by PEC sampling.

Let $\gamma$ denote the circuit-level quasiprobability norm. The factor $\gamma^2$ controls the amplification of estimator variance and, consequently, the number of circuit samples required to attain the same estimation precision as an unmitigated expectation-value estimate~\cite{Temme2017,Endo2018}. We therefore refer to $\gamma^2$ as the PEC variance-amplification or quasiprobability-sampling overhead. The explicit dependence of $\gamma$ on the circuit depth and physical error rates under local depolarizing noise is derived in the following subsection.

\subsection{Cost Inflation via PEC}
\label{sec:pec_cost_inflation}
We now quantify the resource overhead introduced by probabilistic error cancellation (PEC) relative to the noiseless reference model.

From~\eqref{eq:n_pec} we have that the optimization cost remains essentially unchanged. The difference between the ideal and PEC settings therefore arises from the number of measurement shots required per expectation-value evaluation.

In the noiseless reference model, the total number of circuit executions (shots) required to reach empirical margin error $\varepsilon_{\mathrm{mar}}'$:
\[
\mathrm{Shots}_{\mathrm{noiseless}}
=
N_{\mathrm{noiseless}}
\times N_{\mathrm{sh}},
\]
where $N_{\mathrm{sh}}$ denotes the number of shots used to estimate each expectation value to fixed accuracy $\varepsilon_{\mathrm{eval}}$ (Eq.~\ref{eq:shots_ideal}).

A single PEC sample takes values in $[-\gamma,\gamma]$. Therefore, to attain the same additive precision $\varepsilon_{\mathrm{eval}}$,
\begin{equation}
N_{\mathrm{sh}}^{\mathrm{PEC}} =
\gamma^2N_{\mathrm{sh}},
\label{eq:shots_pec_recall}
\end{equation}
as mentioned in the background section (Eq.~\ref{eq:shots_pec}). Here $\gamma$ denotes the quasi-probability norm associated with the inverted noise channel. Because our comparison is theoretical, we use the proven sufficient shot bounds themselves as the resource quantities entering the analysis.

Substituting $N_{\mathrm{sh}}^{\mathrm{PEC}}$ into the shot-counting expression yields
\begin{equation}
\mathrm{Shots}_{\mathrm{PEC}}
=
\mathrm{Shots}_{\mathrm{noiseless}}
\times \gamma^2
\label{eq:cost_pec_final}
\end{equation}

Thus, PEC inflates the total resource cost multiplicatively through variance amplification, while leaving the optimization depth unchanged.

\subsection{Sampled PEC Overhead under Gate-Local Depolarizing Noise}
\label{sec:pec_depol}

We now specialize the PEC sampling overhead to gate-local depolarizing noise. Here, ``gate-local'' means that a depolarizing channel acts independently at each noisy gate location and only on the qubits in the support of that gate.

For a gate acting on $k\in\{1,2\}$ qubits, define
\begin{equation}
\mathcal D_{k,p}(\rho)
=
(1-p)\rho
+
p\,\operatorname{Tr}(\rho)\frac{I_{2^k}}{2^k},
\qquad
p\in[0,1).
\label{eq:local_depolarizing_channel}
\end{equation}
Consider an ideal gate channel $\mathcal U_g$ acting on $k_g$ qubits. Under the noise model adopted in this subsection, its physical implementation is
\begin{equation}
\widetilde{\mathcal U}_g
=
\mathcal D_{k_g,p}\circ\mathcal U_g.
\label{eq:noisy_gate_local_depolarizing}
\end{equation}
PEC represents the ideal gate as a signed linear combination of implementable noisy operations. For the depolarizing channel \eqref{eq:local_depolarizing_channel}, the resulting gate-level quasiprobability $1$-norm is~\cite{Temme2017,Endo2018}
\begin{equation}
\gamma_k(p)
=
\frac{
1+\left(1-\frac{2}{4^k}\right)p
}{
1-p
}.
\label{eq:depolarizing_gate_qpr_norm}
\end{equation}
Consequently, the quasiprobability norms associated with one noisy single-qubit gate and one noisy two-qubit gate are
\begin{equation}
\gamma_1(p)
=
\frac{1+p/2}{1-p},
\qquad
\gamma_2(p)
=
\frac{1+7p/8}{1-p}.
\label{eq:single_two_qubit_qpr_norms}
\end{equation}
Here we assume single- and two-qubit gate noise are same $p_{\rm 1q}=p_{\rm 2q}=p$ for simplicity. On hardware, these strengths are not necessarily equal and may instead satisfy $p_{\rm 1q}\ll p_{\rm 2q}$. For unequal strengths, the same formulas apply separately as $\gamma_1(p_{\rm 1q})$ and $\gamma_2(p_{\rm 2q})$.

For a fixed architecture family, let $G_1(n,d)$ and $G_2(n,d)$ be the numbers of noisy one- and two-qubit gate locations in a circuit of width $n$ and depth $d$. Under the gate-local Markovian Assumption A1 (Section~\ref{ass:a1}), PEC applies an independent gate-level quasiprobability decomposition at each location. The coefficient associated with a sampled circuit is therefore the product of the coefficients selected at the individual gate locations. Accordingly, the circuit-level quasiprobability norm is:
\begin{equation}
\gamma(n,d,p)
=
\gamma_1(p)^{G_1(n,d)}\gamma_2(p)^{G_2(n,d)}.
\label{eq:circuit_qpr_norm_product}
\end{equation}
Substituting Eq.~\eqref{eq:single_two_qubit_qpr_norms} gives
\begin{equation}
\gamma(n,d,p)
=
\left(\frac{1+p/2}{1-p}\right)^{G_1(n,d)}
\left(\frac{1+7p/8}{1-p}\right)^{G_2(n,d)}.
\label{eq:gamma_depol}
\end{equation}
Using the general formula for PEC shots calculation (\ref{eq:cost_pec_final}):
\begin{equation}
\mathrm{Shots}_{\mathrm{PEC}}^{depol}
=
\mathrm{Shots}_{\mathrm{noiseless}}
\times \gamma_{\mathrm{circ}}^2
\label{eq:cost_pec_final_depol}
\end{equation}

\section{Boosting as an Alternative to Cost-Inflated PEC}
\label{sec:boosting_resources}
\subsection{Empirical Margin Target and Cost of Boosting}
\label{sec:boosting_cost_analysis}

The boosting strategy trains noisy PQCs directly under the physical noise model and combines their deterministic mean prediction scores into an ensemble. Unlike PEC, boosting does not attempt to reconstruct the corresponding noiseless expectation values.

Recall from Section~\ref{subsec:generalization_bounds} that
\begin{equation}
A_T
:=
\sum_{t=1}^{T}\alpha_t,
\qquad
\widetilde F_T
:=
\frac{F_T}{A_T},
\label{eq:boosting_normalized_ensemble_recall}
\end{equation}
with $A_T>0$. The normalized ensemble has the same classification decisions as $F_T$ but fixes the scale on which its margins are measured.

The empirical quantity entering the common expected-risk comparison is margin loss. Accordingly, the boosting target is
\begin{equation}
\widehat R_{S,\mu}
\left(
\widetilde F_T
\right)
\leq
\varepsilon_{\mathrm{mar}},
\label{eq:boosting_margin_target}
\end{equation}
where the margin threshold $\mu$ and target $\varepsilon_{\mathrm{mar}}$ are the same as those fixed for the ideal and PEC reference procedures in Section~\ref{sec:assumptions}.

The boosting resource cost is determined by two quantities: the number of rounds required to attain Eq.~\eqref{eq:boosting_margin_target}, and the number of expectation-value evaluations required to train and evaluate the weak learner at each round. To obtain theoretical bounds on these two quantities, we introduce the following assumptions on the weak-learning edge, its finite-shot estimation, and the cost of training the weak learners.

\paragraph{Assumption B1: uniform weak learnability.}
\label{ass:B1}
To state the weak-learning condition uniformly for the two algorithms considered below, let $g_t$ denote the deterministic mean score used to compute the edge of a boosting round and update the distribution $D^t$:
\begin{equation}
g_t
=
\begin{cases}
h_t, & \text{for Real AdaBoost},\\[2mm]
\bar h_t, & \text{for AdaBoost}\mathbb R.
\end{cases}
\label{eq:generic_boosting_update_score}
\end{equation}
Here, $h_t$ is the mean noisy-PQC score, whereas $\bar h_t=h_t/h_t^\star$ is the training-sample-normalized score used by AdaBoost$\mathbb R$, where $h_t^\star=\max_{i\in[m]}|h_t(x_i)|$ is the maximum absolute score on the training sample. Using the distribution $D^t$ over the fixed training sample defined in the boosting framework, the corresponding round edge (\ref{eq:edge_condition}) is $r_t:=\sum_{i=1}^{m}D^t(i)y_i g_t(x_i)$. We assume that there exists a constant $\eta>0$ such that every round satisfies
\begin{equation}
r_t\geq\eta,
\qquad
t=1,\ldots,T.
\label{eq:uniform_weak_edge_assumption}
\end{equation}
Thus, over any fixed sequence of $T$ boosting rounds, $\eta$ is a lower bound on the smallest round edge,
\[
\eta\leq\min_{1\leq t\leq T} r_t.
\]
No monotonic behavior of the edges with $t$ is assumed: as boosting progresses, the distributions $D^t$ may concentrate on increasingly difficult examples, so the round edges may decrease. Recall from Section~\ref{pac_learning} that efficient weak learning requires an advantage that is at least inverse-polynomial in the problem size. Accordingly, we require $\eta(n)\geq 1/q(n)$ for some polynomial $q$, rather than merely $\eta>0$. An exponentially small edge guarantee, such as $\eta=e^{-n}$, is insufficient for a polynomial-time boosting guarantee: amplifying such a small advantage may require exponentially many rounds.

Assumption~B1 is the weak-learning condition used in the boosting convergence analysis. Establishing a positive uniform lower bound $\eta$ is generally difficult, because the edge at round $t$ depends on the weak-learner optimization under the adaptively reweighted distribution $D^t$. For a specific circuit family and noise model, deriving such a bound would constitute a separate theoretical problem. In the present work, we therefore treat $r_t\geq\eta>0$ as an explicit assumption and estimate the realized round edges numerically.

\paragraph{Assumption B2: finite-shot edge accuracy.}
The boosting convergence results are formulated in terms of the deterministic mean update scores $g_t$. In practice, these scores are estimated from a finite number of circuit executions. Let $\widehat g_t(x_i)$ denote the finite-shot estimate of $g_t(x_i)$. For an additive expectation-value precision $\varepsilon_{\mathrm{eval}}>0$, define
\begin{equation}
\mathcal E_{\varepsilon_{\mathrm{eval}}}^{\mathrm B}
:=
\bigcap_{(t,i)\in\mathcal J_{\mathrm B}}
\left\{
\left|
\widehat g_t(x_i)-g_t(x_i)
\right|
\leq
\varepsilon_{\mathrm{eval}}
\right\},
\label{eq:boosting_finite_shot_event}
\end{equation}
where $\mathcal J_{\mathrm B}$ contains the round--example pairs whose finite-shot scores are used to evaluate the boosting edges.

We choose the shot allocation so that all finite-shot scores used in the edge evaluations satisfy the accuracy requirement simultaneously with probability at least $1-\delta_{\mathrm{est}}^{\mathrm B}$:
\begin{equation}
\Pr
\left(
\mathcal E_{\varepsilon_{\mathrm{eval}}}^{\mathrm B}
\right)
\geq
1-\delta_{\mathrm{est}}^{\mathrm B},
\label{eq:boosting_finite_shot_probability}
\end{equation}
where $\delta_{\mathrm{est}}^{\mathrm B}$ is the allowed failure probability of this finite-shot edge-estimation guarantee. For the deterministic mean edge $r_t=\sum_{i=1}^{m}D^t(i)y_i g_t(x_i)$ and its finite-shot estimate $\widehat r_t=\sum_{i=1}^{m}D^t(i)y_i\widehat g_t(x_i)$, the event $\mathcal E_{\varepsilon_{\mathrm{eval}}}^{\mathrm B}$ implies
\begin{align}
\left|
\widehat r_t-r_t
\right|
&\leq
\sum_{i=1}^{m}
D^t(i)
\left|
\widehat g_t(x_i)-g_t(x_i)
\right| \leq
\varepsilon_{\mathrm{eval}},
\label{eq:boosting_edge_deviation}
\end{align}
where we used $D^t(i)\geq0$ and $\sum_iD^t(i)=1$. Consequently, the finite-shot condition
\begin{equation}
\widehat r_t
\geq
\eta+\varepsilon_{\mathrm{eval}}
\label{eq:boosting_buffered_edge_condition}
\end{equation}
implies, on
$\mathcal E_{\varepsilon_{\mathrm{eval}}}^{\mathrm B}$,
\begin{equation}
r_t
\geq
\widehat r_t-\varepsilon_{\mathrm{eval}}
\geq
\eta.
\label{eq:boosting_mean_edge_bound}
\end{equation}
Thus, sufficiently accurate finite-shot edge estimates allow the mean-edge condition in Assumption~B1 to be inferred from the measured scores.

\paragraph{Assumption B3: average weak-learner evaluation budget.}
\label{ass:B3}
For boosting round $t$, let $N_{\mathrm{eval},t}^{\mathrm{weak}}$ denote the total number of expectation-value evaluations required by the round-$t$ weak learner to obtain a deterministic mean predictor satisfying Assumption~B1. The count includes the expectation-value evaluations included in the specified training and round-evaluation protocol.

We assume that the average weak-learner evaluation cost over the $T$ boosting rounds does not exceed the noiseless reference budget:
\begin{equation}
\frac{1}{T}
\sum_{t=1}^{T}
N_{\mathrm{eval},t}^{\mathrm{weak}}
\leq
N_{\mathrm{noiseless}},
\label{eq:Nweak_vs_Nideal}
\end{equation}
where $N_{\mathrm{noiseless}}$ is the reference evaluation budget introduced in Section~\ref{sec:resource_ideal}.

An individual boosting round only needs to produce a weak hypothesis with positive edge, whereas the noiseless reference procedure must reach the final empirical margin-loss target $\varepsilon_{\mathrm{mar}}$. At the same time, later boosting rounds may be more difficult because the adaptive distributions $D^t$ concentrate weight on examples that are poorly handled by the current ensemble. Assumption B3  therefore constrains only the average cost across rounds rather than every round individually. This condition is examined numerically.

The cumulative number of expectation-value evaluations used by boosting through round $T$ is
\begin{equation}
N_{\mathrm{eval}}^{\mathrm{boost}}(T)
:=
\sum_{t=1}^{T}
N_{\mathrm{eval},t}^{\mathrm{weak}}.
\label{eq:boosting_evaluation_cost}
\end{equation}
If each expectation-value evaluation uses $N_{\mathrm{sh}}$ physical circuit executions, then
\begin{equation}
\mathrm{Shots}_{\mathrm{boost}}(T)
=
N_{\mathrm{eval}}^{\mathrm{boost}}(T)
N_{\mathrm{sh}}.
\label{eq:boosting_shot_cost_general}
\end{equation}
Under Assumption B3,
\begin{equation}
\mathrm{Shots}_{\mathrm{boost}}(T)
\leq
T
N_{\mathrm{noiseless}}
N_{\mathrm{sh}}
=
T\,\mathrm{Shots}_{\mathrm{noiseless}}.
\label{eq:cost_boost_steps}
\end{equation}

Thus, the remaining algorithm-specific problem is to determine a sufficient number of rounds $T$ for which the normalized ensemble satisfies the common margin-loss target in Eq.~\eqref{eq:boosting_margin_target}. This target depends on the particular boosting rule used to form the ensemble. We therefore derive the required round conditions separately for Real AdaBoost and AdaBoost$\mathbb R$ below.

The empirical $0$--$1$ classification-error bounds are retained as secondary consequences, but they are not used as substitutes for the positive-margin quantity appearing in the expected-risk comparison.

\subsection{Real AdaBoost Margin-Loss Bound}
\label{subsec:real_adaboost_analysis}

We apply the standard boosting margin analysis ~\cite{Mohri2018,SchapireSinger1999} to obtain a sufficient number of Real AdaBoost rounds for attaining the common empirical margin-loss target $\widehat R_{S,\mu}(\widetilde F_T)\leq\varepsilon_{\mathrm{mar}}$. We use the Real AdaBoost definitions and exponential-potential relations established in Section~\ref{subsec:real_adaboost_schapire_singer}, in particular Eqs.~\eqref{eq:Zt_potential_ratio}, \eqref{eq:Zt_edge_bound}, and \eqref{eq:alpha_closed_form}. Recall also the normalized ensemble used in the generalization analysis:
\[
\widetilde F_T
=
\frac{F_T}{A_T},
\qquad
A_T
=
\sum_{t=1}^{T}\alpha_t
=
\|\boldsymbol{\alpha}\|_1,
\]
where the last equality follows from the positive-edge convention. By Eq.~\eqref{eq:classification_error_margin_relation}, the empirical margin loss is upper-bounded by the corresponding empirical margin error:
\begin{align}
\widehat R_{S,\mu}
\left(
\widetilde F_T
\right)
&\leq
\widehat R^{\mathrm{mar}}_{S,\mu}
\left(
\widetilde F_T
\right)
\nonumber=
\frac{1}{m}
\sum_{i=1}^{m}
\mathbb I
\left\{
y_i\widetilde F_T(x_i)\leq\mu
\right\}
=
\frac{1}{m}
\sum_{i=1}^{m}
\mathbb I
\left\{
y_iF_T(x_i)\leq\mu A_T
\right\}.
\end{align}
Thus, the empirical margin error is used as an intermediate upper bound on the empirical margin loss appearing in the generalization analysis.

Using $\mathbb I\{z\leq0\}\leq e^{-z}$ with $z=y_iF_T(x_i)-\mu A_T$, together with the Real AdaBoost exponential-potential relation established in Section~\ref{subsec:real_adaboost_schapire_singer}, gives
\begin{equation}
\widehat R_{S,\mu}
\left(
\widetilde F_T
\right)
\leq
e^{\mu A_T}
\frac{1}{m}
\sum_{i=1}^{m}
e^{-y_iF_T(x_i)}
\nonumber=
e^{\mu A_T}
\prod_{t=1}^{T}
Z_t(\alpha_t)
\nonumber\leq
\prod_{t=1}^{T}
e^{\mu\alpha_t}
\sqrt{1-r_t^2},
\end{equation}
where the final inequality follows from Eq.~\eqref{eq:Zt_edge_bound}.

Using the coefficient expression Eq.~\eqref{eq:alpha_closed_form}, each factor in this product bound can be written as
\begin{align}
e^{\mu\alpha_t}
\sqrt{1-r_t^2}
&=
\left(
\frac{1+r_t}{1-r_t}
\right)^{\mu/2}
\sqrt{(1-r_t)(1+r_t)}
\nonumber=
\sqrt{
(1-r_t)^{1-\mu}
(1+r_t)^{1+\mu}
}.
\end{align}
Define
\begin{equation}
b_\mu(r)
:=
\sqrt{
(1-r)^{1-\mu}
(1+r)^{1+\mu}
}.
\label{eq:real_adaboost_round_factor}
\end{equation}
We therefore obtain
\begin{equation}
\widehat R_{S,\mu}
\left(
\widetilde F_T
\right)
\leq
\prod_{t=1}^{T}
b_\mu(r_t).
\label{eq:margin_bound_real_ab}
\end{equation}

To express this bound through the uniform weak-edge parameter $\eta$ from Assumption~B1, note that
\begin{equation}
\frac{\mathrm d}{\mathrm dr}
\log b_\mu(r)
=
\frac{\mu-r}{1-r^2}.
\label{eq:real_adaboost_b_derivative}
\end{equation}
Hence $b_\mu(r)$ is decreasing for $r>\mu$. If $r_t\geq\eta>\mu$ for all $t$, then $b_\mu(r_t)\leq b_\mu(\eta)$. Define
\begin{equation}
B_{\mathrm{real}}(\mu,\eta)
:=
b_\mu(\eta)
=
\sqrt{
(1-\eta)^{1-\mu}
(1+\eta)^{1+\mu}
}.
\label{eq:real_adaboost_contraction_factor}
\end{equation}
Equation~\eqref{eq:margin_bound_real_ab} then implies
\begin{equation}
\widehat R_{S,\mu}
\left(
\widetilde F_T
\right)
\leq
\left[
B_{\mathrm{real}}(\mu,\eta)
\right]^T.
\label{eq:margin_bound_real_ab_uniform}
\end{equation}

For this upper bound to decrease geometrically with the number of boosting rounds, we require
\begin{equation}
B_{\mathrm{real}}(\mu,\eta)<1.
\label{eq:real_adaboost_contraction_condition}
\end{equation}
From Eq.~\eqref{eq:real_adaboost_contraction_factor}, this condition is equivalent to
\begin{equation}
\mu
<
\frac{
-\log(1-\eta^2)
}{
\log
\left(
\frac{1+\eta}{1-\eta}
\right)
}.
\label{eq:real_adaboost_margin_threshold_condition}
\end{equation}
Thus, a positive uniform edge alone does not guarantee geometric decay at an arbitrary margin threshold. The margin threshold $\mu$, fixed as part of the common comparison target in Section~\ref{sec:assumptions}, and the weak-edge lower bound $\eta$ must satisfy Eq.~\eqref{eq:real_adaboost_margin_threshold_condition}. A simpler sufficient condition for this compatibility relation is
\begin{equation}
\mu<\frac{\eta}{2}.
\label{eq:real_adaboost_simple_margin_condition}
\end{equation}
This is not an additional assumption; it is a convenient sufficient condition under which the exact contraction condition Eq.~\eqref{eq:real_adaboost_margin_threshold_condition} is guaranteed.

Whenever $B_{\mathrm{real}}(\mu,\eta)<1$, the common margin-loss target $\varepsilon_{\mathrm{mar}}$ is guaranteed if
\begin{equation}
\left[
B_{\mathrm{real}}(\mu,\eta)
\right]^T
\leq
\varepsilon_{\mathrm{mar}}.
\end{equation}
Accordingly, the sufficient Real AdaBoost round budget
\begin{equation}
T_{\mathrm{real}}^{(\mathrm{mar})}
(\eta,\mu,\varepsilon_{\mathrm{mar}})
:=
\left\lceil\frac{
\log(1/\varepsilon_{\mathrm{mar}})
}{
-\log B_{\mathrm{real}}(\mu,\eta)
}\right\rceil
\label{eq:Tstar_real_adaboost}
\end{equation}

The exact exponent governing this round bound is
\begin{equation}
-\log B_{\mathrm{real}}(\mu,\eta)
=
\frac{1}{2}
\left[
-\log(1-\eta^2)
-
\mu
\log
\left(
\frac{1+\eta}{1-\eta}
\right)
\right].
\label{eq:real_adaboost_exact_margin_exponent}
\end{equation}
This expression shows explicitly how the convergence rate depends jointly on the weak edge and the prescribed normalized margin threshold.

To extract the small-edge scaling of the round bound, consider a family of settings in which the margin threshold is chosen proportionally to the weak edge,
\begin{equation}
\mu=c\eta,
\qquad
0<c<\frac{1}{2}.
\end{equation}
This proportional scaling is not required for the exact bound above; it specifies an asymptotic regime in which the contraction condition continues to hold as $\eta\to0$. In this regime,
\begin{equation}
-\log
B_{\mathrm{real}}(c\eta,\eta)
=
\left(
\frac{1}{2}-c
\right)\eta^2
+
O(\eta^4).
\label{eq:real_adaboost_small_edge_exponent}
\end{equation}
Consequently, under this small-edge scaling regime:
\begin{equation}
T_{\mathrm{real}}^{(\mathrm{mar})}
=
O
\left(
\frac{1}{\eta^2}
\log
\frac{1}{\varepsilon_{\mathrm{mar}}}
\right).
\label{eq:real_adaboost_margin_scaling}
\end{equation}
Thus, Real AdaBoost has an explicit sufficient resource bound for attaining the common empirical margin-loss target. This is the Real AdaBoost quantity used in the subsequent comparison with PEC.

\subsection{AdaBoost$\mathbb{R}$ Margin-Loss Bound}
\label{subsec:adaboostr_analysis}

We next derive the corresponding margin-loss condition for AdaBoost$\mathbb R$. The convergence result~\cite{NockNielsen2007} that we use is formulated in terms of a logistic transformation of the ensemble margin. We therefore first relate this logistic-margin bound to the normalized linear margin used in the generalization analysis. We use the AdaBoost$\mathbb R$ definitions established in Section~\ref{subsec:adaboostr}. In particular, the round edge $r_t$ is computed from the training-sample-normalized weak score $\bar h_t$, and the resulting ensemble is
\[
F_T(x)
=
\sum_{t=1}^{T}\alpha_t h_t(x).
\]
As in the preceding analysis, we consider the normalized ensemble
\begin{equation}
\widetilde F_T
=
\frac{F_T}{A_T},
\qquad
A_T
:=
\sum_{t=1}^{T}\alpha_t
=
\|\boldsymbol{\alpha}\|_1,
\label{eq:adaboostr_normalized_ensemble}
\end{equation}
where the last equality follows from the positive-edge convention.

The AdaBoost$\mathbb R$ analysis associates with $F_T$ the logistic margin
\begin{equation}
\vartheta_T(x,y)
:=
\tanh
\left(
\frac{yF_T(x)}{2}
\right).
\label{eq:adaboostr_logistic_margin}
\end{equation}
Recall that the initial training distribution is uniform, $D^1(i)=1/m$. For the ensemble $F_T$ after $T$ boosting rounds, define the empirical logistic-margin error at threshold $\tau\in[0,1)$ by
\begin{equation}
\begin{aligned}
\omega_{D^1,F_T,\tau}:=\sum_{i=1}^{m}D^1(i)\,
\mathbb I\left\{\vartheta_T(x_i,y_i)\leq\tau\right\}=
\frac{1}{m}\sum_{i=1}^{m}
\mathbb I\left\{\vartheta_T(x_i,y_i)\leq\tau\right\}.
\end{aligned}
\label{eq:adaboostr_logistic_margin_error}
\end{equation}
This error is evaluated under the fixed uniform distribution $D^1$, not the reweighted distribution $D^T$ used during training. Applying the AdaBoost$\mathbb R$ margin bound~\cite[Theorem~1]{NockNielsen2007} to our deterministic-label setting, in which the empirical Bayes error is zero, gives, for $\tau\in[0,1)$,
\begin{equation}
\omega_{D^1,F_T,\tau}
\leq
\frac{1+\tau}{1-\tau}
\exp\left(-\frac12\sum_{t=1}^{T}r_t^2\right).
\label{eq:adaboostr_logistic_margin_bound}
\end{equation}
We now translate this result to the common empirical margin-loss target $\varepsilon_{\mathrm{mar}}$. By Eq.~\eqref{eq:classification_error_margin_relation},
\begin{align}
\widehat R_{S,\mu}
\left(
\widetilde F_T
\right)
&\leq
\widehat R^{\mathrm{mar}}_{S,\mu}
\left(
\widetilde F_T
\right)
\nonumber=
\frac{1}{m}
\sum_{i=1}^{m}
\mathbb I
\left\{
y_i\widetilde F_T(x_i)\leq\mu
\right\}
\nonumber=
\frac{1}{m}
\sum_{i=1}^{m}
\mathbb I
\left\{
y_iF_T(x_i)\leq\mu A_T
\right\}.
\end{align}
Because $\tanh(\cdot)$ is strictly increasing, define
\begin{equation}
\tau_{\mu,T}
:=
\tanh
\left(
\frac{\mu A_T}{2}
\right).
\label{eq:adaboostr_tau_mu}
\end{equation}
Then
\[
y_iF_T(x_i)\leq\mu A_T
\quad\Longleftrightarrow\quad
\vartheta_T(x_i,y_i)
\leq
\tau_{\mu,T},
\]
and therefore
\begin{equation}
\widehat R_{S,\mu}(\widetilde F_T)
\leq
\omega_{D^1,F_T,\tau_{\mu,T}}.
\label{eq:adaboostr_margin_to_logistic}
\end{equation}
Applying Eq.~\eqref{eq:adaboostr_logistic_margin_bound} at $\tau=\tau_{\mu,T}$ gives
\begin{equation}
\widehat R_{S,\mu}
\left(
\widetilde F_T
\right)
\leq
\frac{
1+\tau_{\mu,T}
}{
1-\tau_{\mu,T}
}
\exp
\left(
-\frac{1}{2}
\sum_{t=1}^{T}r_t^2
\right).
\label{eq:adaboostr_normalized_margin_bound}
\end{equation}
Using
\begin{equation}
\frac{
1+\tanh(z/2)
}{
1-\tanh(z/2)
}
=
e^z,
\end{equation}
with $z=\mu A_T$, we obtain
\begin{equation}
\widehat R_{S,\mu}
\left(
\widetilde F_T
\right)
\leq
\exp
\left(
\mu A_T
-
\frac{1}{2}
\sum_{t=1}^{T}r_t^2
\right).
\label{eq:adaboostr_normalized_margin_explicit}
\end{equation}
For convenience, define the accumulated squared edge
\begin{equation}
E_T
:=
\sum_{t=1}^{T}r_t^2
\label{eq:adaboostr_accumulated_edge}
\end{equation}
and the corresponding normalized-margin progress quantity
\begin{equation}
Q_T(\mu)
:=
E_T-2\mu A_T.
\label{eq:adaboostr_margin_progress}
\end{equation}
Equation~\eqref{eq:adaboostr_normalized_margin_explicit} can then be written as
\begin{equation}
\widehat R_{S,\mu}
\left(
\widetilde F_T
\right)
\leq
\exp
\left(
-\frac{Q_T(\mu)}{2}
\right).
\label{eq:adaboostr_margin_progress_bound}
\end{equation}
Hence the common empirical margin-loss target $\varepsilon_{\mathrm{mar}}$ is guaranteed whenever
\begin{equation}
Q_T(\mu)
\geq
2
\log
\left(
\frac{1}{\varepsilon_{\mathrm{mar}}}
\right),
\label{eq:adaboostr_margin_target_condition}
\end{equation}
or, equivalently,
\begin{equation}
E_T-2\mu A_T
\geq
2
\log
\left(
\frac{1}{\varepsilon_{\mathrm{mar}}}
\right).
\label{eq:adaboostr_margin_target_explicit}
\end{equation}
We therefore define the trajectory-dependent sufficient round index:
\begin{equation}
T_{\mathbb R}^{(\mathrm{mar})}
(\mu,\varepsilon_{\mathrm{mar}})
:=
\min
\left\{
T\in\mathbb N:
E_T-2\mu A_T
\geq
2
\log
\left(
\frac{1}{\varepsilon_{\mathrm{mar}}}
\right)
\right\}.
\label{eq:T_adaboostr_margin}
\end{equation}
The dependence on the realized coefficient norm $A_T$ is important. From the AdaBoost$\mathbb R$ coefficient definition introduced in Section~\ref{subsec:adaboostr},
\begin{equation}
A_T
=
\sum_{t=1}^{T}
\alpha_t
=
\sum_{t=1}^{T}
\frac{
\operatorname{arctanh}(r_t)
}{
h_t^\star
},
\label{eq:adaboostr_AT_realized}
\end{equation}
where $h_t^\star=\max_{i\in[m]}|h_t(x_i)|$. At this stage, $A_T$ is determined exactly by the realized boosting trajectory. The resulting margin condition is nevertheless sufficient rather than exact, because the derivation first upper-bounds the empirical margin loss by the empirical margin error and then applies the logistic-margin bound~\cite{NockNielsen2007}. Assumption~B1 implies
\begin{equation}
E_T
=
\sum_{t=1}^{T}r_t^2
\geq
T\eta^2,
\label{eq:adaboostr_edge_lower_bound}
\end{equation}
but this lower bound alone is insufficient to control the positive-margin term in
Eq.~\eqref{eq:adaboostr_normalized_margin_explicit}. Indeed, the condition $r_t\geq\eta$ does not provide an upper bound on $A_T$: the realized edges may approach one, and the normalization factors $h_t^\star$ may become small.

Consequently, unlike the Real AdaBoost result, Assumption~B1 alone does not yield an a priori AdaBoost$\mathbb R$ round budget in terms of the uniform lower bound $\eta$ of the edge together with the fixed comparison parameters $\mu$ and $\varepsilon_{\mathrm{mar}}$. For a realized AdaBoost$\mathbb R$ trajectory, however, both $E_T$ and $A_T$ are known after each round. Hence Eq.~\eqref{eq:adaboostr_margin_target_explicit} is a trajectory-dependent sufficient condition for the empirical margin loss target under its stated assumptions; failing this condition does not imply that the target has not been reached.

These assumptions should control the coefficient growth. For example, if $r_t\leq\rho<1$ and $h_t^\star\geq\kappa>0$ uniformly over the boosting rounds, then
\[
A_T
\leq
T\frac{\operatorname{arctanh}(\rho)}{\kappa}.
\]
We do not impose these additional assumptions in the present analysis. Instead, the AdaBoost$\mathbb R$ margin condition is evaluated from the realized boosting trajectory.

We have now derived the analytical resource conditions needed for PEC, Real AdaBoost, and AdaBoost$\mathbb R$. In the next section, we collect these results into a common resource comparison and identify the quantities that remain to be determined numerically. In particular, the numerical analysis will evaluate the realized boosting edges and weak-learner evaluation costs, and, for AdaBoost$\mathbb R$, the trajectory-dependent quantities $E_T$ and $A_T$ entering Eq.~\eqref{eq:adaboostr_margin_target_explicit}.

\section{Theoretical Comparative Analysis}
\label{sec:comparative_analysis}

The preceding sections provided the derivations needed to compare probabilistic error cancellation (PEC) and boosting at a common statistical target. The quantity of interest is the expected $0$--$1$ classification risk, while the empirical quantity used in the comparison is the margin loss $\widehat R_{S,\mu}(f)$. We now identify a common hypothesis class for the generalization bound, summarize the assumptions entering the resource analysis, and combine the method-specific results of Sections~\ref{sec:pec_effect_risk}--\ref{subsec:adaboostr_analysis} into a single resource comparison.

\subsection{Common Hypothesis Class and Expected-Risk Bound}
\label{subsec:common_reference_class}

The general comparison principle of Section~\ref{subsec:generalization_bounds} requires PEC and boosting to be bounded using the same reference hypothesis class. We therefore define
\begin{equation}
\mathcal H_{\mathrm{ref},\pm}
:=
\mathcal H_{d,\pm}^{\mathrm{noiseless}}.
\label{eq:reference_signed_class}
\end{equation}
For PEC, Eq.~\eqref{eq:pec_mean_hypothesis_class} gives
\begin{equation}
\mathcal H_d^{\mathrm{PEC,mean}}
=
\mathcal H_d^{\mathrm{noiseless}}.
\label{eq:pec_reference_class_relation}
\end{equation}
Hence the deterministic PEC mean predictor belongs directly to the noiseless reference class.

For boosting, we use the structural result established in Appendix~\ref{app:pauli_noise_complexity}. Under the Pauli-noise and circuit architecture conditions stated there,
\begin{equation}
\mathcal H_d^{\mathrm{noisy}}
\subseteq
\operatorname{conv}
\left(
\mathcal H_d^{\mathrm{noiseless}}
\right).
\label{eq:noisy_class_reference_containment}
\end{equation}
Because the boosting procedure may reverse the sign of a weak hypothesis, the relevant signed noisy class is correspondingly contained in $\operatorname{conv}(\mathcal H_{\mathrm{ref},\pm})$. The normalized boosting ensemble is a convex combination of these signed weak hypotheses and therefore satisfies
\begin{equation}
\widetilde F_T
\in
\operatorname{conv}
\left(
\mathcal H_{\mathrm{ref},\pm}
\right).
\label{eq:boosting_reference_class_membership}
\end{equation}
Importantly, Eq.~\eqref{eq:noisy_class_reference_containment} is an inclusion, not an equality of the noisy and noiseless hypothesis classes. The purpose of introducing $\mathcal H_{\mathrm{ref},\pm}$ is instead to place both strategies under the same upper bound on the complexity contribution. Define
\begin{equation}
\Delta_{\mathrm{gen}}
(\mu,m,\delta)
:=
\frac{2}{\mu}
\mathfrak R_m
\left(
\mathcal H_{\mathrm{ref},\pm}
\right)
+
\sqrt{
\frac{\log(1/\delta)}{2m}
}.
\label{eq:common_generalization_term}
\end{equation}
By Theorem~\ref{thm:ensemble_margin_bound}, both the deterministic PEC mean predictor and the normalized boosting ensemble can therefore be bounded as
\begin{equation}
R(f,c)
\leq
\widehat R_{S,\mu}(f)
+
\Delta_{\mathrm{gen}}
(\mu,m,\delta).
\label{eq:common_expected_risk_bound}
\end{equation}
Consequently, if either strategy attains the common empirical target $\widehat R_{S,\mu}(f)\leq\varepsilon_{\mathrm{mar}}$, then
\begin{equation}
R(f,c)
\leq
\varepsilon_{\mathrm{mar}}
+
\Delta_{\mathrm{gen}}
(\mu,m,\delta).
\label{eq:section6_common_expected_risk_target}
\end{equation}
Thus, the resource comparison below concerns the cost required to attain the same upper bound on expected classification risk, rather than equality of the unknown expected risks themselves.

\subsection{Assumptions Entering the Comparison}
\label{subsec:comparison_assumptions}

For clarity, Tables~\ref{tab:common_assumptions}--\ref{tab:boosting_assumptions} summarize the assumptions introduced in Sections~\ref{sec:assumptions}--\ref{sec:boosting_cost_analysis}. The hypothesis-class containment in Eq.~\eqref{eq:noisy_class_reference_containment} is not listed as an additional learning assumption: under the Pauli-noise and circuit architecture conditions used in this work, it is the structural result proved in Appendix~\ref{app:pauli_noise_complexity}.
\begin{table}[!htbp]
\centering
\small
\renewcommand{\arraystretch}{1.2}
\begin{tabular}{p{0.10\linewidth} p{0.48\linewidth} p{0.34\linewidth}}
\hline
\textbf{Label} &
\textbf{Assumption} &
\textbf{Role in the comparison}
\\
\hline

A1 &
The physical circuit noise is described by local Markovian CPTP channels acting at prescribed circuit locations. &
Defines the physical noise model used in the resource analysis. The common-reference-class result additionally specializes this setting to the Pauli-noise and circuit conditions of Appendix~\ref{app:pauli_noise_complexity}.
\\

A2 &
For the realized training sample $S$, fixed margin threshold $\mu$, and selected PQC architecture, there exists $h_\theta\in\mathcal H_d^{\mathrm{noiseless}}$ such that
$\widehat R_{S,\mu}(h_\theta)\leq\varepsilon_{\mathrm{mar}}$. &
Ensures that the common empirical margin-loss target is attainable by the ideal reference class.
\\

A3 &
The ideal training procedure reaches the target $\widehat R_{S,\mu}\leq\varepsilon_{\mathrm{mar}}$ within a finite expectation-value evaluation budget $N_{\mathrm{noiseless}}$ with probability at least $1-\delta_{\mathrm{opt}}$. &
Defines the ideal reference resource budget against which PEC and boosting are compared.
\\

\hline
\end{tabular}
\caption{Assumptions common to the ideal reference problem and the
PEC--boosting comparison.}
\label{tab:common_assumptions}
\end{table}
\begin{table}[!htbp]
\centering
\small
\renewcommand{\arraystretch}{1.2}
\begin{tabular}{p{0.10\linewidth} p{0.48\linewidth} p{0.34\linewidth}}
\hline
\textbf{Label} &
\textbf{PEC-specific assumption} &
\textbf{Role in the comparison}
\\
\hline

P1 &
The elementary noisy operations are known sufficiently accurately to construct the quasiprobability representation used by PEC. The cost of obtaining these channel descriptions is excluded. &
Ensures exact PEC unbiasedness within the theoretical model.
\\

P2 &
Finite-shot PEC prediction scores are estimated to additive precision $\varepsilon_{\mathrm{eval}}$ with the probability specified in Section~\ref{sec:pec_effect_risk}. Consequently, the finite-shot empirical margin loss differs from that of the deterministic PEC mean predictor by at most $\varepsilon_{\mathrm{eval}}/\mu$. &
Relates the finite-shot PEC implementation to the deterministic margin-loss quantity entering the generalization bound.
\\

P3 &
The additional sampling randomness introduced by PEC does not require more expectation-value evaluations than the ideal reference budget to reach the deterministic target $\widehat R_{S,\mu}\leq\varepsilon_{\mathrm{mar}}$. &
Separates the PEC sampling overhead from changes in the optimization evaluation count.
\\

\hline
\end{tabular}
\caption{Additional assumptions specific to probabilistic error
cancellation.}
\label{tab:pec_assumptions}
\end{table}
\begin{table}[!htbp]
\centering
\small
\renewcommand{\arraystretch}{1.2}
\begin{tabular}{p{0.10\linewidth} p{0.48\linewidth} p{0.34\linewidth}}
\hline
\textbf{Label} &
\textbf{Boosting-specific assumption} &
\textbf{Role in the comparison}
\\
\hline

B1 &
For every accepted boosting round, the deterministic mean update score has edge
$r_t\geq\eta>0$ with respect to the current reweighting $D^t$. &
Provides the weak-learning condition required by the boosting convergence results.
\\

B2 &
The finite-shot edge estimate satisfies $|\widehat r_t-r_t|\leq\varepsilon_{\mathrm{eval}}$ on the stated high-probability event. In particular, $\widehat r_t\geq\eta+\varepsilon_{\mathrm{eval}}$ implies $r_t\geq\eta$. &
Relates finite-shot edge measurements to the deterministic mean-edge condition in B1.
\\
B3 &
Averaged over the $T$ boosting rounds, $\frac{1}{T}\sum_{t=1}^{T}N_{\mathrm{eval},t}^{\mathrm{weak}}\leq N_{\mathrm{noiseless}}$, where $N_{\mathrm{eval},t}^{\mathrm{weak}}$ includes the training and round-evaluation operations specified in Eq.~\eqref{eq:Nweak_vs_Nideal}. &
Provides a sufficient upper bound on the cumulative boosting resource.
\\

\hline
\end{tabular}
\caption{Additional assumptions specific to boosting.}
\label{tab:boosting_assumptions}
\end{table}
\FloatBarrier

\subsection{Expected-Risk-Aligned Resource Comparison}
\label{subsec:resource_comparison}

Recall from Section~\ref{sec:resource_ideal} that:
\begin{equation}
\mathrm{Shots}_{\mathrm{noiseless}}
=
N_{\mathrm{noiseless}}
N_{\mathrm{sh}}.
\label{eq:section6_ideal_shots}
\end{equation}
We can now state the central comparison result.
\begin{theorem}[\textbf{Expected-risk-aligned resource comparison for PEC
and boosting}]
\label{thm:resource_comparison}

Fix
$\mu\in(0,1]$,
$\varepsilon_{\mathrm{mar}}\in(0,1)$,
$m\in\mathbb N$,
and
$\delta\in(0,1)$.
Assume A1--A3 and the common-reference-class condition Eq.~\eqref{eq:noisy_class_reference_containment}.

Let $f_{\mathrm{PEC}}$ denote the deterministic PEC mean predictor and let $f_{\mathcal A}$ denote the normalized ensemble produced by a boosting algorithm $\mathcal A\in\{\mathrm{\text{Real Adaboost}},\text{Adaboost}\mathbb R\}$.

If
\begin{equation}
\widehat R_{S,\mu}(f_{\mathrm{PEC}})
\leq
\varepsilon_{\mathrm{mar}},
\qquad
\widehat R_{S,\mu}(f_{\mathcal A})
\leq
\varepsilon_{\mathrm{mar}},
\label{eq:theorem_common_margin_targets}
\end{equation}
then, with probability at least $1-\delta$ over the draw of the training sample,
\begin{equation}
R(f_{\mathrm{PEC}},c)
\leq
\varepsilon_{\mathrm{mar}}
+
\Delta_{\mathrm{gen}}(\mu,m,\delta),
\label{eq:theorem_pec_expected_risk}
\end{equation}
and
\begin{equation}
R(f_{\mathcal A},c)
\leq
\varepsilon_{\mathrm{mar}}
+
\Delta_{\mathrm{gen}}(\mu,m,\delta).
\label{eq:theorem_boost_expected_risk}
\end{equation}

Hence both strategies attain the same high-probability upper bound on expected $0$--$1$ classification risk.

Under P1--P3 and the matched per-evaluation precision used in Sections~\ref{sec:resource_ideal} and~\ref{sec:pec_cost_inflation}, a sufficient PEC shot budget is
\begin{equation}
\mathrm{Shots}_{\mathrm{PEC}}^{\mathrm{suff}}
\leq
\gamma^2
N_{\mathrm{noiseless}}
N_{\mathrm{sh}}
=
\gamma^2
\mathrm{Shots}_{\mathrm{noiseless}}.
\label{eq:theorem_pec_resource}
\end{equation}

For boosting algorithm $\mathcal A$, let $T_{\mathcal A}^{(\mathrm{mar})}$ be a sufficient round index for attaining Eq.~\eqref{eq:theorem_common_margin_targets}, and define
\begin{equation}
N_{\mathrm{eval}}^{\mathcal A}
\left(
T_{\mathcal A}^{(\mathrm{mar})}
\right)
:=
\sum_{t=1}^{T_{\mathcal A}^{(\mathrm{mar})}}
N_{\mathrm{eval},t}^{\mathrm{weak}}.
\label{eq:theorem_boost_evaluation_count}
\end{equation}
Its corresponding shot cost is
\begin{equation}
\mathrm{Shots}_{\mathcal A}
=
N_{\mathrm{eval}}^{\mathcal A}
\left(
T_{\mathcal A}^{(\mathrm{mar})}
\right)
N_{\mathrm{sh}}.
\label{eq:theorem_boost_resource_general}
\end{equation}
Under Assumption B3 (\ref{ass:B3}),
\begin{equation}
\mathrm{Shots}_{\mathcal A}
\leq
T_{\mathcal A}^{(\mathrm{mar})}
\mathrm{Shots}_{\mathrm{noiseless}}.
\label{eq:theorem_boost_resource_simple}
\end{equation}
\end{theorem}

\begin{proof}
For PEC, Eq.~\eqref{eq:pec_mean_hypothesis_class} implies that the deterministic mean predictor belongs to $\mathcal H_d^{\mathrm{noiseless}}$ and hence to $\operatorname{conv}(\mathcal H_{\mathrm{ref},\pm})$.

For boosting, Eq.~\eqref{eq:noisy_class_reference_containment} places every noisy base predictor in the convex hull of the noiseless class. Allowing the sign reversal used by the boosting procedures places every accepted signed weak hypothesis in $\operatorname{conv}(\mathcal H_{\mathrm{ref},\pm})$. Because the normalized ensemble is itself a convex combination of these hypotheses, Eq.~\eqref{eq:boosting_reference_class_membership} follows.

Theorem~\ref{thm:ensemble_margin_bound} can therefore be applied to both predictors using the same reference complexity term Eq.~\eqref{eq:common_generalization_term}. Combining that result with Eq.~\eqref{eq:theorem_common_margin_targets} yields Eqs.~\eqref{eq:theorem_pec_expected_risk} and \eqref{eq:theorem_boost_expected_risk}.

For PEC, Assumption~P3 bounds the number of expectation-value evaluations by the ideal reference budget, while PEC multiplies the required number of shots per evaluation by $\gamma^2$ at matched additive precision and confidence. This gives Eq.~\eqref{eq:theorem_pec_resource}.

For boosting, summing the weak-learner evaluation counts through the first sufficient round gives Eq.~\eqref{eq:theorem_boost_evaluation_count}. Multiplication by $N_{\mathrm{sh}}$ gives Eq.~\eqref{eq:theorem_boost_resource_general}, and Assumption B3 (\ref{ass:B3}) yields Eq.~\eqref{eq:theorem_boost_resource_simple}.
\end{proof}

Theorem~\ref{thm:resource_comparison} separates the statistical statement from the resource statement. Equality of the empirical margin-loss targets does not imply equality of the unknown expected risks. Rather, because both methods are bounded using the same reference class, it gives the same high-probability \emph{upper bound} on expected classification risk. The resource comparison then asks how many physical circuit executions each method requires to attain that common empirical target.

The most direct resource criterion is obtained before applying Assumption B3 (\ref{ass:B3}). For either boosting algorithm $\mathcal A$, the boosting evaluation count is smaller than the PEC reference-schedule cost whenever
\begin{equation}
N_{\mathrm{eval}}^{\mathcal A}
\left(
T_{\mathcal A}^{(\mathrm{mar})}
\right)
<
\gamma^2
N_{\mathrm{noiseless}}.
\label{eq:general_resource_crossover}
\end{equation}
This is the form used most naturally in numerical experiments, where the expectation-value evaluations can be counted directly. Assumption B3 (Section~\ref{ass:B3}) gives the simpler sufficient comparison
\begin{equation}
T_{\mathcal A}^{(\mathrm{mar})}
<
\gamma^2.
\label{eq:simplified_resource_crossover}
\end{equation}

\subsection{Algorithm-Specific Forms of the Crossover}
\label{subsec:algorithm_specific_crossover}

The general criterion Eq.~\eqref{eq:general_resource_crossover} takes different forms for Real AdaBoost and AdaBoost$\mathbb R$ because their positive-margin guarantees have different structures.

\paragraph{Real AdaBoost.}
Under Assumption~B1 and the contraction condition $B_{\mathrm{real}}(\mu,\eta)<1$, Section~\ref{subsec:real_adaboost_analysis} gives
\begin{equation}
T_{\mathrm{real}}^{(\mathrm{mar})}
(\eta,\mu,\varepsilon_{\mathrm{mar}})
=
\left\lceil\frac{
\log(1/\varepsilon_{\mathrm{mar}})
}{
-\log B_{\mathrm{real}}(\mu,\eta)
}\right\rceil.
\label{eq:section6_real_round_recall}
\end{equation}
The direct resource crossover is therefore
\begin{equation}
N_{\mathrm{eval}}^{\mathrm{real}}
\left(
T_{\mathrm{real}}^{(\mathrm{mar})}
\right)
<
\gamma^2
N_{\mathrm{noiseless}}.
\label{eq:real_resource_crossover_general}
\end{equation}
Under Assumption B3 (Section~\ref{ass:B3}), the simpler sufficient condition becomes
\begin{equation}
\left\lceil\frac{
\log(1/\varepsilon_{\mathrm{mar}})
}{
-\log B_{\mathrm{real}}(\mu,\eta)
}\right\rceil
<
\gamma^2.
\label{eq:real_resource_crossover_simple}
\end{equation}

Thus, for Real AdaBoost, the weak-edge lower bound produces an explicit a priori positive-margin resource condition once $\mu$, $\eta$, and $\varepsilon_{\mathrm{mar}}$ are fixed.

\paragraph{AdaBoost$\mathbb R$.}
Section~\ref{subsec:adaboostr_analysis} gives the trajectory-dependent sufficient round index
\begin{equation}
T_{\mathbb R}^{(\mathrm{mar})}
(\mu,\varepsilon_{\mathrm{mar}})
=
\min
\left\{
T\in\mathbb N:
E_T-2\mu A_T
\geq
2\log
\left(
\frac{1}{\varepsilon_{\mathrm{mar}}}
\right)
\right\},
\label{eq:section6_adaboostr_round_recall}
\end{equation}
whenever this set is nonempty. The corresponding direct resource crossover is
\begin{equation}
N_{\mathrm{eval}}^{\mathbb R}
\left(
T_{\mathbb R}^{(\mathrm{mar})}
\right)
<
\gamma^2
N_{\mathrm{noiseless}}.
\label{eq:adaboostr_resource_crossover_general}
\end{equation}
Under Assumption B3 (Section~\ref{ass:B3}), a sufficient comparison of the resource bounds is $T_{\mathbb R}^{(\mathrm{mar})}<\gamma^2$. Unlike Real AdaBoost, $T_{\mathbb R}^{(\mathrm{mar})}$ is not determined by the lower-edge parameter $\eta$ alone. The realized quantities
\[
E_T
=
\sum_{t=1}^{T}r_t^2
\qquad\text{and}\qquad
A_T
=
\sum_{t=1}^{T}\alpha_t
\]
must also be taken into account. We do not impose additional assumptions solely to upper-bound $A_T$. Instead, the AdaBoost$\mathbb R$ condition is evaluated from the realized boosting trajectory.

\subsection{Interpretation of the Comparison}
\label{subsec:comparison_interpretation}

The preceding analysis identifies the quantities that determine the resource comparison between PEC and boosting. The central distinction is that the PEC sampling overhead is determined by the circuit and noise model, whereas the boosting cost also depends on the learning task and on the resulting optimization trajectory. For a fixed task, noise model, and training procedure, this trajectory may still vary because of initialization, finite-shot sampling, and other internal training randomness.

For PEC, we apply the theoretical quasiprobability overhead to the measured resource of the noiseless reference procedure:
\begin{equation}
C_{\mathrm{PEC}}
=
\gamma^2\,C_{\mathrm{ref}}^{\mathrm{alg}}.
\label{eq:pec_projected_resource_interpretation}
\end{equation}
Here $\gamma^2$ is fixed by the circuit structure, noise strength, and chosen PEC decomposition. Thus, within the noise model considered in this work, the PEC contribution to the resource comparison can be calculated once the reference resource has been measured.

For boosting, the conditions in Section~\ref{subsec:algorithm_specific_crossover} identify the quantities that remain to be determined. Applying the Real AdaBoost round bound requires a uniform edge lower bound $\eta$ satisfying its contraction condition. The AdaBoost$\mathbb R$ criterion additionally depends on the realized quantities $E_T$ and $A_T$. For the noisy PQC family and weak-learner training procedure considered here, we do not derive an analytic expression for a positive lower bound $\eta$ that holds uniformly across boosting rounds. Establishing such a bound requires controlling the weak learner's performance under the adaptively reweighted distributions $D^t$, which we leave to future work. We therefore retain the uniform-edge condition as an assumption and evaluate the realized edges and weak-learner costs along the training trajectories, rather than predicting them from physical noise strength alone.

The most direct comparison for either boosting method is therefore the resource criterion derived in Section~\ref{subsec:resource_comparison}:
\begin{equation}
    C_{\mathrm{boost}}^{\mathrm{alg}}
    <
    C_{\mathrm{PEC}}
    =
    \gamma^2\,
    C_{\mathrm{ref}}^{\mathrm{alg}},
\label{eq:direct_numerical_resource_criterion}
\end{equation}
where $C_{\mathrm{boost}}^{\mathrm{alg}}$ is the algorithm-required circuit resource used by the boosting trajectory to reach the same empirical margin-loss target as the reference procedure.

This observation motivates the numerical study that follows. For each classification task, circuit depth, and physical noise strength, we first measure the resource required by the noiseless reference learner to reach the prescribed empirical margin-loss target. Equation \eqref{eq:pec_projected_resource_interpretation} then determines the corresponding PEC resource.  We separately run the noisy boosting procedures and record their realized weak-learner behavior and algorithm-required circuit resources up to the same target.  Comparing these quantities determines whether the rapid growth of the PEC sampling overhead is offset by the task-dependent cost of constructing a boosted ensemble.

The numerical experiments therefore do not assume a particular dependence of the weak-learner edge on circuit depth or physical noise.  Instead, they measure the resulting boosting resource directly and determine the depth--noise regimes in which the ordering of $C_{\mathrm{PEC}}$ and $C_{\mathrm{boost}}^{\mathrm{alg}}$ changes.

\section{Numerical Experiments}
\label{sec:numerics}

The theoretical analysis compares two distinct approaches to learning with noisy parameterized quantum circuits (PQCs). We are interested in how the resources required by PEC and boosting change with the physical noise strength and circuit depth, and whether these different dependences give rise to regimes in which one method requires fewer circuit executions than the other.

We investigate this question using a teacher--student binary-classification setting. \footnote{The code for the numerical experiments is available at \url{https://github.com/NastyaVicodin/boosting-noisy-qml}.} For each task, a fixed ``teacher'' PQC generates the binary labels, while the compared learning procedures train separate ``student'' PQCs from a finite sample of those labeled inputs. This construction gives us a controlled family of classification problems for which the target labeling rule is known and can be varied systematically across circuit depths and random teacher instances. We refer to each such teacher-defined classification problem as a \emph{teacher task}. In the theory-based comparison, the reference learner and both boosting methods are required to reach the same task-specific empirical margin-loss target,
\begin{equation}
    \widehat R_{S_j,\mu}\!\left(f_j^{\mathrm{method}}\right)
    \le
    \varepsilon_{\rm mar}^{(j)},
    \label{eq:numerics_common_target_intro}
\end{equation}
where $f_j^{\mathrm{method}}$ denotes the predictor produced by the method being evaluated on teacher task $j$ and $\widehat R_{S_j,\mu}$ is the empirical margin loss introduced in Section~\ref{subsec:generalization_bounds}. The target $\varepsilon_{\rm mar}^{(j)}$ is fixed independently for each teacher task $j$ before the PEC--boosting resource comparison and is identical for all methods on that task. Consequently, the primary numerical quantity is the number of algorithm-required circuit executions needed to produce a hypothesis satisfying Eq.~\eqref{eq:numerics_common_target_intro}.

We also perform a separate resource comparison in which the methods are required to reach the same teacher-specific empirical \(0\)–\(1\) classification error on the validation set, derived from the noiseless reference. This target and the corresponding comparison are described in Section~\ref{sec:resource_crossover}.

We compare two boosting methods, AdaBoost$\mathbb{R}$ and Real AdaBoost, against two numerical realizations of PEC. On the complete depth--noise grid we use \emph{Projected PEC}: the algorithmic cost of the corresponding noiseless reference learner is multiplied by the PEC sampling overhead $\gamma^2$. This provides the full-grid PEC resource estimate used in the main comparison. In addition, we use \emph{Sampled PEC} on a smaller preselected subset of depth--noise settings, where the quasiprobability sampling procedure is simulated explicitly. The Sampled-PEC experiments serve as a direct numerical check of the projected PEC resource model and are kept distinct from the full-grid Projected-PEC results.

The empirical margin-loss target determines only the resource-to-target comparison. Nevertheless, satisfying the training margin condition is not by itself the ultimate objective of the learning task. The quantity of interest is the expected $0$--$1$ classification risk, i.e., the probability of misclassification on a fresh input drawn from $\mathcal D$:
\begin{equation}
    R_{\mathcal D}(f,c)
    =
    \Pr_{x\sim\mathcal D}
    \bigl[c(x)f(x)\le 0\bigr],
\end{equation}
where $c:\mathcal X\to\{-1,+1\}$ is the teacher-defined target labeling function and $f:\mathcal X\to\mathbb R$ is the learned prediction score.

We therefore additionally report the empirical training classification error and the classification error on an independent test set. These quantities do not affect the resource comparison; they are used as generalization checks to verify that reaching the common training margin target with fewer resources is not accompanied by a deterioration of predictive performance on unseen data.

\subsection{Teacher--student classification tasks}
\label{sec:numerics_teacher_student}

\begin{figure*}[htbp]
    \centering
    \includegraphics[width=0.90\textwidth,keepaspectratio]
        {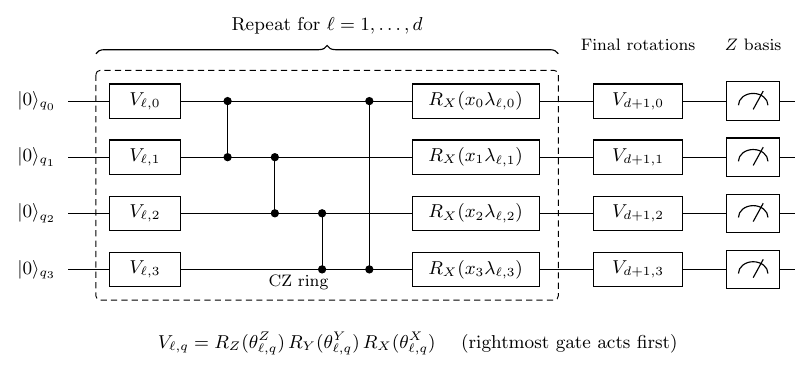}
    \caption{\textbf{Circuit architecture shared by teachers and students.}
    Starting from $|0\rangle^{\otimes4}$, each of the $d$ repeated layers applies a local block $V_{\ell,q}$ to every qubit, controlled-$Z$ gates on $(0,1),(1,2),(2,3),(3,0)$, and then $R_X(x_q\lambda_{\ell,q})$, where $x_q$ is an input feature and $\lambda_{\ell,q}$ its scale parameter. A final local block $V_{d+1,q}$ lies outside the repeated region. Measurements in the $Z$ basis yield a parity outcome with expectation
    $\langle Z_0Z_1Z_2Z_3\rangle$.}
    \label{fig:pqc_architecture}
\end{figure*}

We construct binary-classification tasks using four-qubit teacher circuits at depths $d\in\{1,\ldots,5\}$. Each teacher supplies the labels for a dataset, and student circuits are trained to predict these labels from the inputs. Teachers and students share the data-reuploading architecture shown in Fig.~\ref{fig:pqc_architecture}. For the corresponding width $n=4$ architecture, the numbers of one- and two-qubit gate locations are
\begin{equation}
    G_1(n,d)=n(4d+3),
    \qquad
    G_2(n,d)=nd.
    \label{eq:teacher_student_gate_counts}
\end{equation}

At a fixed depth, let $U_j(x)$ denote the circuit of teacher $j\in\{0,1,2\}$ evaluated at input $x\in[0,2\pi]^4$. The exact noiseless parity expectation defines its score,
\begin{equation}
    s_j(x)=
    \langle 0^{\otimes4}|
    U_j(x)^\dagger O\,U_j(x)
    |0^{\otimes4}\rangle,
    \qquad O=Z_0Z_1Z_2Z_3 ,
    \label{eq:teacher_score}
\end{equation}
and the corresponding label is $c_j(x)=\operatorname{sign}s_j(x)$, with $\operatorname{sign}(0)=+1$. These labels are computed from exact teacher expectations and remain fixed throughout the learning experiments.

We generate inputs by drawing their four coordinates independently and uniformly from $[0,2\pi]$. We keep an input only if $|s_j(x)|\ge0.2$, excluding examples for which the teacher score is close to zero. We denote the distribution of the retained inputs by $\mathcal D^X_j$; it is uniform over the region that passes this check:
\begin{equation}
    \mathcal D^X_j
      =\operatorname{Unif}\!\left(
          \left\{x\in[0,2\pi]^4:\ |s_j(x)|\ge0.2\right\}
        \right).
    \label{eq:conditional_teacher_distribution}
\end{equation}
We use $\mathcal D_j$ for the distribution of the corresponding labeled examples $(x,c_j(x))$.

To generate candidate teachers, we draw their rotation angles and input scales independently from a normal distribution with mean zero and standard deviation one. We check that their scores vary sufficiently across inputs and that both classes contain enough examples. To check whether students can learn each task, we use five preliminary student training runs per teacher. A teacher passes this check if the median validation accuracy is at least $90\%$ at depths $1$ and $2$, $80\%$ at depths $3$ and $4$, and $75\%$ at depth $5$. We choose three teachers at each depth from those meeting these criteria, giving $15$ tasks in total. Appendix~\ref{app:teacher_task_details} describes the screening criteria and the final teacher selection in more detail.

For each teacher, we generate $1536$ training examples, $384$ validation examples, and $1024$ test examples using the sampling rule above. We do not force the two classes to contain equal numbers of examples. The class counts are given in Figure~\ref{fig:teacher_class_balance} in the Appendix~\ref{app:teacher_task_details}.

\subsection{Noise model and training procedures}
\label{sec:numerics_noise}
\label{sec:numerics_training}

We apply the depolarizing channels defined in Section~\ref{sec:pec_depol} after every one- and two-qubit gate of the student architecture. Each channel acts only on the qubits involved in that gate. We denote the corresponding noise strengths by $p_{\rm 1q}$ and $p_{\rm 2q}$ and set $p_{\rm 1q}=p_{\rm 2q}=p$, with $p\in\{0.005,0.01,0.02\}$ at every depth $d=1,\ldots,5$. The reference circuit uses $p=0$.

For an input $x$ and circuit parameters $\theta$, we calculate the final density matrix $\rho_{\theta,p}(x)$ and its mean prediction $h_{\theta,p}(x)=\operatorname{Tr}[O\rho_{\theta,p}(x)]$, with $O$ defined in Section~\ref{sec:numerics_teacher_student}. Thus $h_{\theta,0}(x)$ is the ideal circuit mean without physical gate noise. Reference and boosting predictions use $N_{\rm sh}=25$ measurement shots per circuit and input:
\begin{equation}
    K\sim\operatorname{Binomial}\!\left(
        N_{\rm sh},\frac{1+h_{\theta,p}(x)}{2}
    \right),
    \qquad
    \widehat h_{\theta,p}(x)=\frac{2K}{N_{\rm sh}}-1.
    \label{eq:finite_shot_expectation}
\end{equation}
The reference therefore retains measurement fluctuations even though it has no physical gate noise.

We fit the circuits using a smooth hinge loss evaluated at the sampled prediction $z$:
\begin{equation}
    \begin{aligned}
    \ell(y,z)&=\Psi_G(yz;\sigma), \qquad \sigma=0.5,\\
    \Psi_G(a;\sigma)
      &=(1-a)\Phi_{\mathcal N}\!\left(\frac{1-a}{\sigma}\right)
        +\sigma\phi_{\mathcal N}\!\left(\frac{1-a}{\sigma}\right).
    \end{aligned}
    \label{eq:numerics_training_smooth_hinge}
\end{equation}
Here $\Phi_{\mathcal N}$ and $\phi_{\mathcal N}$ are the cumulative distribution function and density of a normal distribution with mean zero and standard deviation one. This fitting loss is separate from the criteria used to decide whether a learning target has been reached. We use this loss for fitting because its derivative remains nonzero for misclassified examples, whereas the margin loss used for comparison is constant there.

We initialize the student rotation angles independently from a normal distribution with mean zero and standard deviation $0.1$. The input scales are initialized independently with mean one and the same standard deviation. These distributions give small initial rotation angles and input scales close to one, corresponding to little initial rescaling of the inputs. Each epoch processes the $1536$ training examples in $24$ minibatches of $64$. We use Adam~\cite{KingmaBa2015} with learning rate $0.01$, clipping the full gradient to Euclidean norm at most one before applying the optimizer to limit the influence of unusually large gradients without restricting the parameter values.

The parameter update combines the loss derivative at the sampled prediction with the derivative of the circuit mean, calculated by JAX automatic differentiation~\cite{jax2018github}. We differentiate the ideal mean for the reference and explicit PEC, and the noisy mean for the boosting weak learners. Measurement fluctuations therefore affect the parameter update through the sampled loss derivative, even though the circuit-mean derivative is calculated without measurement sampling. Appendix~\ref{app:numerics_gradient_details} specifies this derivative convention.

To decide when to stop fitting an individual circuit, we track the mean of its minibatch losses in each epoch. An epoch counts as an improvement when this mean is strictly below its best recorded value. The reference stops after $100$ consecutive epochs without improvement, while each boosting weak learner uses a patience of $20$ epochs. Both allow at most $500$ epochs per fit. When fitting ends through either of these limits, we restore the parameters saved at the best epoch.

A separate rule determines when to stop adding weak learners. After each accepted round, we evaluate the normalized ensemble's margin loss on the full training set at $\mu=0.1$, using exact noisy circuit means. Boosting stops after $20$ accepted rounds without an improvement greater than $10^{-5}$ over the best recorded value, or after $500$ accepted rounds. This rule tracks the progress of the whole ensemble, whereas the epoch-based rule controls the fitting of each individual circuit. Target-based termination, shot-budget limits, and resource costs are specified in Section~\ref{sec:numerics_target_stopping}.

\subsection{Comparison protocol and resource costs}
\label{sec:numerics_comparison}
\label{sec:numerics_margin_target}
\label{sec:numerics_target_stopping}
\label{sec:numerics_resource_accounting}
\label{sec:numerics_aggregation}

We compare the number of circuit shots required to reach two learning targets, considered separately. The first uses the empirical margin loss defined in Eq.~\eqref{eq:empirical_margin_loss}, with $\mu=0.1$, and therefore penalizes both incorrect predictions and correct predictions with small positive margins. The second uses validation classification accuracy. For teacher $j$ at depth $d$, we consider a training margin loss no greater than $\varepsilon_{\rm mar}^{(d,j)}$ and a validation accuracy of at least $a^\star_{d,j}$:
\begin{equation}
    \begin{aligned}
    \widehat R_{S_{d,j}^{\rm train},\mu}(\widehat f)
        &\leq \varepsilon_{\rm mar}^{(d,j)},
        \qquad \mu=0.1,\\
    \widehat A_{\rm val}(\widehat f)
        &\geq a^\star_{d,j}.
    \end{aligned}
    \label{eq:num_common_target}
\end{equation}
Here $\widehat f$ is the finite-shot prediction of the reference circuit or the normalized boosting ensemble, and $\widehat A_{\rm val}$ is its validation accuracy. Both thresholds are set for each depth using separate calibration runs and kept fixed across methods, repetitions and noise strengths for the same task. Appendix~\ref{app:numerics_targets} gives their construction, numerical values and the details of the accuracy check.

For each learning target, we retain the first confirmed model and the resources accumulated up to its confirmation. Training can continue to seek the other target, so the two comparisons may select different reference parameters or different boosting ensembles from the same run.

We count resources in circuit shots, with one circuit execution and its measurement outcome contributing one shot. The primary cost includes measurements used for fitting and, for boosting, the full-training predictions needed to determine the ensemble coefficients and the next round's example weights:
\begin{equation}
    \begin{aligned}
    C_{\rm ref}^{\rm alg}
        &= C_{\rm ref}^{\rm fit},\\
    C_{\rm boost}^{\rm alg}
        &= C_{\rm boost}^{\rm fit}
           + C_{\rm boost}^{\rm edge}.
    \end{aligned}
    \label{eq:numerics_algorithm_cost}
\end{equation}
These quantities accumulate through the first confirmation of the target being considered. Boosting includes all fitting and edge acquisitions up to that point, including those for attempted weak learners that are not accepted into the ensemble. At $25$ shots per input, one fitting update costs $64\times25=1600$ shots, and one full-training evaluation of a weak learner costs $1536\times25=38\,400$ shots. 

For projected PEC, the resource assigned to a teacher task is the matched reference cost multiplied by the PEC sampling overhead:
\begin{equation}
    C_{{\rm PEC},d,p,j}^{\rm proj}
        =
    \gamma^2(n=4,d,p)\,
    C_{{\rm ref},d,j}^{\rm alg}.
    \label{eq:num_pec_alg_cost}
\end{equation}
Here the first argument of $\gamma^2$ (defined earlier in Eq.~\eqref{eq:gamma_depol}) specifies the four-qubit circuit width. The projection is applied separately to each learning target, using the reference cost for the same teacher and target. Sampled PEC instead counts each sampled corrected-circuit execution directly, without applying an additional $\gamma^2$ multiplier. Its separate experiments and the interpretation of the projection are discussed in Section~\ref{sec:sampled_pec_validation}.

A run that ends without confirming a target is recorded as unsuccessful for that target, together with its consumed resources. An unconfirmed target in an unfinished run remains unresolved; a target already confirmed on that run retains its selected model and cost.

In the main reference and boosting comparison, each task has three training run repetitions. For each method and learning target, we report a teacher's cost when at least two repetitions have confirmed successes, taking the median of all confirmed first-hit costs. These medians describe successful runs and are accompanied by the corresponding success counts. Test data do not enter target construction, fitting or model selection; predictive performance is considered separately in Section~\ref{sec:classification_generalization}.

\subsection{Resource requirements and predictive performance}
\label{sec:numerics_resources_performance}
\label{sec:resource_crossover}
\label{sec:classification_generalization}
\label{sec:numerics_generalization_check}
\begin{figure*}[!htbp]
    \centering
    \includegraphics[width=0.83\textwidth]{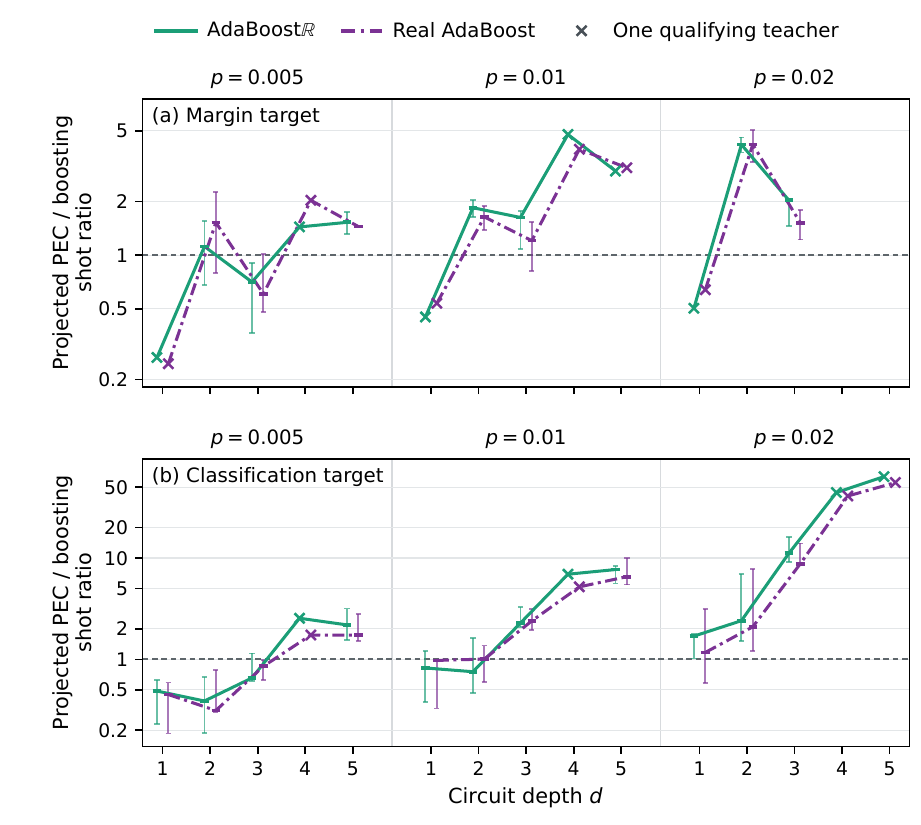}
    \caption{\textbf{Resources required to reach the two learning targets.}
    The ratio of Projected-PEC to boosting cost is shown for the margin target above and the classification target below. Each teacher's ratio uses median costs over successful repetitions. Lines connect medians across teachers meeting the two-out-of-three success requirement for both the reference and boosting. At $p=0.02$ and $d=4,5$, neither boosting method confirms the margin target in any of its nine repetitions, so no cost to reach that target is available. Vertical bars give the minimum and maximum; short ticks mark the medians. Crosses indicate one qualifying teacher, for which no range is shown. Ratios above one favour boosting. The vertical axes are logarithmic. Projected PEC is calculated from the reference cost.}
    \label{fig:n25_resource_ratios}
\end{figure*}

Figure~\ref{fig:n25_resource_ratios} compares the shots required by AdaBoost$\mathbb R$ and Real AdaBoost with the Projected-PEC cost defined in Section~\ref{sec:numerics_comparison}. A teacher enters the comparison when both the reference and the relevant boosting method reach the target in at least two of three repetitions. Results based on one qualifying teacher are marked separately from summaries across two or three teachers.

Boosting becomes more favourable relative to Projected PEC as the physical noise increases. For the classification target, the advantage also tends to grow with depth, although the depth dependence is not monotonic. Boosting requires fewer shots in $39$ of the $56$ displayed comparisons across methods, depths, noise strengths and targets.

\begin{figure*}[!htbp]
    \centering
    \includegraphics[width=0.84\textwidth]{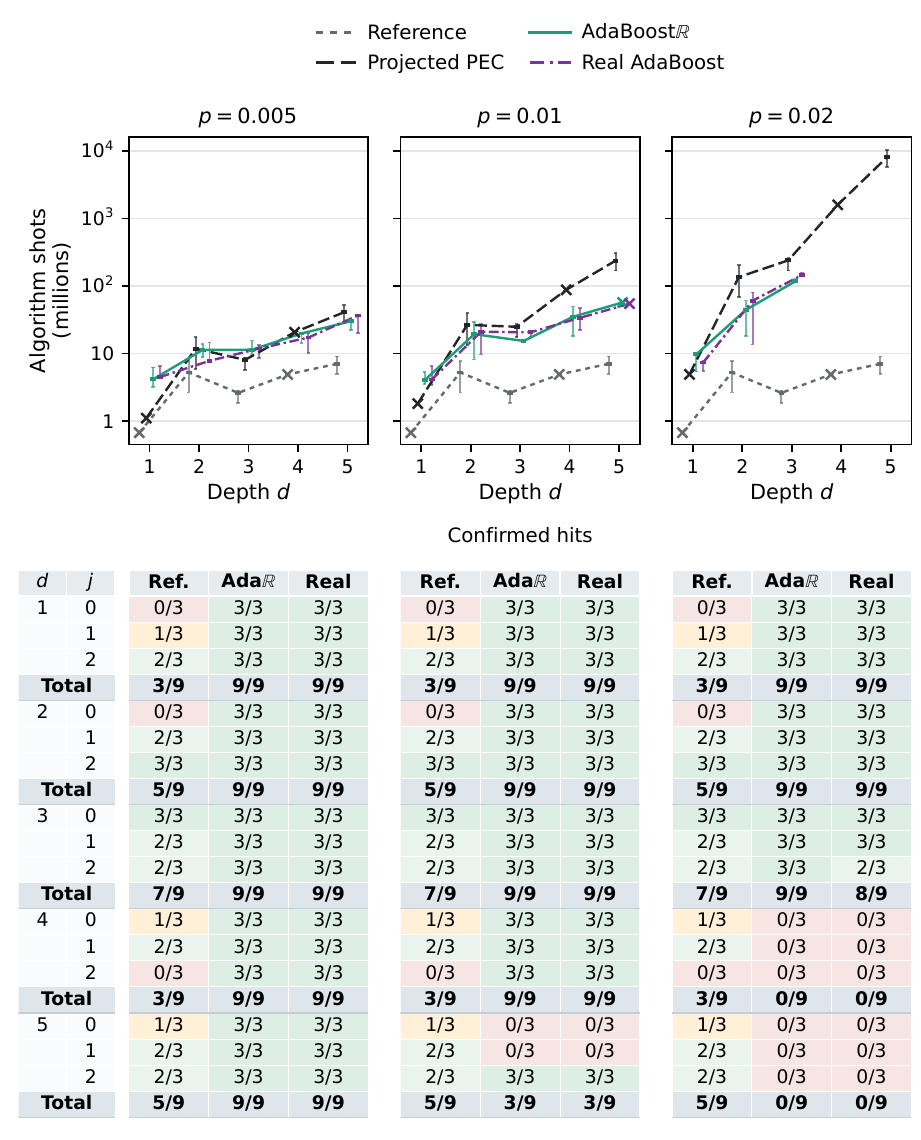}
    \caption{\textbf{Margin-target costs and successful repetitions.}
    Teachers qualify separately for each method after at least two of three repetitions succeed. Each teacher's cost is the median over successful repetitions. Lines and ticks show medians across teachers, bars span their minimum and maximum. A cross marks a cost based on one qualifying teacher; its value is the median over that teacher's successful repetitions. Tables show all confirmed hit counts. Green, yellow and red cells indicate at least two, one and zero successes, respectively. The reference is reused across noise settings. Projected PEC uses reference costs. Cost axes are logarithmic.}
    \label{fig:n25_margin_costs_hits}
\end{figure*}

\begin{figure*}[!htbp]
    \centering
    \includegraphics[width=0.84\textwidth]{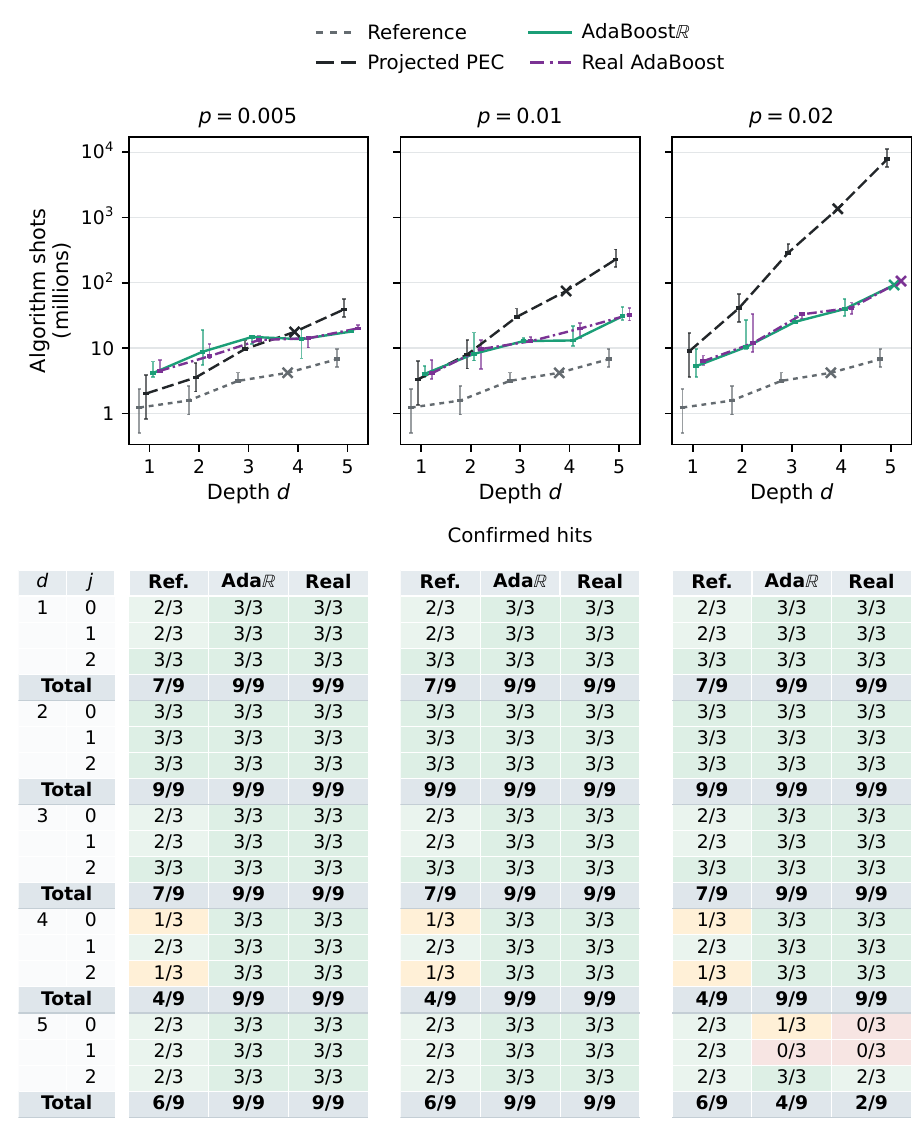}
    \caption{\textbf{Classification-target costs and successful repetitions.}
    Costs, ranges and hit counts follow the conventions of Fig.~\ref{fig:n25_margin_costs_hits}. A cost is omitted only if no teacher meets the success requirement.}
    \label{fig:n25_classification_costs_hits}
\end{figure*}

The ratio plot describes costs on tasks where both methods reach the target sufficiently often. Figures~\ref{fig:n25_margin_costs_hits} and~\ref{fig:n25_classification_costs_hits} show the absolute costs together with all successful-repetition counts, including tasks that cannot enter a paired cost comparison. They show that boosting can also reach targets consistently where the reference fails on the two-out-of-three requirement. At $d=4$, for example, both boosting methods reach the classification target in every repetition at all three noise strengths, whereas the reference meets the requirement for only one of the three teachers. The ratios at this depth therefore describe one teacher, despite consistent boosting success on all three. There are also limits to this improvement: under the stopping rules used here, neither boosting method reaches the margin target at $d=4$ or $5$ when $p=0.02$.

\begin{figure*}[!htbp]
    \centering
    \includegraphics[width=0.9\textwidth]{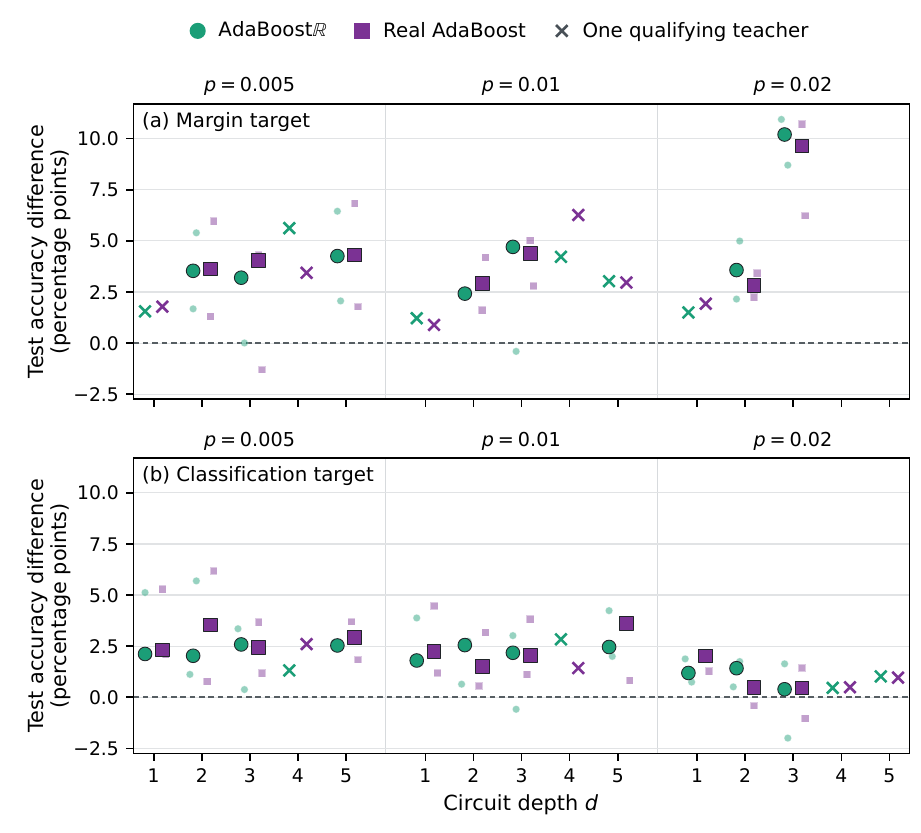}
    \caption{\textbf{Test accuracy of the succeeded models relative to the reference.}
    The upper and lower panels correspond to the margin and classification targets. A teacher contributes when both the reference and the corresponding boosting method confirm that target in at least two of three training repetitions, as in Figure~\ref{fig:n25_resource_ratios}. Small symbols show the difference between the boosting and reference median test accuracies within each teacher, in percentage points; large symbols give the median difference across two or three qualifying teachers. When only one teacher qualifies, its result is shown by a cross. The missing margin comparisons at $p=0.02$, $d=4,5$ arise because neither boosting method confirms that target in any of its nine repetitions at either depth. Positive values favour boosting; the dashed line marks equal accuracy.}
    \label{fig:n25_full_grid_test_accuracy_differences}
\end{figure*}

We next evaluate the models selected at the first confirmed attainment of each target on the test set, which is not used for fitting or model selection. For each method and teacher, we then take the median over successful training repetitions and subtract the reference median from the boosting median. Figure~\ref{fig:n25_full_grid_test_accuracy_differences} shows these differences for the same teacher subsets as Figure~\ref{fig:n25_resource_ratios}: $36$ comparisons include at least two qualifying teachers, and a further $20$ include one qualifying teacher.

The median test-accuracy difference is positive in all $36$ comparisons, with gains ranging from about $0.4$ to $10.2$ percentage points. In particular, all $25$ comparisons in this subset where boosting has a lower training cost than Projected PEC also have a positive median test-accuracy difference. Some individual teacher comparisons favour the reference, even when the median favours boosting. The reference test results describe the predictors associated with the Projected-PEC cost benchmark; the performance of explicitly sampled PEC is examined in Section~\ref{sec:sampled_pec_validation}.

\subsection{Projected vs Sampled PEC cost}
\label{sec:sampled_pec_validation}

Sampled PEC estimates the noiseless circuit score by sampling the gatewise corrections and averaging the measured outcomes with their prescribed signs and weights (Section~\ref{subsec:pec}). The additional Pauli corrections are ideal under the noise model used here. We use the sampled PEC score in the fitting loss, retaining the simulator-supplied derivatives described in Section~\ref{sec:numerics_training}. Each sampled circuit counts as one raw shot, so the recorded PEC cost already includes its sampling overhead.

We use these experiments to assess how well Projected PEC describes the resources needed for learning, and whether fewer shots per prediction can preserve consistent target attainment at a lower total cost. For comparison with the $25$-shot reference, the usual PEC variance bound gives an allocation scale of $25\times\gamma^2(4,d,p)$. This scale is related to the accuracy of an individual prediction; the learning targets may still be reached with fewer shots. Within each PEC run, the same allocation is used for fitting and target checks.

\begin{figure*}[!htbp]
    \centering
    \includegraphics[width=\textwidth]{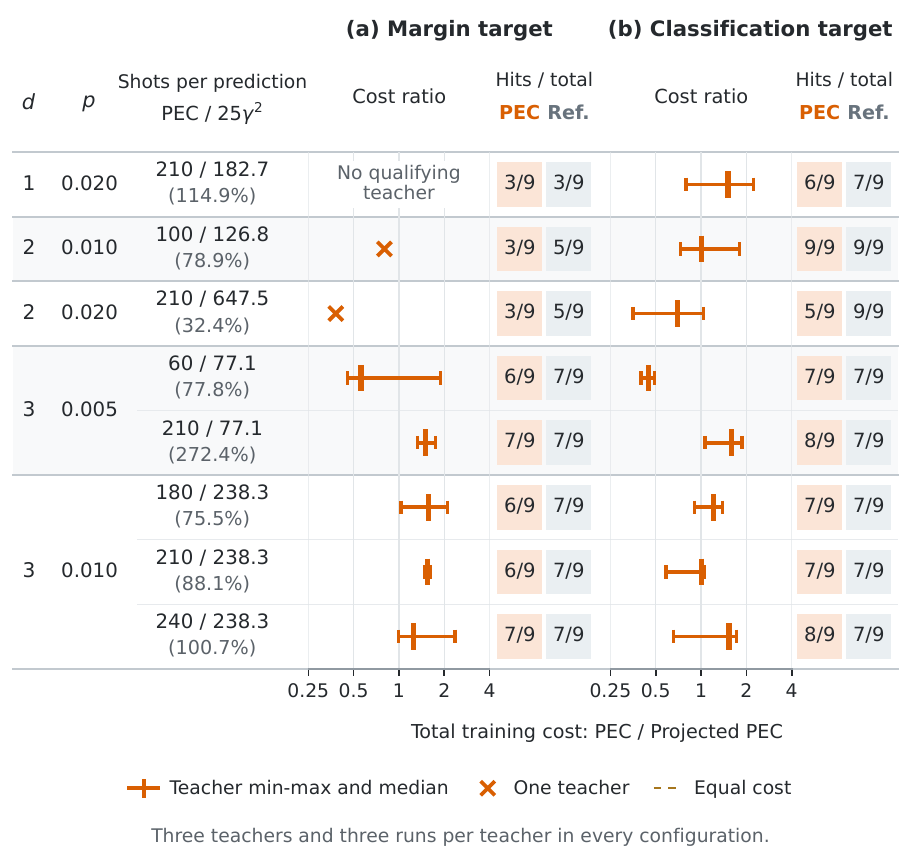}
    \caption{\textbf{Sampled PEC resources and target attainment.}
    Each configuration has three teachers and three repetitions per teacher. The allocation column lists the actual PEC shots per prediction and the projected allocation $25\gamma^2$. Percentages below these values give $100N_{\mathrm{PEC}}/(25\gamma^2)$. For each teacher, the resource ratio is the median successful PEC cost divided by $\gamma^2$ times the median successful reference cost. Both procedures must reach the target in at least two of that teacher's three repetitions. Thick vertical bars show the median across qualifying teachers, and horizontal segments span their minimum and maximum ratios. These ranges are not confidence intervals. Crosses denote one qualifying teacher. Ratios below one favour sampled PEC; the dashed line marks equal costs. A missing ratio means no teacher qualifies for both procedures.}
    \label{fig:n25_explicit_pec_resources_hits}
\end{figure*}

Figure~\ref{fig:n25_explicit_pec_resources_hits} compares the sampled PEC first-hit cost with the projected cost for the same teacher, together with the number of completed runs reaching each target. Reducing the shots per prediction can save resources, but additional optimizer updates can offset that saving.

\begin{figure*}[!htbp]
    \centering
    \includegraphics[width=\textwidth]{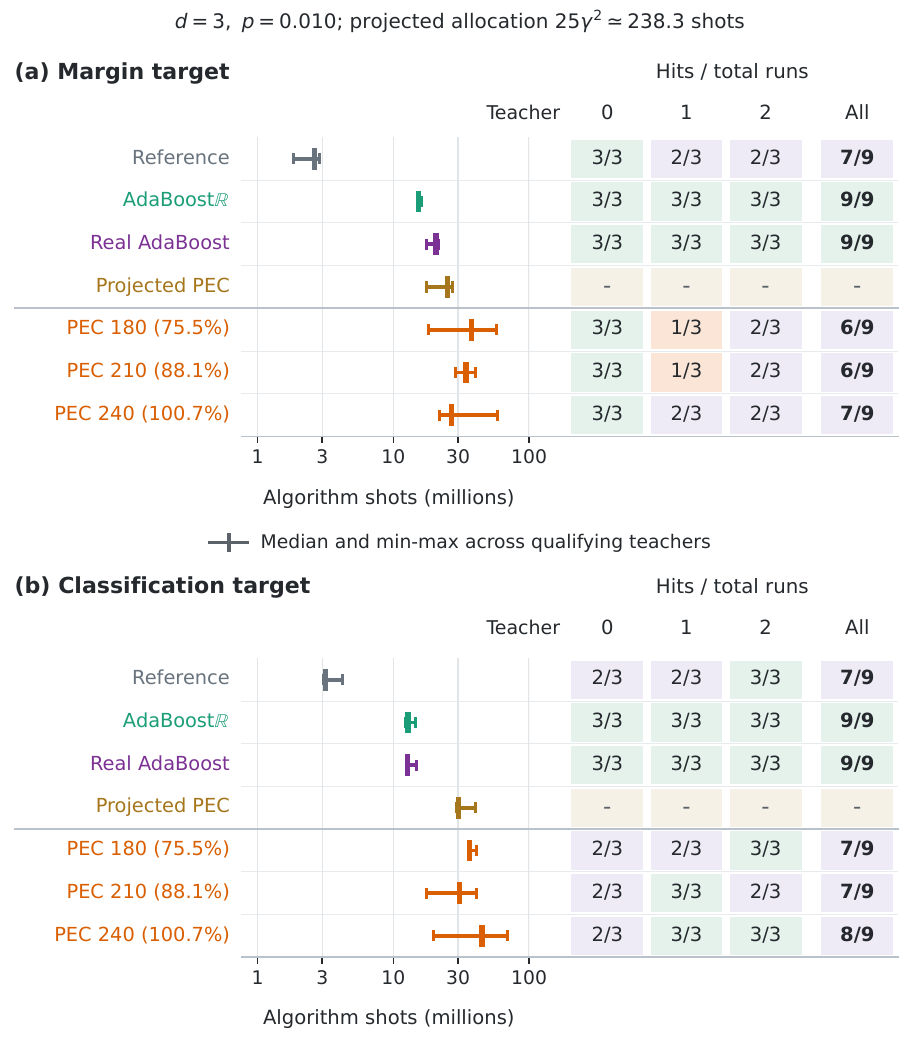}
    \caption{\textbf{Boosting and sampled PEC at $d=3$, $p=0.01$.}
    Each learning procedure has three completed repetitions for each of three teachers. For each teacher, the cost is the median algorithm-shot count to the first confirmed target hit. A teacher contributes when at least two of its three runs succeed. Thick vertical bars show the median across qualifying teachers, and horizontal segments show their minimum and maximum costs; these ranges are not confidence intervals. The tables report hits for each teacher. Projected PEC multiplies the reference costs by $\gamma^2$ and has no independently observed hit count. Sampled PEC shots allocation is specified per experiment and its percentage of the unrounded projection $25\gamma^2$, approximately $238.3$ shots here.}
    \label{fig:n25_all_methods_pec_comparison}
\end{figure*}

We next compare sampled PEC with the reference and both boosting methods at $d=3$ and $p=0.01$ (Figure~\ref{fig:n25_all_methods_pec_comparison}). All methods use the same teacher tasks and learning targets. The figure places the two learning targets together and reports each teacher's successful-repetition count beside the resource costs.

Both boosting methods reach both targets in all nine runs, while each sampled PEC allocation includes runs that do not reach the target (Figure~\ref{fig:n25_all_methods_pec_comparison}). Comparing costs within the same qualifying teachers before taking medians, the PEC-to-boosting cost ratios range from approximately $1.5$ to $2.4$ for the margin target and from $2.1$ to $3.5$ for classification, across the three PEC allocations and the two boosting methods. The effect of the PEC allocation differs between targets: the $240$-shot setting has the lowest reported PEC median cost for the margin target but the highest for classification. Reducing the shots used for each prediction therefore does not consistently reduce the total training cost.

\subsection{Boosting trajectories and theoretical bounds}
\label{sec:numerics_boosting_trajectories}

We finally compare the theoretical round counts with the number of rounds needed to reach the margin target in the experiments. Let $T_{\rm num}$ denote the first round at which this target is confirmed. For Real AdaBoost, we evaluate Eq.~\eqref{eq:Tstar_real_adaboost} using the smallest measured edge up to that round:
\begin{equation}
    \begin{aligned}
    \widehat\eta_{\min}
        &=\min_{1\leq t\leq T_{\rm num}}\widehat r_t,\\
    \widehat T_{\rm real}^{(\rm mar)}
        &=\left\lceil
          \frac{\log(1/\varepsilon_{\rm mar})}
          {-\log B_{\rm real}(\mu,\widehat\eta_{\min})}
          \right\rceil,
          \qquad \mu=0.1.
    \end{aligned}
    \label{eq:n25_calculated_rounds}
\end{equation}
Here $B_{\rm real}$ is defined in Eq.~\eqref{eq:real_adaboost_contraction_factor}. We report a finite count when $\widehat\eta_{\min}>\mu$ and $B_{\rm real}(\mu,\widehat\eta_{\min})<1$. These calculations use the recorded training histories. They do not establish that the same lower edge persists in later rounds, and applying the guarantees to the implemented finite-shot updates requires further analysis.

\begin{figure*}[!htbp]
    \centering
    \includegraphics[width=\textwidth]{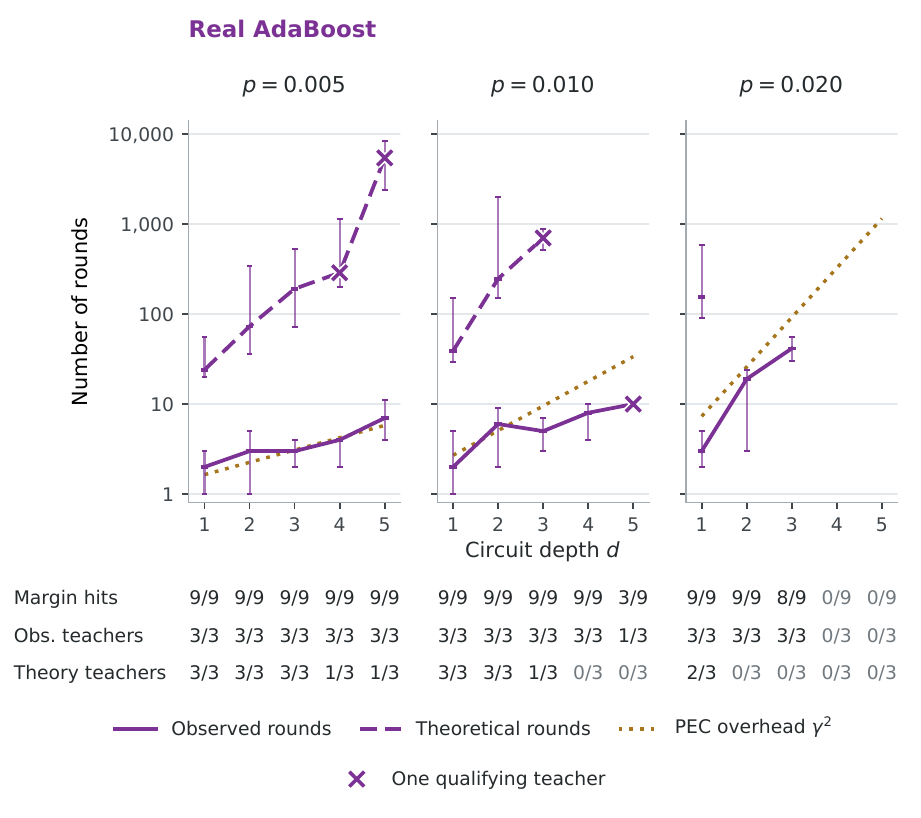}
    \caption{\textbf{Observed and calculated boosting rounds for Real AdaBoost.}
    Solid purple lines show the number of rounds at the attainment of the margin target; dashed purple lines show the theoretical count from Eq.~\eqref{eq:n25_calculated_rounds}, evaluated using the recorded edges. At each depth and noise strength, the observed curve includes the successful runs of every teacher for which at least two of three runs reach the target, including runs without a finite theoretical count. The theoretical curve requires at least two runs of a teacher to both reach the target and give a finite theoretical count. The two curves can therefore show different runs and teachers. Lines connect medians over the runs included in each curve, and vertical bars show their minimum and maximum, not confidence intervals. A cross ($\times$) marks a result supported by only one teacher on the corresponding curve. The ochre curve shows the PEC sampling overhead $\gamma^2$, not a PEC round count. The vertical axis is logarithmic. Below each panel, the rows give all confirmed margin hits out of nine runs. A missing summary means that no teacher meets the inclusion rule for that curve.}
    \label{fig:n25_boosting_rounds}
\end{figure*}

Real AdaBoost reaches the margin target in $110$ of its $135$ runs. Figure~\ref{fig:n25_boosting_rounds} shows the round counts for every teacher with at least two successful runs out of three, including cases where the theoretical expression gives no finite round count. Among the $57$ runs contributing to the theoretical curve, the median observed count is $3$ and the median theoretical count is $120$. The additional observed results show that learning can also succeed in settings where no teacher meets the condition for inclusion in the theoretical curve. Thus, the sufficient-round formula is substantially more demanding than the observed training. The number of qualifying teachers also decreases in many of the deeper or noisier settings, even where the measured margin target is reached consistently.

For AdaBoost$\mathbb R$, the corresponding condition in Eq.~\eqref{eq:T_adaboostr_margin} is $E_T-2\mu A_T\geq2\log(1/\varepsilon_{\rm mar})$, where $E_T$ is the sum of squared round edges and $A_T$ is the sum of the ensemble coefficients. This condition is not satisfied at any recorded round, although $111$ of the $135$ runs reach the measured margin target. It therefore gives no sufficient round index within these training histories.

To identify where the bounds become loose, we evaluate their intermediate expressions at the first confirmed margin hit of each successful run. This gives $111$ AdaBoost$\mathbb R$ and $110$ Real AdaBoost models from tasks satisfying the two-of-three margin-success rule. For each model, let $u_i=y_i\widehat F_T(x_i)/A_T$ be the normalized signed margin and let $R_{S,\mu}$ denote its empirical margin loss. On the same predictions, we evaluate the empirical margin error and exponential expression,
\begin{equation}
    \begin{aligned}
    R_{S,\mu}^{\rm mar}
        &=\frac{1}{m}\sum_{i=1}^{m}\mathbf 1\{u_i\leq\mu\},\\
    \mathcal B_T^{\exp}
        &=\frac{1}{m}\sum_{i=1}^{m}
          \exp\!\left[A_T(\mu-u_i)\right].
    \end{aligned}
    \label{eq:n25_bound_intermediates}
\end{equation}
Together with the final edge expression $\mathcal B_T^M$ from Eq.~\eqref{eq:margin_bound_real_ab} for Real AdaBoost or Eq.~\eqref{eq:adaboostr_normalized_margin_explicit} for AdaBoost$\mathbb R$, these give the sequence
\begin{equation}
    R_{S,\mu}
    \leq R_{S,\mu}^{\rm mar}
    \leq \mathcal B_T^{\exp}
    \leq \mathcal B_T^M.
    \label{eq:n25_bound_chain}
\end{equation}
All three inequalities hold numerically for all $221$ reconstructed models. Table~\ref{tab:n25_bound_stages} compares their values relative to the margin target.

\begin{table}[H]
    \centering
    \small
    \renewcommand{\arraystretch}{1.15}
    \setlength{\tabcolsep}{8pt}
    \begin{tabular}{@{}lrr@{}}
        \toprule
        Quantity divided by $\varepsilon_{\rm mar}^{(j)}$ & AdaBoost$\mathbb R$ & Real AdaBoost \\
        \midrule
        Ramp loss $R_{\rm ramp}$
            & $0.896$ & $0.927$ \\
        Margin error $R_{\rm margin}$
            & $1.227$ & $1.260$ \\
        Exponential expression $\mathcal B_T^{\exp}$
            & $6.033$ & $6.265$ \\
        Final edge expression $\mathcal B_T^M$
            & $6.943$ & $6.897$ \\
        \bottomrule
    \end{tabular}
    \caption{\textbf{Successive upper bounds at the observed margin target.}
    Each quantity is divided by its task's margin target before
    taking the median over $111$ AdaBoost$\mathbb R$ and $110$
    Real AdaBoost models, respectively. Values below one satisfy
    the target. The table includes all first-hit models from tasks with
    at least two successful repetitions out of three.}
    \label{tab:n25_bound_stages}
\end{table}

The increase from margin error to the exponential expression is the largest of the three increases in $206$ of the $221$ models. This comparison is made separately for each model, before summarizing the results. Most of this increase comes from examples whose margins already exceed $\mu$: their empirical margin loss and margin error are zero, but their exponential contribution remains positive. Thus, reaching the empirical margin target does not require the exponential expression to be small. This gap is already present before replacing the recorded edge sequence by its minimum.

The minimum-edge substitution adds further conservatism to the Real AdaBoost round count. When its contraction factor is close to one, the denominator in Eq.~\eqref{eq:n25_calculated_rounds} becomes small, explaining the sensitivity of the largest calculated counts. The calculated round count depends on how quickly the bound decreases. When $B_{\rm real}(\mu,\widehat\eta_{\min})$ is close to one, the theoretical decrease per round is small. A bound that is only several times above the target can therefore require tens or hundreds of times as many calculated rounds to fall below it.

A tighter treatment of the empirical margin loss or the distribution of margins could reduce the observed gap. A predictive round guarantee would additionally require control of how these quantities evolve during training and how finite-shot predictions affect the updates.

The ochre curve in Figure~\ref{fig:n25_boosting_rounds} relates this comparison to the sufficient resource condition $T_{\mathcal A}^{(\rm mar)}<\gamma^2$ in Eq.~\eqref{eq:simplified_resource_crossover}. In addition to the convergence assumptions, this comparison requires the average weak-learner cost through the sufficient round budget to be no greater than the reference cost, as in Assumption~B3 (Section~\ref{ass:B3}). None of the displayed calculated counts falls below $\gamma^2$, so this calculation does not certify the resource advantages observed in the experiments. Nevertheless, within the simulator-assisted protocol used here, the direct shot comparison in Section~\ref{sec:numerics_resources_performance} identifies settings where both boosting methods reach the same learning target at lower cost than Projected PEC. These observations support the central idea that combining noisy weak learners can require fewer measurements than compensating for noise through PEC's sampling overhead.

\FloatBarrier

\section{Discussion}
\label{sec:discussion}

A noisy quantum circuit may retain useful information for classification even when its expectation values differ substantially from their noiseless values. Earlier work on quantum ensembles has demonstrated that this information can be used to improve predictive accuracy~\cite{LiEtAl2024Ensemble,WangEtAl2024QuantumAdaBoost,ChenEtAl2025QuantumEnsemble}. Whether this also saves circuit executions is less immediate, since training and combining several classifiers introduces a cost of its own. Our analysis and experiments make this trade-off explicit: under the stated assumptions and in the regimes identified numerically, combining noisy classifiers can cost less than using PEC to reach the same learning target.

Boosting provides a concrete realization of this task-aware alternative. It does not require the individual noisy PQCs to approximate the corresponding ideal PQC accurately. What matters is whether they retain predictive information that can be combined across boosting rounds. This changes the relevant resource question from \emph{how accurately can the ideal quantum computation be reconstructed?} to \emph{how much task-relevant information remains accessible after noise?}

The resource comparison shows that the theoretical estimates of the required boosting rounds can be pessimistic. In Section~\ref{sec:numerics_boosting_trajectories}, evaluating the Real AdaBoost bound on the $57$ training histories included in the theoretical comparison gives a median of $120$ rounds, compared with $3$ observed rounds.

A first direction for further work is to obtain tighter bounds on the number of boosting rounds needed to reach the margin target. Section~\ref{sec:numerics_boosting_trajectories} identifies a source of overestimation: the exponential expression assigns positive contributions to examples that already satisfy the margin requirement and therefore contribute zero to the empirical margin loss. Using the distribution of margins more directly may give a tighter loss bound. To turn this improvement into a bound on the required number of rounds, one would also need to control how the margins change during boosting.

A complementary theoretical question is how useful lower bounds on the weak-learner edge can be derived directly from properties of a noisy PQC family. In the present analysis, the existence of a positive edge is a conditional weak-learning assumption.  A substantially stronger result would connect the edge to circuit architecture, physical noise, the data distribution, and the adaptive distributions generated during boosting. Together with a tighter loss bound, such a result could turn the present resource comparison into a predictive criterion rather than one that must be evaluated numerically.

A second question is whether the two strategies should really be treated as competitors. An intermediate regime may be more useful in practice: modest error mitigation could improve the quality of each noisy weak learner, while boosting could remove the need to mitigate it to high precision.  Determining the optimal amount of mitigation per weak learner would define a continuous trade-off between mitigation cost and ensemble complexity. A further direction is to combine boosting with adaptive shot allocation to concentrate measurements on the examples or rounds where additional precision is most valuable. The present results concern four-qubit simulations with gate-local depolarizing noise and simulator-assisted gradients. Establishing the same resource comparison on hardware would require accounting for gradient estimation, noise characterization, and measurement errors.

A third direction is to identify other application objectives for which exact observable reconstruction is stronger than necessary. Classification is a particularly clear example, but similar opportunities may arise for ranking, threshold detection, hypothesis testing, control decisions, or other tasks whose output depends only on a coarse property of a quantum observable. The relevant question is then whether task-aware post-processing can exploit information that remains useful even after the underlying quantum state or expectation value has been substantially distorted.

Another interesting question is whether boosting can remain effective in regimes where noisy PQCs become increasingly difficult to optimize as strong classifiers but still retain sufficient trainability to produce weak hypotheses with positive edge. This distinction may be relevant to barren-plateau phenomena: vanishing gradients can make high-accuracy optimization increasingly costly~\cite{McClean2018Barren,Wang2021NoiseBarren}, whereas boosting requires each constituent PQC only to achieve a weak-learning advantage. Establishing whether an intermediate regime of weak trainability persists under barren-plateau-inducing circuit families and noise models requires a dedicated scaling analysis and is left for future work.

Overall, the present work establishes a resource comparison between two ways of using noisy quantum circuits. A stronger next step would combine task-aware robustness to quantum noise with a learning problem for which the relevant quantum features are also difficult to reproduce classically. Finding a regime in which the task is simultaneously classically hard, quantumly learnable, and robust enough for noisy weak learners would connect the resource advantage studied here to a more complete notion of practical quantum learning advantage.

\section{Conclusions}
\label{sec:conclusion}

This work reframes the treatment of noise in quantum machine learning as a task-dependent resource problem. For supervised classification, the relevant goal is not necessarily to reconstruct the output of an ideal quantum circuit, but to preserve enough information to make reliable predictions. This distinction motivates boosting noisy PQCs as an alternative strategy: rather than correcting each circuit toward its noiseless counterpart, one can combine imperfect quantum predictors at the level of the learning task.

Our theoretical framework makes this comparison explicit through a common empirical margin loss criterion and a shared generalization bound under the stated assumptions. For PEC, the chosen quasiprobability representation determines the sampling overhead, while the boosting resource depends on the weak-learner edges and the cost of constructing the ensemble. The finite-shot numerical experiments identify a depth--noise trend that generally shifts the resource comparison in favour of boosting relative to Projected PEC. Within identified regimes, AdaBoost$\mathbb{R}$ and Real AdaBoost can reach the same task-specific target using substantially fewer algorithmic circuit executions.

The resource advantage is accompanied by favourable observed test performance. For matched classification, all $22/22$ method--depth--noise cells supported by at least two qualifying teachers have positive teacher-paired median boosting-minus-reference test accuracy differences, with a median across these cells of $+2.15$ percentage points. For empirical margin loss, all $14/14$ cells meeting this reporting requirement have positive differences, with a median across these cells of $+3.84$ percentage points. The reference predictors provide the performance associated with Projected PEC. Thus, in the cells where boosting requires fewer resources, its resource advantage is accompanied by higher observed teacher-paired median test accuracy, rather than obtained by accepting worse predictive performance. These findings concern successful-run summaries on the comparable teacher tasks; they do not establish lower expected risk for every boosting model.

The explicitly sampled PEC experiments suggest that Projected PEC provides a useful approximation to the scale of the training resources required in the tested settings. We also examined whether using fewer shots per prediction than suggested by the PEC variance bound could reduce the total cost while preserving reliable target attainment. The smaller shot allocations still sometimes reached the learning targets, but were less stable, and successful training often required more resources than projected. In this comparison, boosting retained both lower observed training costs and more consistent target attainment.

The broader lesson is that quantum error mitigation should not always be treated as a mandatory preprocessing step before learning. When the application requires only a task-specific property of the quantum output, there may be substantial value in designing the learning procedure around the information that survives the noise rather than reconstructing information that the task ultimately does not use. The present resource results remain conditional on the simulator-assisted training protocol and the stated measurement-accounting convention. Establishing when this principle extends to hardware-scale problems, more realistic noise, and learning tasks with an independently established quantum-over-classical advantage is the natural next step.

\section*{Acknowledgments}

V.D. acknowledges the support of the Dutch Research Council (NWO) through the NWO/NWA project Divide and Quantum (Project No. 1389.20.241). V.D. also acknowledges the support of the European Union’s Horizon Europe program through the ERC CoG BeMAIQuantum (Grant No. 101124342). This work was also supported by the Dutch National Growth Fund (NGF) as part of the QDNL program.

This publication is part of the “Quantum Inspire–the Dutch Quantum Computer in the Cloud” project (Project No. NWA.1292.19.194) of the NWA research program “Research on Routes by Consortia (ORC),” which is funded by the Netherlands Organization for Scientific Research (NWO).

The authors used ChatGPT and Codex (OpenAI; [GPT-5.6 Sol]) to assist with drafting and revising the manuscript, and with developing code for the numerical experiments and data analysis. Access to ChatGPT was provided through OpenAI's ChatGPT for Academic Researchers programme.

\clearpage
\bibliographystyle{alpha}
\bibliography{sample}

\clearpage
\appendix

\section{Hypothesis class complexity under Pauli noise}
\label{app:pauli_noise_complexity}

In this appendix, we establish a common upper bound on the Rademacher-complexity contribution for noiseless PQC mean predictors, exact PEC mean predictors and normalized ensembles of noisy PQC mean predictors. Under input-independent Pauli noise and the circuit-parameter assumptions stated below, the noisy mean-prediction class is contained in the convex hull of the corresponding noiseless class. This permits the strategies to be compared using the same reference complexity term; it does not require their actual hypothesis-class complexities to be equal.

\begin{figure}
\centering
\includegraphics[width=0.5\linewidth]{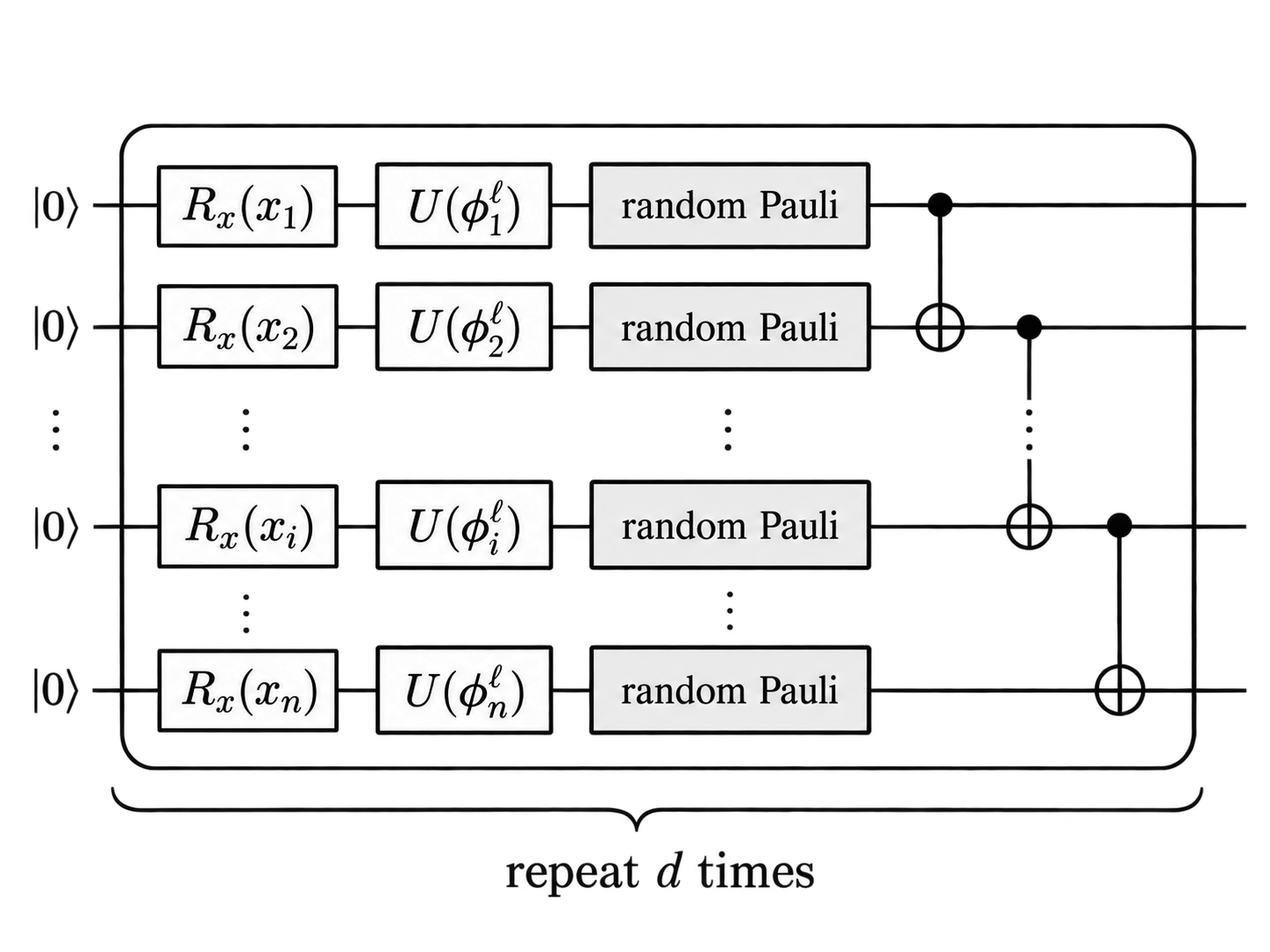}
\caption{\label{fig:circuit}Circuit structure for deriving the Rademacher complexity of a noisy PQC.}
\end{figure}

Consider a PQC architecture consisting of interleaving layers of single-qubit data-encoding rotations and trainable single-qubit unitaries of the form
\begin{equation}
U(\phi_1,\phi_2,\phi_3) = R_z(\phi_1) R_y(\phi_2) R_z(\phi_3),
\label{eq:single_qubit_universal}
\end{equation} possibly interspersed with fixed entangling gates from the Clifford group (e.g., CNOT or CZ) as in Fig. \ref{fig:circuit}.
Such architectures are standard in variational quantum algorithms and are universal on each single-qubit subsystem up to a global phase.

For the following argument, the trainable single-qubit blocks have independent parameters and can each realize an arbitrary single-qubit unitary up to global phase. The data-encoding gates are $R_x(x_q\lambda_{\ell,q})$, and the allowed parameter values permit each replacement $\lambda_{\ell,q}\mapsto-\lambda_{\ell,q}$ independently of the other encoding gates. We also assume a final layer of independently trainable single-qubit unitaries on every qubit, after the last data-encoding and entangling operations. These conditions ensure that the required reparameterizations remain within the same noiseless hypothesis class.

Assume that each noisy circuit location acts on a support $S_\ell$ of size $k_\ell\in\{1,2\}$ and is followed by a Pauli channel of the form
\begin{equation}
    \mathcal N_\ell(\rho)
    =
    \sum_{P\in\mathcal P_{S_\ell}}
    p_{\ell,P}\,
    P\rho P,
    \qquad
    p_{\ell,P}\geq 0,
    \qquad
    \sum_{P\in\mathcal P_{S_\ell}}p_{\ell,P}=1,
    \label{eq:appendix_pauli_channel}
\end{equation}
where $\mathcal P_{S_\ell}=\{I,X,Y,Z\}^{\otimes k_\ell}$ denotes the Pauli operators supported on the qubits acted on at location $\ell$. Each Pauli operator is extended by the identity on qubits outside $S_\ell$. The channel probabilities are fixed independently of the input $x$. Consequently, expanding the channel sequence gives mixture weights that are the same for every input.

The gate-local depolarizing channels used in the numerical experiments are a special case of Eq.~\eqref{eq:appendix_pauli_channel}.  Indeed, for a support $S$ of size $k$,
\begin{equation}
    \mathcal D_{S,p}(\rho)
    =
    (1-p)\rho
    +
    \frac{p}{4^k}
    \sum_{P\in\mathcal P_S}
    P\rho P ,
    \label{eq:appendix_depolarizing_pauli}
\end{equation}
which is equivalent to replacing the subsystem on $S$ by the maximally mixed state with probability $p$.

Fix an initial state $\rho_0$ and a Hermitian observable $O$ with $\lVert O\rVert_\infty\leq1$. Let $\mathcal H$ denote the class of noiseless expectation-value functions obtained by varying the admissible circuit parameters, with the architecture, initial state and observable held fixed. For a fixed parameter vector $\theta$, expanding all channels in Eq.~\eqref{eq:appendix_pauli_channel} gives
\begin{equation}
    h_{\rm noisy}(x)
    =
    \sum_{\omega}
    \pi(\omega)\,
    h_{\omega}(x),
    \qquad
    \pi(\omega)\geq 0,
    \qquad
    \sum_{\omega}\pi(\omega)=1,
    \label{eq:appendix_pauli_trajectories}
\end{equation}
where $\omega$ specifies one Pauli insertion at every noisy circuit location and $h_\omega$ is the noiseless predictor obtained for that fixed Pauli-insertion trajectory. For $\omega=(P_1,\ldots,P_G)$, where $G$ is the number of noisy gate locations, the weight is $\pi(\omega)=\prod_{g=1}^{G}p_{g,P_g}$. In particular, $\pi(\omega)$ does not depend on $x$.

It remains to show that these Pauli insertions can be absorbed into the parameters of the same noiseless circuit architecture.  Every two-qubit Pauli operator is a tensor product of single-qubit Paulis, and the controlled-$Z$ entanglers used here are Clifford gates.  Hence conjugating a Pauli string through the entangling layer produces another Pauli string.

For the data-reuploading rotations, a single-qubit Pauli satisfies
\begin{equation}
    P R_x(\alpha)
    =
    R_x(s_P\alpha)P,
    \qquad
    s_P\in\{+1,-1\},
    \label{eq:appendix_pauli_rx}
\end{equation}
up to an irrelevant global phase.  Since the data angle is $\alpha=x_q\lambda_{\ell,q}$ and the coefficients $\lambda_{\ell,q}$ are trainable, the possible sign change in Eq.~\eqref{eq:appendix_pauli_rx} can be absorbed by replacing $\lambda_{\ell,q}$ with $-\lambda_{\ell,q}$.

The remaining local Pauli factors can be absorbed into the trainable single-qubit blocks. An incoming Pauli multiplied into such a block produces another single-qubit unitary, which can be represented by a new choice of that block's parameters. Pauli insertions between the component rotations of a trainable block can be treated in the same way, since the complete product is still a single-qubit unitary. The final trainable layer absorbs any remaining local Pauli factors without changing the measured observable.

Consequently, for every fixed original parameter vector $\theta$ and Pauli trajectory $\omega$, there is an admissible parameter vector $\theta_\omega$, independent of $x$, such that
\begin{equation}
    h_\omega(x)=h_{\theta_\omega,0}(x)
    \qquad\text{for every }x.
\end{equation}
Here, the subscript $0$ denotes evaluation without physical noise. Thus $h_\omega\in\mathcal H$ for every trajectory appearing in Eq.~\eqref{eq:appendix_pauli_trajectories}.

It follows that the hypothesis class induced by noisy PQCs satisfies
\begin{equation}
\mathcal{H}_{\mathrm{noisy}} \subseteq \mathrm{conv}(\mathcal{H}),
\label{eq:noisy_subset_convex}
\end{equation}
where $\mathrm{conv}(\mathcal{H})$ denotes the convex hull of the noiseless hypothesis class.

Rademacher complexity is monotone under set inclusion and invariant under taking convex hulls. Therefore, for any sample size $m$,
\begin{equation}
\mathfrak{R}_m(\mathcal{H}_{\mathrm{noisy}})
\;\le\;
\mathfrak{R}_m(\mathrm{conv}(\mathcal{H}))
\;=\;
\mathfrak{R}_m(\mathcal{H})
\label{eq:radamacher_bound_pauli_noise}
\end{equation}

Equation~\eqref{eq:radamacher_bound_pauli_noise} establishes an upper bound on the complexity of the noisy mean-prediction class; it does not assert equality with the noiseless complexity.

For boosting, we also account for the classical sign reversal of weak learners. Using the signed class $\mathcal H_\pm=\mathcal H\cup(-\mathcal H)$ defined in Eq.~\eqref{eq:signed_base_class}, the normalized mean ensemble satisfies
\begin{equation}
    \widetilde F_T
    \in
    \operatorname{conv}
    \left(
        \mathcal H_{\rm noisy}\cup(-\mathcal H_{\rm noisy})
    \right)
    \subseteq
    \operatorname{conv}(\mathcal H_\pm).
\end{equation}
The inclusion follows because each signed noisy predictor is a convex combination of signed noiseless predictors.

For PEC with an exact quasiprobability representation, the mean prediction equals the corresponding noiseless expectation, so its mean-prediction class is $\mathcal H$. Therefore noiseless predictors, PEC mean predictors and normalized noisy boosting ensembles can all use the common complexity upper bound $\mathfrak R_m(\mathcal H_\pm)$ in the margin-based generalization analysis. This is a shared upper bound, not a claim that the strategies have identical actual complexities or expected risks.

These statements concern mean prediction functions. Finite-shot prediction scores, including the reweighted Sampled-PEC estimates, require the separate measurement-error analysis described in the main text.

This argument relies on input-independent Pauli noise, Clifford entanglers and the parameter freedom specified above. It does not establish the same containment for general noise channels, such as amplitude damping, or for architectures whose parameter constraints prevent the required reparameterizations. Other noise models must be assessed separately. Some coherent errors can already be absorbed when they act directly next to sufficiently expressive trainable blocks, but this does not imply the same result for arbitrary coherent errors throughout the circuit.

\section{Teacher selection and datasets}
\label{app:teacher_task_details}

We first check that each candidate teacher produces scores that vary across inputs and that both labels occur sufficiently often. We then check whether students can learn the resulting task. At each depth, candidates are generated in a fixed order of random seeds. Their rotation angles and input scales are drawn independently from a normal distribution with mean zero and standard deviation one.

For each candidate, we evaluate the exact noiseless teacher score on a fixed set of $M=58{,}880$ inputs in $[0,2\pi]^4$. These inputs are generated by Latin hypercube sampling, which spreads the sampled values across the range of each input coordinate. Writing $s_i$ for the score at the $i$th input, we require its standard deviation over these inputs to be at least $0.18$:
\begin{equation}
    \sqrt{\frac{1}{M}\sum_{i=1}^{M}(s_i-\bar{s})^2}\ge0.18,
    \qquad
    \bar{s}=\frac{1}{M}\sum_{i=1}^{M}s_i.
    \label{eq:teacher_std_screen}
\end{equation}
At least $10\%$ of the scores must be below $-0.2$, and at least $10\%$ must be greater than or equal to $0.2$. No more than $75\%$ may lie in the interval $[-0.2,0.2)$. We also require the fifth percentile to be below $-0.2$ and the $95$th percentile to be above $0.2$, using linear interpolation to calculate the percentiles. There must be at least $1472$ distinct inputs with scores below $-0.2$ and at least $1472$ with scores greater than or equal to $0.2$. Together, these conditions exclude teachers with nearly constant scores and require enough examples of both labels to remain after removing scores close to zero. They provide an initial check that the teacher defines a useful classification task.

For each teacher passing these checks, we complete five preliminary student training runs. Training stops after $20$ consecutive epochs without an improvement in the average training loss (patience criteria), or after maximum training limit of $500$ epochs. We retain the parameters from the epoch with the lowest average training loss and evaluate accuracy on the full validation set using sampled circuit outputs. The median accuracy across the five runs must be at least $90\%$ at depths $1$ and $2$, $80\%$ at depths $3$ and $4$, and $75\%$ at depth $5$. These preliminary student runs use $1000$ measurement shots per circuit and input, with depolarizing noise strengths $0.001$ after one-qubit gates and $0.01$ after two-qubit gates. These runs are separate from the comparison experiments and are excluded from their reported training costs. At each depth, we keep the three candidates that pass all checks. This choice is made before the methods comparison runs.

\begin{figure*}[htbp]
    \centering
    \includegraphics[width=0.7\textwidth]{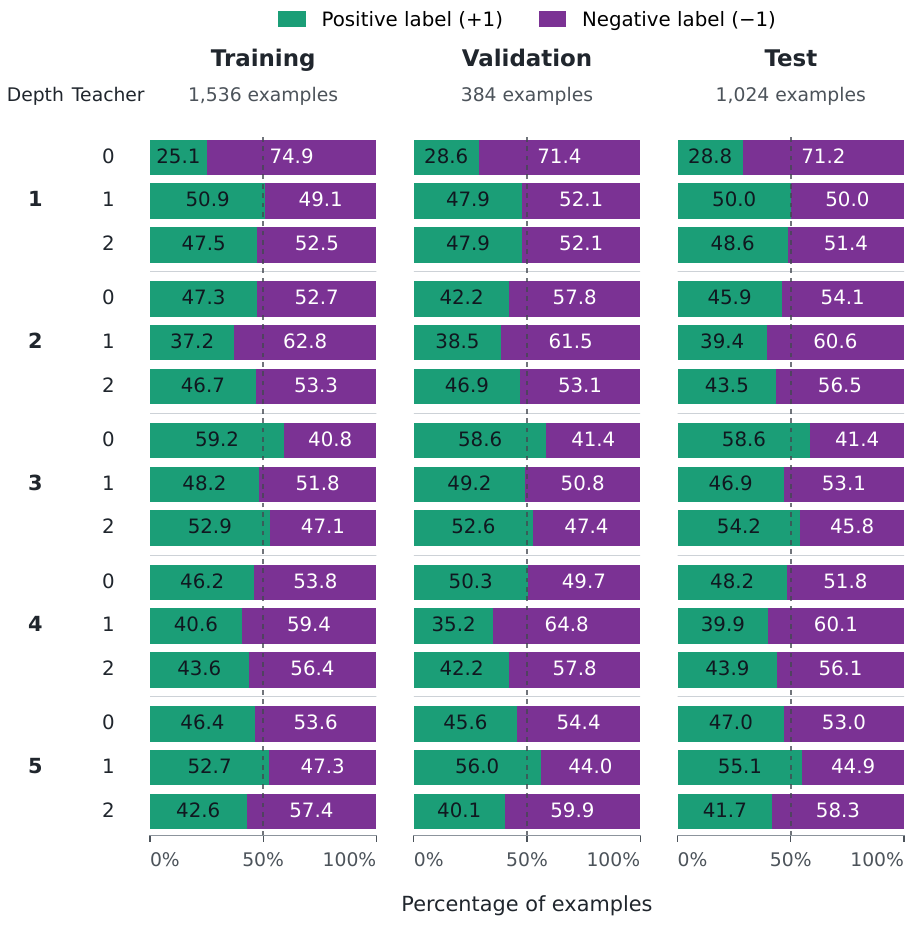}
    \caption{\textbf{Class proportions in the teacher--student datasets.}}
    \label{fig:teacher_class_balance}
\end{figure*}

For the comparison, we generate new training, validation, and test examples. We draw inputs uniformly from $[0,2\pi]^4$ and keep those with $|s_j(x)|\ge0.2$.  Each dataset uses independent random draws, separate from those used to check the teachers. We do not force equal class sizes or remove repeated inputs. The larger class contains between $50.91\%$ and $74.87\%$ of the training examples, depending on the teacher. Figure~\ref{fig:teacher_class_balance} shows the proportions of both classes in the training, validation, and test sets.

\section{Details of circuit training}
\label{app:numerics_gradient_details}

The loss is evaluated at a sampled prediction, while the derivative of the underlying circuit mean is obtained from the simulator. Let $m_\theta(x)=h_{\theta,0}(x)$ for the reference and sampled PEC, and $m_\theta(x)=h_{\theta,p}(x)$ for a noisy boosting weak learner. For a minibatch $\mathcal B$, the quantity supplied to gradient-norm clipping and Adam is
\begin{equation}
    g_{\mathcal B}
      =\sum_{i\in\mathcal B}
       w_{i,\mathcal B}\,y_i\,
       \Psi_G'\!\left(y_i\widehat h_i;0.5\right)
       \nabla_\theta m_\theta(x_i).
    \label{eq:straight_through_gradient}
\end{equation}
Here $\widehat h_i$ is the sampled prediction and $w_{i,\mathcal B}$ is the weight multiplying its minibatch loss. The reference and sampled PEC use $w_{i,\mathcal B}=1/|\mathcal B|$. For boosting, we first convert the incoming training weights to 32-bit floating-point numbers, replace values below $2^{-52}$ by $2^{-52}$, and divide all weights by their sum over the training set. After selecting a minibatch, we apply the same lower limit again and divide the selected weights by their minibatch sum. These calculations use 32-bit floating-point arithmetic, in which $2^{-52}$ is exactly representable. The loss derivative is $\Psi_G'(a;\sigma)=-\Phi_{\mathcal N}((1-a)/\sigma)$.

Within each weak-learner epoch, each minibatch loss is evaluated before its corresponding parameter update. When the mean of the $24$ losses is strictly lower than its best previous value, we save the parameters and optimizer state after the final update of that epoch. This mean therefore contains losses evaluated at successive parameter values during the epoch.

In this derivative rule, the sampled outcomes are held fixed and the score before clipping is assigned the derivative of $m_\theta(x_i)$. This is commonly called a straight-through derivative convention. In general, it gives neither the derivative of the loss evaluated at the exact mean nor an unbiased gradient of the expected finite-shot loss. It therefore specifies a simulator-assisted training procedure, rather than a measured-gradient implementation on quantum hardware.

\section{Construction of the learning targets}
\label{app:numerics_targets}
\label{app:classification_targets}

For each teacher task, the margin and classification targets are constructed from two separate sets of five preliminary circuit fits. These fits use the same training and validation splits as the main comparison. Both procedures use $1000$ forward shots per circuit and input, minibatches of $64$ examples, Adam with learning rate $0.01$, gradient clipping at Euclidean norm one, and the smooth hinge loss with $\sigma=0.5$. The maximum fitting length is $500$ epochs. Their stopping and target-evaluation procedures differ as described below.

The circuits used to define the margin target are fitted under gate-local physical noise $p_{\rm 1q}=0.001$ and $p_{\rm 2q}=0.01$. Fitting stops after $20$ consecutive epochs without a strict improvement in the mean minibatch fitting loss, and the parameters saved at the best epoch are restored. We then evaluate each saved circuit on the $1536$ training examples using exact noiseless circuit means. The target $\varepsilon_{\rm mar}^{(d,j)}$ is the median of the five resulting unweighted empirical margin losses at $\mu=0.1$. Thus, physical noise enters the fitting of these calibration models, while their contributions to the margin target are evaluated without physical noise or measurement sampling.

The classification target uses five separate reference fits with zero physical gate noise. During these fits, the training-margin target is checked after each optimizer update. Each fit contributes its first confirmed margin-target model. If the margin target is not reached, it instead contributes the terminal parameter state at the $500$-epoch limit. We evaluate each selected model on the $384$ validation examples using a fresh sampled prediction pass with the shot allocation stated above. All five validation accuracies enter the median, including those from fits that did not reach the margin target. These accuracies are calculated from sampled predictions, whereas the margin-target calibration uses exact circuit means.

For each task, the classification target $a^\star_{d,j}$ is the median validation accuracy of the five reference models. A model meets the classification criterion when its validation accuracy is at least this value. Table~\ref{tab:numerics_learning_targets} reports the required validation accuracies as percentages, together with the margin-loss targets. The calibration models are separate from the $25$-shot training runs used to measure resource costs, and the targets remain fixed throughout those comparisons.
\begin{table}[htbp]
    \centering
    \small
    \setlength{\tabcolsep}{8pt}
    \renewcommand{\arraystretch}{1.15}
    \begin{tabular}{rr r r r}
        \toprule
        Depth $d$ &
        Teacher $j$ &
        \shortstack{Margin-loss\\threshold} &
        \shortstack{Required accuracy\\(\%, approximate)} \\
        \midrule
        1 & 0 & 0.021365 & 98.96 \\
        1 & 1 & 0.123781 & 88.28 \\
        1 & 2 & 0.057489 & 95.05 \\
        \addlinespace[3pt]
        2 & 0 & 0.024486 & 97.40 \\
        2 & 1 & 0.053064 & 92.19 \\
        2 & 2 & 0.139771 & 88.80 \\
        \addlinespace[3pt]
        3 & 0 & 0.187082 & 84.11 \\
        3 & 1 & 0.192292 & 83.07 \\
        3 & 2 & 0.233321 & 77.08 \\
        \addlinespace[3pt]
        4 & 0 & 0.201659 & 83.33 \\
        4 & 1 & 0.218824 & 80.47 \\
        4 & 2 & 0.160409 & 85.42 \\
        \addlinespace[3pt]
        5 & 0 & 0.199141 & 80.21 \\
        5 & 1 & 0.260832 & 77.86 \\
        5 & 2 & 0.089587 & 90.63 \\
        \bottomrule
    \end{tabular}
    \caption{\textbf{Learning targets for the main comparison.}
    Teacher indices are $0,1,2$ within each depth.
    The training-margin target uses $\mu=0.1$.}
    \label{tab:numerics_learning_targets}
    \label{tab:appendix-classification-targets}
\end{table}
\end{document}